\documentclass[twocolumn]{aa} % for a paper on 2 column -> from the original file
\usepackage{graphicx}
\usepackage[varg]{txfonts}
\usepackage{color}
\usepackage{xcolor}
\usepackage[colorlinks=black, linkcolor=black, citecolor=black]{hyperref} % color for links, and references
\usepackage{booktabs}      % to be able to use commands toprule, midrule and bottomrule in the tables, so tables look better
\usepackage{afterpage}     % to use in combination con float to force a position after Here, but after the page is finished
\usepackage{lscape}        % to set landscape for figures and tables 
\usepackage{pifont}        % for empty start
\usepackage{eqnarray,amsmath}
\usepackage{rotating}

\begin{document} 
 
%%%%%%%%%%%%%%%
% DEFINITIONS %  
%%%%%%%%%%%%%%%
\newcommand{\emptystar}{\ding{73}}
\newcommand{\filledstar}{\ding{72}}

%%%%%%%%%
% UNITS %
%%%%%%%%%
\def\KMS{km s$^{-1}$}
\def\VLSR{V$_{lsr}$}
\def\DELTAV{$\Delta$V(FWHM)}
\def\INTEG{$I$}
\def\TA{T$_{A}^{*}$}
\def\TAA{T$_{A}^{'}$}
\def\TPEAK{T$_{peak}$}
\def\TPEAKK{T$_{peak}^{mb}$}
\def\TKIN{T$_{kin}$}
\def\TEX{T$_{ex}$}
\def\DEG{$^\circ$}
\def\TREC{T$_{REC}$}
\def\TRMS{T$_{A,rms}^{*}$}
\def\TRMSS{T$_{rms}$}
\def\TRMSSS{T$_{rms}^{mb}$}
\def\TMB{T$_{mb}$}
\def\RATIO{R$_{i/CO(4-3)}$}
\def\SOLAR{$_{\odot}$}
\def\HII{\mbox{H \footnotesize{II}}\normalsize}
\def\OABUNDANCE{\mbox{[$^{18}$O/$^{17}$O]}}
\def\OABUNDANCEAVG{\mbox{$\overline{[^{18}\rm{O}/^{17}\rm{O}]}$}}
\def\NABUNDANCE{\mbox{[C/N]}}

%\rm{}

%%%%%%%%%%%%%%%%%%%%%%
%%%%%% MOLECULES %%%%%
%%%%%%%%%%%%%%%%%%%%%%
% MAIN MOLECULES %
\def\COLINEAI{\mbox{CO(1-0)}}
\def\COLINEAII{\mbox{CO(2-1)}}
\def\COLINEAIII{\mbox{CO(3-2)}}
\def\COLINEAIV{\mbox{CO(4-3)}}
\def\COLINEAV{\mbox{CO(5-4)}}
\def\COLINEAVI{\mbox{CO(6-5)}}
\def\COLINEAVII{\mbox{CO(7-6)}}
\def\HCNLINEA{\mbox{HCN(1-0)}}
\def\HCOLINEA{\mbox{HCO$^{+}$(1-0)}}
% ISOTOPOLOGUES %
\def\COOLINEAI{\mbox{C$^{18}$O(1-0)}}
\def\COOLINEAII{\mbox{C$^{18}$O(2-1)}}
\def\COOOLINEAI{\mbox{C$^{17}$O(1-0)}}
\def\CCOLINEAI{\mbox{$^{13}$CO(1-0)}}
\def\HCCNLINEA{\mbox{H$^{13}$CN(1-0)}}
\def\HCCOLINEA{\mbox{H$^{13}$CO$^{+}$(1-0)}}
% OTHERS %
\def\METHANOL{\mbox{CH$_{3}$OH}}
\def\WATER{\mbox{H$_{2}$O}}
\def\AMONIA{\mbox{NH$_{3}$(1,1)}}
\def\DIAZENYLIUM{\mbox{N$_{2}$H$^{+}$(1-0)}}
\def\CS{\mbox{CS(2-1)}}
% QM NOTATION %
\def\COLINEAIl{\mbox{CO(J$=$1-0)}}
\def\COLINEAIIl{\mbox{CO(J$=$2-1)}}
\def\COLINEAIIIl{\mbox{CO(J$=$3-2)}}
\def\COLINEAIVl{\mbox{CO(J$=$4-3)}}
\def\COLINEAVl{\mbox{CO(J$=$5-4)}}
\def\COLINEAVIl{\mbox{CO(J$=$6-5)}}
\def\COLINEAVIIl{\mbox{CO(J$=$7-6)}}
\def\HCNLINEAl{\mbox{HCN(J$=$1-0)}}
\def\HCOLINEAl{\mbox{HCO$^{+}$(J$=$1-0)}}
\def\CCOLINEAIl{\mbox{$^{13}$CO(J$=$1-0)}}
\def\COOLINEAIl{\mbox{C$^{18}$O(J$=$1-0)}}
\def\HCCNLINEAl{\mbox{H$^{13}$CN(J$=$1-0)}}
\def\HCCOLINEAl{\mbox{H$^{13}$CO$^{+}$(J$=$1-0)}}

%%%%%%%%%%%%%%%
%%%%%% ATOMS %%%%
%%%%%%%%%%%%%%%
\def\CILINEAI{\mbox{[CI](1-0)}}
\def\CILINEAII{\mbox{[CI](2-1)}}
\def\NII{\mbox{[NII]}}
\def\CII{\mbox{[CII]}}
\def\CCII{\mbox{[${^{13}}$CII]}}

% QM NOTATION %
\def\CILINEAIl{\mbox{[C\footnotesize{I}\normalsize] ${^{3}}$P$_{1}$ - ${^{3}}$P$_{0}$}}
\def\CILINEAIIl{\mbox{[C\footnotesize{I}\normalsize] ${^{3}}$P$_{2}$ - ${^{3}}$P$_{1}$}}
\def\NIIl{\mbox{[N\footnotesize{II}\normalsize] ${^{3}}$P$_{1}$ - ${^{3}}$P$_{0}$}}
\def\CIIl{\mbox{[C\footnotesize{II}\normalsize] ${^{2}}$P$_{3/2}$ - ${^{2}}$P$_{1/2}$}}
\def\CCIIl{\mbox{[$^{13}$C\footnotesize{II}\normalsize] F $=$ 2$\rightarrow$1, 1$\rightarrow$0, 1$\rightarrow$1}}

\title{Warm ISM in the Sgr A Complex. II.}
\subtitle{The [C/N] abundance ratio traced by [CII] 158 $\mu m$ and [NII] 205 $\mu m$ observations toward the Arched Filaments at the Galactic center\footnote{Partially based on observations carried out under project number 020-16 with the IRAM 30m Telescope. IRAM is supported by INSU/CNRS (France), MPG (Germany) and IGN (Spain).}}

\author{P. Garc\'ia \inst{1,2}
\and
N. Abel\inst{3}
\and 
M. R\"ollig\inst{4}
\and 
R. Simon\inst{4}
\and
J. Stutzki\inst{4}
}
\institute{Chinese Academy of Sciences South America Center for Astronomy, National Astronomical Observatories, CAS, 
Beijing 100101, China.
\email{pablo.garcia@nao.cas.cn}
\and
Instituto de Astronom\'ia, Universidad Cat\'olica del Norte, Av. Angamos 0610, Antofagasta 1270709, Chile.
\and
University of Cincinnati, Clermont College, Batavia, OH 45103, USA.
\and
I. Physikalisches Institut der Universit\"at zu K\"oln, D-50937 Cologne, Germany.
}

\date{Received xxxxxx xx, XXXX; Accepted xxxxxx xx, XXXX}

\abstract
    {The Arches Cluster - Arched Filaments (AF) system is our Galaxy's prime example of the complexity involved in the interaction between the strong radiation field of numerous OB 
    stars and their surrounding ISM in extremely harsh environments such as the Galactic center (GC) of the Milky Way. It offers a unique opportunity to study the close relationship 
    between photon-dominated regions (PDRs) and \HII\textrm{ }regions and their relative contributions to the observed \CII\textrm{ }emission.}
    {We aim to investigate the I(\CII) versus I(\NII) integrated intensity behavior in the AF region in order to assess the \CII\textrm{ }emission contribution from the \HII\textrm{ }region, 
    which is traced by \NII\textrm{ }line observations, and PDR components in the high-metallicity environment of the GC.}
    {We used [CII] 158 $\mu m$ and [NII] 205 $\mu m$ fine structure line observations of the AF in the literature to compare their observational integrated intensity distribution 
    to semi-theoretical predictions for the contribution of \HII\textrm{ }regions and adjacent PDRs to the observed \CII\textrm{ }emission. We explored variations in the [C/N] elemental abundance 
    ratio to explain the overall behavior of the observed relationship. Based on our models, the \HII\textrm{ }region and PDR contributions to the observed \CII\textrm{ }emission is calculated 
    for a few positions within and near to the AF. Estimates for the [C/N] abundance ratio and [N/H] nitrogen elemental abundance in the AF can then be derived.}
    {The behavior of the I(\CII) versus I(\NII) relationship in the AF can be explained by model results satisfying 0.84 $<$ [C/N]$_{\textrm{AF}}$ $<$ 1.41, with model metallicities ranging from 
    1 Z\SOLAR\textrm{ }to 2 Z\SOLAR, hydrogen volume density $\log$ n(H) $=$ 3.5, and ionization parameters $\log$ U from $-$1 to $-$2. A least-squares fit to the model data points yields 
    $\log$ I(\CII) $=$ 1.068$\times \log$ I(\NII) $+$ 0.645 to predict the \CII\textrm{ }emission arising from the \HII\textrm{ }regions in the AF. The fraction of the total observed \CII\textrm{ }emission 
    arising from within PDRs varies between $\sim$ 0.20 and $\sim$ 0.75. Our results yield average values for the carbon-to-nitrogen ratio and nitrogen elemental abundances of 
    [C/N]$_{\textrm{AF}}$ $=$ 1.13 $\pm$ 0.09 and [N/H]$_{\textrm{AF}}$ $=$ 6.21$\times$10$^{-4}$ for the AF, respectively. They are a factor of $\sim$ 0.4 smaller and $\sim$ 7.5 larger than 
    their corresponding Galactic disk values.}
    {The large spatial variation of the fraction of \CII\textrm{ }emission arising either from \HII\textrm{ }regions or PDRs suggests that both contributions must be disentangled before any modeling 
    attempt is made to explain the observed \CII\textrm{ }emission in the AF. We suggest that secondary production of nitrogen from low- to intermediate-mass stars in the Galactic bulge is a 
    plausible mechanism to explain the large abundance differences between the GC and the Galactic disk. The mass loss of such stars would enrich the GC ISM with nitrogen as the gas falls 
    into the inner GC orbits where the AF are located. Overall, our results show that tight constraints are needed on the [C/N] abundance ratio for the GC, significantly tighter than 
    previous abundance measurements have discerned.} 


   \maketitle

\section{Introduction}\label{intro}

The spatial locations of photon-dominated regions (PDRs) - which are defined as regions where the far-ultraviolet field (FUV) has photon energies ranging between 6.0 eV $<$ $h\nu$ $<$ 13.6 eV \citep{roellig2013} - and \HII\textrm{ }regions surrounding stellar nurseries with photon energies 13.6 eV $\leq$ $h\nu$ \citep{draine2011} are closely related. As a consequence, and given the low ionization potential of neutral carbon of $\sim$ 11.2 eV, a fraction of the observed \CII\textrm{ }158 $\mu m$ emission does not originate within PDRs but arises from their adjacent \HII\textrm{ }regions instead, where ionized carbon can be collisionally excited by electrons \citep{goldsmith2012}. Moreover, \citet{abel2006a} showed that between 10\% and 60\% of the observed \CII\textrm{ }158 $\mu m$ emission can arise from \HII\textrm{ }regions. This can significantly alter the derived energy balance in PDRs by attributing more energy cooling to them via the \CII\textrm{ }158 $\mu m$ line than what they truly emit. More recently, \citet{goldsmith2015} showed that, along several lines of sight (LOS) in the Galactic plane, velocity-resolved \NII\textrm{ }205 $\mu m$ line emission is highly correlated with the \CII\textrm{ }158 $\mu m$ line emission in the sense that any significant detection of the former is always accompanied by a detection of the latter, though relative line intensities can vary dramatically. This strongly suggests a physical connection between the observed emission of both species on large scales.\\

The contribution of a \HII\textrm{ }region to the observed \CII\textrm{ }emission toward PDRs can be traced by \NII\textrm{ }line observations at 205 $\mu m$, as shown by semi-theoretical work in the literature \citep{heiles1994,abel2006a}. Given the large nitrogen ionization potential  ($\sim$ 14.6 eV), \NII\textrm{ }emission can only originate from within \HII\textrm{ }regions, removing the origin ambiguity present in \CII\textrm{ }line observations. On the other hand, the correlation between the \CIIl\textrm{ }and \NIIl\textrm{ }fine structure line transitions within \HII\textrm{ }regions allows us to estimate the \CII\textrm{ }emission contribution of \HII\textrm{ }regions to the total observed emission, assuming both transitions emit at the same radial velocities. The intensity of the \CII\textrm{ }line inside a  \HII\textrm{ }region is mainly influenced by the electron density n$_{e,}$ which controls the excitation of ionized carbon, the size of the \HII\textrm{ }region with respect to the PDR's size constraining the volume extent from within which \CII\textrm{ }emission can arise, and the intensity of the stellar continuum controlling the [CIII] $\rightarrow$ \CII\textrm{ }line transition \citep{abel2006a}. Therefore, a semiempirical relationship between I(\CII) and I(\NII) can be established to predict the \CII\textrm{ }emission contribution from \HII\textrm{ }regions to PDRs across massive star forming regions \citep{abel2006a}. The total observed \CII\textrm{ }emission toward PDRs can be then corrected accordingly, yielding a more accurate energy balance determination in such regions. In this sense, the interplay between the extremely massive and luminous Arches Cluster \citep{cotera1996} and the large scale structures known as the Arched Filaments (AF) \citep{pauls1976} offers a unique possibility to test the validity of such an approach to determine the relative \CII\textrm{ }emission contribution between PDRs and \HII\textrm{ }regions in the extremely harsh and high-metallicity environment of the Galactic center (GC). In Figure \ref{intro:filaments_figure}, the 20 cm continuum emission view of the region containing the Arches Cluster and the AF is shown \citep{yusef1987}. Further discussion on the figure is featured later in the text. \\

%The Arches Cluster + %The Arched-Filaments
The Arches Cluster, first discovered by \citet{cotera1996} and  located at a projected distance of $\sim$ 27 pc from Sgr A$^{\star}$, is tightly packed, young, contains a mass of $\sim$ 2$\times$10$^{4}$ M\SOLAR\textrm{ }within a radius of $\sim$ 0.19 pc, and has an estimated age between 2 $-$ 2.5 Myr \citep{clark2018}. The bolometric luminosity of its $\sim$ 160 OB stars is L$_{bol}$ $\sim$ 10$^{8}$ L\SOLAR, producing a Lyman continuum flux $Q_{Lyc}$ $\sim$ 1 - 4$\times$10$^{51}$ s$^{-1}$ \citep{figer2008,simpson2007}. The measured local standard of rest (LSR) velocity of the cluster is $+$95 $\pm$ 8 \KMS\textrm{ }\citep{figer2002}, while its proper motion is 172 $\pm$ 15 \KMS\textrm{ }\citep{clarkson2012}, yielding a 3D velocity $\sim$ 232 $\pm$ 30 \KMS\textrm{ }\citep{stolte2008}. The large 3D velocity of the cluster suggests that it must be in a noncircular orbit if the cluster is to be found within the X$_{2}$ orbits \citep{stolte2008}. Given the short orbital time ($\lesssim$ 1 Myr) for a stellar cluster to orbit the GC, it is likely that the cluster is no longer spatially related to its parent molecular cloud \citep{simpson2007}. The interaction of the stellar radiation field of the Arches Cluster with the surrounding ISM has given rise to arc-like structures known as the AF  \citep{simpson2007}, shown in Figure \ref{intro:filaments_figure}. Their radio continuum emission originates mainly in a thermal plasma of electron temperature T$_{e}$ $\sim$ 7000 K, which is derived from hydrogen recombination lines and radio continuum observations \citep{pauls1976,lang2001}. \citet{garcia2016} showed that the brightest \CII\textrm{ }and \NII\textrm{ }emission around this region is found within the AF, at negative radial velocities. These structures are closely associated to a system of dense molecular clouds traced by \CS\textrm{ }emission \citep{bally1987,bally1988}. Previous work by \citet{genzel1990} used low spectral (67 \KMS) and spatial (55'') \CII\textrm{ }line observations in combination with $[$OI$]$ at 63 $\mu$m, mid-, far-infrared (FIR), and radio continuum observations to model the emission from the massive ($\sim$2$\times$10$^{4}$ M\SOLAR) G0.07$+$0.04 region within the AF in the context of PDRs. Their results were consistent with a PDR nature of the emission and they interpreted the \CII\textrm{ }line observations as coming from partially ionized interfaces between fully molecular and fully ionized gas. \\

The main goal of the present work is to assess whether the semi-theoretical predictions in \citet{abel2006a} for the \CII\textrm{ }emission contribution from \HII\textrm{ }regions to nearby PDRs, derived for other star forming regions across the Galactic disk such as the Orion Nebula, are applicable to the AF region in the GC, where the ISM physical conditions are much more extreme \citep{mezger1996,guesten2004}. For this purpose, the behavior of the observed \CII\textrm{ }versus \NII\textrm{ }emission in the AF \citep{garcia2016} is  analyzed. In Section \ref{observations}, a summary of the observations used in the present work is given. In Section \ref{arches:selected_positions_sec}, we discuss the criteria used to overcome the intrinsic difficulties involved in the \CII\textrm{ }line profile interpretation, mainly due to the presence of emission and absorption features along the LOS and opacity uncertainties. In Section \ref{abel_dispersion}, the applicability of the \citet{abel2006a} analysis under the extreme physical conditions in the GC is investigated. We explore how local variations on the carbon-to-nitrogen elemental abundance ratio [C/N] can explain the overall behavior of the emission. In Section \ref{hii_contribution}, we present the derived \CII\textrm{ }emission contribution from the \HII\textrm{ }region component to the observed \CII\textrm{ }emission in the AF at the selected positions in our analysis. Using these results, in Section \ref{c_n_ratio_af} we derive the average [C/N] elemental abundance ratio for the AF combining independent estimates of electron temperatures T$_{e}$ and electron volume densities n$_{e}$ with our observations. Also, by assuming a carbon elemental abundance [C/H]  consistent with values found  toward the GC, an average nitrogen elemental abundance [N/H] is estimated for the AF region and compared with other values in the literature found across the GC. In Section \ref{conclusions}, our main results are summarized. 

%%%%%%%%%%%%%%%%%%%%%%%%%%%%%%%%%%%%%%%%%%%%%%%%%%%%%%%%%%%%%%%%%%%%%%%%%%%%%%%%%%%%%%%%%%%%%%%%%%%%
%%%%%%%%%%%%%%%%%%%%%%%%%%%%%%%%%%%%%%%%%%%%%%%%%%%%%%%%%%%%%%%%%%%%%%%%%%%%%%%%%%%%%%%%%%%%%%%%%%%%
%%%%%%%%%%%%%%%%%%%%%%%%%%%%%%%%%%%%%%%%%%%%%%%%%%%%%%%%%%%%%%%%%%%%%%%%%%%%%%%%%%%%%%%%%%%%%%%%%%%%

\section{Summary of the observations}\label{observations}

We made use of the \NIIl\textrm{ }at 205 $\mu m$ (1462.1 GHz)  and \CIIl\textrm{ }at 158 $\mu m$ (1900.5 GHz) fine structure line observations from \citet{garcia2016} covering the AF region. These observations were part of the Sgr A complex observations obtained with the HIFI heterodyne receiver \citep{roelfsema2012} onboard the Herschel Satellite \citep{pilbratt2010} as part of the Herschel Guaranteed Time HEXGAL Key Program (P.I. R. G\"usten, MPIfR) in which the \CILINEAIl, \CILINEAIIl, \NIIl, and \CIIl\textrm{ }submillimeter lines were measured. The HIFI receiver was a double sideband (DSB) single pixel receiver covering the frequency range from 480 GHz to 1250 GHz and from 1410 up to 1910 GHz.  The  \NIIl\textrm{ }and \CIIl\textrm{ }lines (hereinafter referred to as  \NII\textrm{ }and  \CII) were covered by the receiver bands 6a and 7b, respectively. Two wide-band acousto-optical spectrometers (WBS) were used as backends, measuring vertical and horizontal polarizations. The Herschel beam sizes at the \NII\textrm{ }and \CII\textrm{ }frequencies are 14.5" and 11.2",  respectively, with corresponding spectral resolutions of 0.103 \KMS\textrm{ }and 0.079 \KMS. We used the on-the-fly (OTF) map with load chop and position-switch reference observing mode of Herschel-HIFI to measure the \NII\textrm{ }and \CII\textrm{ }line intensities. This observing mode is very efficient in spending a large fraction of the OTF scanning time ON source, with a total OTF scanning time shorter than the Allan stability time for these high-frequency lines \citep{roelfsema2012}. The downside is that it produces under-sampled maps  in the OTF scan direction as shown in Figure 3.5 by \citet{garcia2015}. The pointing accuracy of the telescope was estimated to be $\sim$2'', and the calibration error in the measured intensities for bands 6a and 7b is $\sim$25\% \citep{roelfsema2012,garcia2016}. For the present work, all data sets have been resampled to a common 1 \KMS\textrm{ }spectral resolution and spatially smoothed to 46'' spatial resolution. This allowed us to further increase the signal-to-noise ratio (S/N) in the \NII\textrm{ }and \CII\textrm{ }data sets, and to match the lowest spatial resolution of the data in  \citet{garcia2016} obtained for the \CILINEAIl\textrm{ }line observations (hereinafter referred to as  \CILINEAI). Given the observing mode used to measure the \NII\textrm{ }and \CII\textrm{ }line transitions, a considerable amount of emission is interpolated in this process. We used the \TA\textrm{ }antenna temperature scale for the reported spectra, which is the best approximation of the true convolved antenna temperature measured in the region given the error beam pickup level in the HIFI 6a and 7b  bands, and the large spatial extension of the emission \citep{garcia2016,garcia2015}. \\

In order to investigate the \CII\textrm{ }line profile and the number of radial velocity components along the LOS at the targeted positions in this work, we included the \CILINEAI\textrm{ }and \CILINEAII\textrm{ }line observations from \citet{garcia2016} in our analysis. For the same purpose, complementary \COOLINEAI\textrm{ }observations obtained with the EMIR\footnote{\url{www.iram.es/IRAMES/mainWiki/EmirforAstronomers}} receiver \citep{carter2012} at the IRAM 30m telescope under project number 020-16 are also used. These data sets were smoothed to the common  1 \KMS\textrm{ }and 46" spectral and spatial resolutions shared by the \CILINEAI, \NII, and \CII\textrm{ }line observations. The \COOLINEAI\textrm{ }data sets are part of a different and ongoing work, so the corresponding technical information will be published in more detail in a future paper. The \CII\textrm{ }and \NII\textrm{ }line intensities are weak compared to what is observed toward other massive star forming regions in the Galactic disk, where \CII\textrm{ }peak line intensities are well above 50 K \citep{ossenkopf2013}. For our selected positions in the AF, the \CII\textrm{ }line is not brighter than $\sim$ 8 K, despite the fact that the AF are the brightest structures in \CII\textrm{ }emission within the Sgr A complex. The low-level emission is particularly troublesome for the \NII\textrm{ }line observations for which the resampling to a 46'' spatial resolution increases the S/N only to yield detections with $\gtrsim$ 3$\sigma$ significance level, while for the \CII\textrm{ }line observations, the S/N increase allows us to study the line profile in more detail. Despite the aforementioned limitations, the \NII\textrm{ }and \CII\textrm{ }data sets in \citet{garcia2016} constitute the most extended velocity-resolved ones available for these transitions in the Sgr A complex at the time of publication.

%%%%%%%%%%%%%%%%%%%%%%%%%%%%%%%%%%%%%%%%%%%%%%%%%%%%%%%%%%%%%%%%%%%%%%%%%%%%%%%%%%%%%%%%%%%%%%%%%%%%
%%%%%%%%%%%%%%%%%%%%%%%%%%%%%%%%%%%%%%%%%%%%%%%%%%%%%%%%%%%%%%%%%%%%%%%%%%%%%%%%%%%%%%%%%%%%%%%%%%%%
%%%%%%%%%%%%%%%%%%%%%%%%%%%%%%%%%%%%%%%%%%%%%%%%%%%%%%%%%%%%%%%%%%%%%%%%%%%%%%%%%%%%%%%%%%%%%%%%%%%%

\section{Complexity of line profiles}\label{arches:selected_positions_sec}
% Position Selection
The overall view of the area containing the Arches Cluster and the AF traced by 20 cm continuum observations \citep{yusef1987} is shown in Figure \ref{intro:filaments_figure}. The region contains several small \HII\textrm{ }regions (labeled H1 to H13)  \citep{lang2001,yusef1987,zhao1993}, and O, B, and WR field stars \citep{dong2012,dong2015}. The selected positions (open circles) for our analysis in Section \ref{hii_contribution} are named after their association with the northern (N) or southern (S) parts of the E1, E2, W1, and W2 AF segments (e.g., E1, E2-N, E2-S, W1, and W2), other known sources (e.g., G0.10$+$0.02 and G0.07$+$0.04), or high-density gas identified in CS rotational line observations (e.g., P1 and P2) \citep{lang2001,timmermann1996,serabyn1987}. The E1, E2-N, E2-S, and G0.10$+$0.0 positions trace the two sub-filaments forming the eastern (E) filament, while the W1, W2, and G0.07$+$0.04 positions trace the western (W) filament. The G0.07$+$0.04 region is particularly interesting since the gas there is thought to be interacting with the Northern Thread filament (NTF), a structure identified in high-resolution 6 cm continuum observations \citep{lang1999b}. The local magnetic field is aligned with the thread, indicating the possible interaction of the ionized, neutral atomic and molecular material with the pervasive large-scale, dipolar magnetic field in the GC. Most of the selected positions are seen as local maxima in the 3.6 cm continuum observations by \citet{lang2001}, while P1 and P2 are located at the edges of the E and W filaments. The criteria for selecting these positions is based on the \CII\textrm{ }and \NII\textrm{ }emission distribution shown in \citet{garcia2016}, which is most intense along the AF and closely follows the 20 cm continuum emission in the region \citep{yusef1987}, and on the location of the high-density gas, where PDR-like emission is mostly expected if the stellar FUV field is the main excitation source of the gas \citep{draine2011}. The measured \CII, \NII, and \COOLINEAI\textrm{ }spectra for all positions are displayed in the corresponding panel insets, while their equatorial coordinates are listed in Table \ref{arches:tab_positions}.  \\

%%%%%%%%%%%%%%%%%%%%%%%%%%%%%%%%%%%%%%%%%%%%%%%%%%%%%%%%%%%%%%%%%%%%%%%%%%%%%%%%%%%%%%%%%%%%%%%%%%%%
\afterpage{
\begin{figure*}
\centering
\includegraphics[angle=-90,width=\hsize]{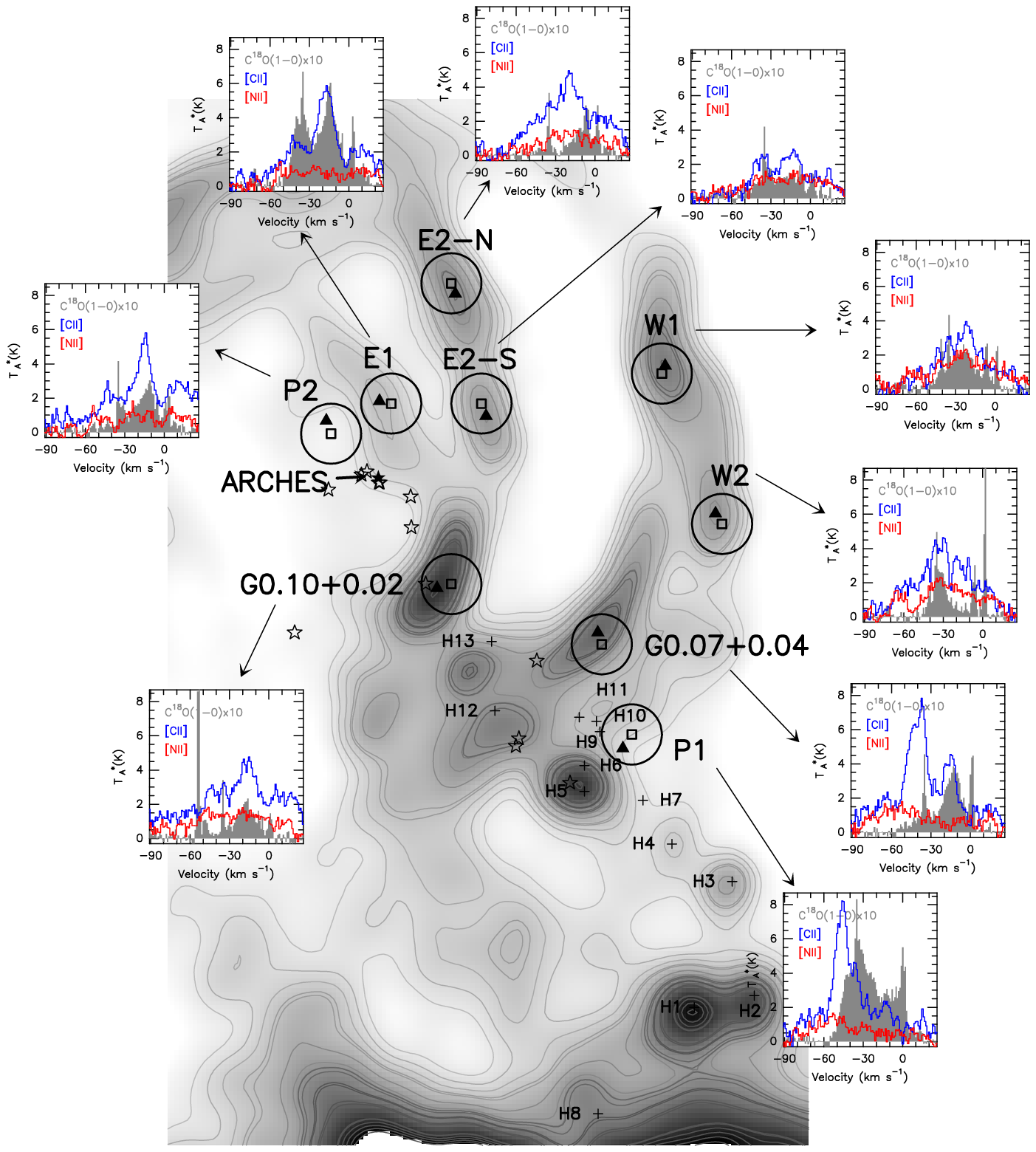}
\caption{AF region outlined in 20 cm continuum emission from \citet{yusef1987}. Selected positions for the I(\CII) versus I(\NII) integrated intensity analysis in this work: E1, E2-N, E2-S, W1, W2, G0.07$+$0.04, G0.10$+$0.02, P1, and P2, are shown as open squares in the \citet{garcia2016} data grid. The 46'' spatial resolution of their data is represented by open circles. Filled triangles depict either the closest continuum emission peak to each position or one of the two \CS\textrm{ }peaks (P1 or P2) in the observations by \citet{serabyn1987}. Panel insets contain the \CII\textrm{ }(blue), \NII\textrm{ }(red), and scaled \COOLINEAI\textrm{ }(filled-gray) spectra at the positions selected for this work, showing very complex line profiles with large variations in line shape and brightness across the region. Black crosses show the positions of several \HII\textrm{ }regions labeled H1 to H13 spanning a region south of the filaments (the so-called H-Region). The \filledstar\textrm{ }symbol shows the position of the Arches Cluster, while \emptystar\textrm{ }symbols indicate the locations of some of the massive (O, B, and WR) field stars scattered across the region \citep{dong2012,dong2015}.\label{intro:filaments_figure}}
\end{figure*}
}
%%%%%%%%%%%%%%%%%%%%%%%%%%%%%%%%%%%%%%%%%%%%%%%%%%%%%%%%%%%%%%%%%%%%%%%%%%%%%%%%%%%%%%%%%%%%%%%%%%%%

%%%%%%%%%%%%%%%%%%%
%%% SELECTED POSITIONS %%%
%%%%%%%%%%%%%%%%%%%

\afterpage{
\begin{table}
\centering
\begin{tabular}{lrrrr}
\toprule
\multicolumn{1}{l}{\textbf{Position}}        &
\multicolumn{1}{c}{\textbf{$\alpha(J2000)$}} &
\multicolumn{1}{c}{\textbf{$\delta(J2000)$}} &
\multicolumn{1}{c}{\textbf{$\Delta\alpha$}}  &
\multicolumn{1}{c}{\textbf{$\Delta\delta$}}  \\

\multicolumn{1}{l}{}                                   &
\multicolumn{1}{c}{\textbf{($^{h}$ $^{m}$ $^{s}$)}}     &
\multicolumn{1}{c}{\textbf{($^{\circ}$ $^{'}$ $^{''}$)}} &
\multicolumn{1}{c}{\textbf{('')}}                      &
\multicolumn{1}{c}{\textbf{('')}}                     \\
\midrule
P1                      & 17\textrm{ }45\textrm{ }35.71 & $-$28\textrm{ }52\textrm{ }44.1 & $+$6.9   & $-$10.7  \\  
P2                      & 17\textrm{ }45\textrm{ }53.25 & $-$28\textrm{ }48\textrm{ }54.1 & $+$3.7   & $+$10.0 \\  
W1                      & 17\textrm{ }45\textrm{ }33.96 & $-$28\textrm{ }48\textrm{ }08.1 & $-$2.5    & $+$6.0   \\
W2                      & 17\textrm{ }45\textrm{ }30.45 & $-$28\textrm{ }50\textrm{ }03.1 & $+$5.0   & $+$8.0   \\ 
G0.10$+$0.02    & 17\textrm{ }45\textrm{ }46.24 & $-$28\textrm{ }50\textrm{ }49.1 & $+$11.0 & $-$3.0    \\  
G0.07$+$0.04    & 17\textrm{ }45\textrm{ }37.47 & $-$28\textrm{ }51\textrm{ }35.1 & $+$3.0   & $+$9.0   \\  
E2-N            & 17\textrm{ }45\textrm{ }46.24 & $-$28\textrm{ }46\textrm{ }59.1 & $-$3.0    & $-$8.0    \\ 
E2-S            & 17\textrm{ }45\textrm{ }44.48 & $-$28\textrm{ }48\textrm{ }31.1 & $-$3.5    & $-$9.5    \\ 
E1                      & 17\textrm{ }45\textrm{ }49.74 & $-$28\textrm{ }48\textrm{ }31.1 & $+$9.0   & $+$2.0   \\
\bottomrule
\end{tabular}
\caption{Selected positions in the AF region shown in Figure \ref{intro:filaments_figure}. Their absolute Equatorial (J2000) coordinates are listed in the first two columns of the table. The $\Delta\alpha$ and $\Delta\delta$ values correspond to their angular offsets with respect to the closest 20 cm continuum emission peak in each case.\label{arches:tab_positions}}
\end{table}
}

% Spectrum Complexity + why 13C+ hyperfine can not be used + Assumptions for interpreting C+ opt thin: C18O + EOF limit.
When observing the GC, the LOS crosses a large fraction of the Galactic plane. Hence, related structures such as spiral arms appear in the measured spectra as either deep absorption or narrow bright emission features, depending on the observed species. Typical LSR radial velocities of spiral arms in the LOS toward the GC are $\sim$ $-$55 \KMS\textrm{ }for the 3-kpc expanding arm, $\sim$ $-$30 \KMS\textrm{ }for the 4-kpc arm, and $\sim$ $-$5 \KMS\textrm{ }for the  local arm \citep{oka1998,jones2012,garcia2016}. On top of this, the emission originating from within the GC volume is extremely broad and is usually detected between $\pm$200 \KMS\textrm{ }at Galactic longitude l $=$ 0\DEG\textrm{ }\citep{garcia2016}, making the identification of individual gas components in the observed spectra particularly troublesome due to source blending, varying opacity effects across the spectra, different gas excitation conditions, etcetera. Moreover, in the case of the \CII\textrm{ }emission this effect is worsened as C$^{+}$ can exist in a plethora of environments given its low ionization potential, while its 1.9 THz fine structure transition can be excited in different phases of the ISM, from diffuse clouds to PDRs \citep{langer2010,langer2016,pineda2014}, making the identification of its exact origin along the LOS in broad, blended, and complex line profiles a very challenging task. In order to illustrate the complexity of the emission detected in the AF, Figure \ref{intro:cii_ci_c18o} shows the \CII, \CILINEAI, and \COOLINEAI\textrm{ }spectra for all positions in Table \ref{arches:tab_positions} tracing the molecular, atomic, and ionized content of the ISM in the GC. From the figure, bright and narrow components at negative radial velocities seen in the optically thin \COOLINEAI\textrm{ }emission (marked by vertical-dotted lines) are superimposed on a much broader emission distribution along the radial velocities. \\

In order to identify the physical components in the GC traced by the \CII\textrm{ }emission, a natural choice would be to target the optically thin \CCIIl\textrm{ }hyperfine triplet. The selection of this isotope would drastically decrease the confusion between different components along the LOS, foreground emission and absorption features, etcetera, as the conditions for sufficient line excitation (mainly warm, high-column-density gas) are not easily met across the entire Galactic plane. Unfortunately, these diagnostic lines are not easily detectable in the GC for two main reasons. Firstly, the small triplet frequency separation, equivalent to $+$11.2 \KMS, $-$65.2 \KMS, and $+$63.2 in radial velocities with regard to the rest frequency of the main isotope at 0 \KMS\textrm{ }(with relative intensities 0.625, 0,250, and 0.125, respectively) are located too close to the rest frequency of the main isotope to avoid strong line blending, given the extremely broad \CII\textrm{ }line profiles shown in Figure \ref{intro:cii_ci_c18o}. Secondly, even if there were a position where the \CII\textrm{ }line became narrow enough, the peak \CII\textrm{ }line intensities in Figure \ref{intro:filaments_figure} are quite low ($<$ 10 K) when compared with measurements in other massive star forming regions in the Galactic disk \citep{guevara2019}. The detection of even the strongest \CCII\textrm{ }hyperfine component would require prohibitively long integration times on the SOFIA Airborne Observatory, which is currently the only astronomical facility capable of observing the \CII\textrm{ }line. On the other hand, \CCII\textrm{ }observations of massive star forming regions outside the GC such as M43, Mon R2, M17 SW, Orion Bar, and NGC 2024 have shown that opacity effects are a major issue when interpreting \CII\textrm{ }line profiles, sometimes with up to 90\% of the line intensity  absorbed away in the measured spectra \citep{graf2012,ossenkopf2013,guevara2019}. This is a major problem for the \CII\textrm{ }data analysis, potentially causing a strong underestimation of the gas cooling rate and the misidentification of radial velocity components along the LOS, which causes misleading model results. Therefore, in dealing with any \CII\textrm{ }data interpretation, this issue must be addressed. Fortunately, this problem can be overcome in the GC given the extreme ISM physical conditions and the low brightness of the \CII\textrm{ }line, in contrast to what it is observed in the Galactic disk. In the absence of \CCII\textrm{ }observations for the AF, other complementary data can be used to assess the impact of opacity effects on the observed \CII\textrm{ }spectra. Experimentally, the \COOLINEAI\textrm{ }line seems to closely follow the \CCII\textrm{ }line profile in massive star forming regions such as M43, Mon R2, and M17 SW in the Galactic disk \citep{guevara2019}. Therefore, this optically thin transition could potentially be used as a template to determine the number of \CII\textrm{ }physical components contained in each spectrum, when no \CCII\textrm{ }observations are available. Figure \ref{intro:cii_ci_c18o} shows a remarkable similarity between the \CILINEAI\textrm{ }and \COOLINEAI\textrm{ }line profiles, except for some of the very narrow spectral features, most likely tracing cold gas in the foreground, where the \CILINEAI\textrm{ }transition is not very bright. Such a similarity implies that opacity effects do not affect the observed \CILINEAI\textrm{ }transition very strongly. At the same time, the \CII\textrm{ }line shape at W2, W1, P2, G0.10$+$0.02, E2S, and E1 closely follows that of the more optically thin \COOLINEAI\textrm{ }and \CILINEAI\textrm{ }lines. This provides empirical support to the idea that opacity effects in the \CII\textrm{ }spectra at the selected positions can not be the dominant factor responsible for the observed line shapes. Moreover, the fact that all these positions, which are several parsecs away from each other, exhibit the same spectral behavior in these transitions suggests this as a common feature across the entire AF region. \\  
 
 %%%%%%%%%%%%%%%%%%%%%%%%%%%%%%%%%%%%%%%%%%%%%%%%%%%%%%%%%%%%%%%%%%%%%%%%%%%%%%%%%%%%%%%%%%%%%%%%%%%%
\afterpage{
\begin{figure}
\centering
\includegraphics[angle=0, width=\hsize]{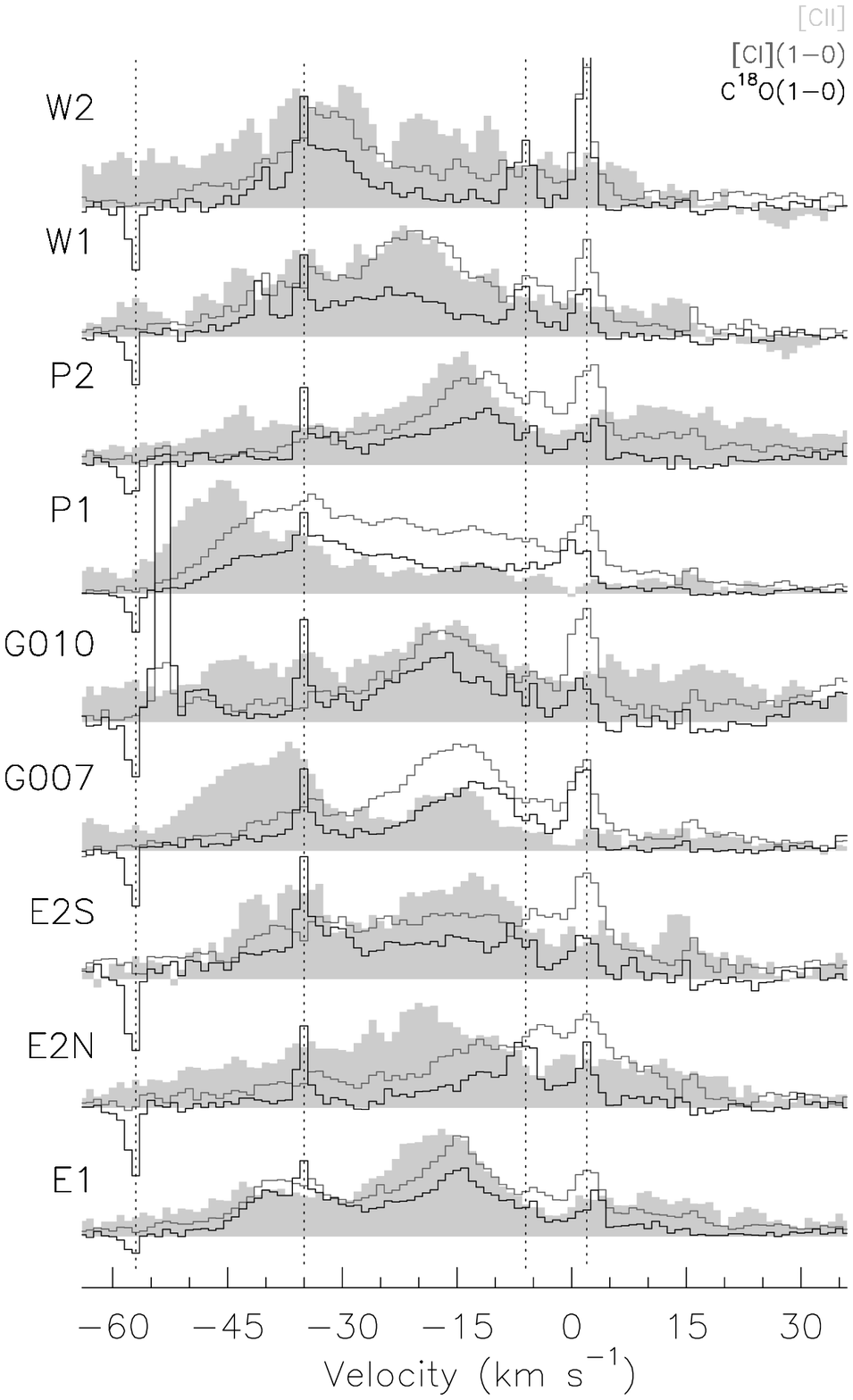}
\caption{Comparison between \CII\textrm{ }(filled light gray), \CILINEAI\textrm{ }(gray), and \COOLINEAI\textrm{ }(black) line profiles for all positions in Table \ref{arches:tab_positions}. Relative intensities among the spectra have been scaled for display purposes. The observations trace different phases of the ISM from ionized, neutral, and molecular gas. Narrow and strong emission and absorption features in the \COOLINEAI\textrm{ }optically thin line common to all positions are marked with vertical dotted lines as they are most likely foreground components located in the Galactic disk somewhere along the LOS and outside the GC. \label{intro:cii_ci_c18o}}
\end{figure}
}
%%%%%%%%%%%%%%%%%%%%%%%%%%%%%%%%%%%%%%%%%%%%%%%%%%%%%%%%%%%%%%%%%%%%%%%%%%%%%%%%%%%%%%%%%%%%%%%%%%%%

At positions P1, G0.07$+$0.04, and E2N, the \CII\textrm{ }line shapes are also similar to those of \COOLINEAI\textrm{ }and \CILINEAI, but they are largely ($>$5 \KMS) blueshifted toward more negative radial velocities. The difference in central radial velocities (obtained from Gaussian fits in Section \ref{intensities} and displayed in Appendix \ref{appendix:gaussian_fits}) between the \CII\textrm{ }and \COOLINEAI\textrm{ }radial velocity components identified in the spectra are $\sim$ $-$13 \KMS\textrm{ }(P1-46\KMS), $\sim$ $-$6 \KMS\textrm{ }(P1-13\KMS), $\sim$ $-$8 \KMS\textrm{ }(G007-41\KMS), and $\sim$ $-$9 \KMS\textrm{ }(E2N-18\KMS). On the other hand, at position E2S,  although a visual inspection of Figure \ref{intro:cii_ci_c18o} shows a similar emission distribution of these two lines, the central radial velocity of the \CII\textrm{ }line shows a large red-shift of $\sim$ $+$7 \KMS\textrm{ }(E2S-17\KMS) with respect to what is obtained for the \COOLINEAI\textrm{ }line. For the rest of the radial velocity components identified at all positions, the difference in central radial velocities between both lines is $<$5 \KMS. A large shift in central radial velocities between ionized and neutral species could be attributed to geometry effects or local interactions with the magnetic field in the GC, rather than local opacity effects. For instance, the G0.07$+$0.04 source is known to exhibit a large radial velocity blueshift of the emission produced by ions (e.g., H$^{+}$ and C$^{+}$) with regard to the line profiles produced by neutral species such as C, CO, and C$^{18}$O  \citep{genzel1990}, which is consistent with what is shown in Figure \ref{intro:cii_ci_c18o}. Two interpretations for the blueshifted emission at G0.07$+$0.04 have been given in the literature. On the one hand, the radial velocity shift between the H92$\alpha$ line, at $\sim$ $-$44 \KMS\textrm{ }\citep{lang2001}, and the molecular emission could be a consequence of the location of the AF ionized gas at the near side of a molecular cloud \citep{serabyn1987,lang2001}. \citet{genzel1990} found that their velocity-unresolved observations of \CII\textrm{ }and H110$\alpha$ recombination lines are blueshifted by 10 to 15 \KMS\textrm{ }with respect to their CO and CS observations, indicating streaming motions between ionized and molecular gas. In that context, the increasing magnitude of the radial velocity blueshift, from the emission originating in neutral molecular/atomic gas to the \CII\textrm{ }emission and further to the H92$\alpha$ recombination line, would reflect a layered spatial distribution along the LOS of the gas at this position. On the other hand, it has been proposed that G0.07$+$0.04 is interacting with the nonthermal NTF, whose intrinsic magnetic field is aligned with the filament \citep{yusef1988,lang1999b}. Such interpretation is mainly based on the brightness discontinuity in 20 cm continuum emission at the location where both regions intersect, as shown in Figure 23 in \citet{lang1999b}. The NTF is expected to have a much stronger magnetic field than its surrounding diffuse ISM \citep{yusef2007}, so the interaction between the ionized gas at G0.07$+$0.04 and the NTF magnetic field could cause the observed radial velocity shift. Despite of the observational evidence supporting this scenario, there is also observational evidence against it. Using H92$\alpha$ and 8.3 GHz continuum observations, \citet{lang2001} found that the intersection region between G0.07$+$0.04 and the NTF shows no discontinuities concerning the rest of the W1 filament in terms of the gas velocity field, line width, or line-to-continuum ratio. Hence, the exact nature of the interaction between both regions is still unclear, if it exists at all \citep{lang1999b,lang2001}. \\

Another strong argument against large opacity effects in the observed \CII\textrm{ }line shape is given by \citet{goldsmith2012}. It is based on the fact that, for \CII\textrm{ }line intensities $\lesssim$ 8 K, the emission falls into their so-called \emph{\emph{effective optically thin}} (EOT) limit where the I(\CII) integrated intensity is proportional to the C$^{+}$ column density along the LOS, irrespectively of the optical depth of the line. This is indeed the case for all positions in the present work, as can be seen from the antenna temperature scale of the spectra in Figure \ref{intro:filaments_figure}. The optically thin behavior of the \CII\textrm{ }line in the GC resembles what is observed in the \COLINEAI\textrm{ }transition when used to trace the large-scale mass distribution in the Milky Way's Galactic disk. Despite the fact that locally the \COLINEAI\textrm{ }line is known to be optically thick, at large scales it behaves as if it were optically thin, tracing most of the molecular mass in giant molecular clouds (GMCs) \citep{dame86,garcia2014}. The most common interpretation for this feature is the presence of a large collection of unresolved small \COLINEAI\textrm{ }clumps within the telescope's beam, moving at differential radial velocities, which when observed yield an effectively optically thin transition. Based on all the aforementioned evidence, in the following we assume that spectral features in the \CII\textrm{ }data are mostly due to the existence of different radial velocity components along the LOS, and opacity effects play a less important role in shaping the observed line profiles compared to what is indeed the case in other regions of the Galactic disk.

\subsection{Estimation of I([CII]) and I([NII]) integrated intensities}\label{intensities}
In order to estimate integrated intensities of the \CII\textrm{ }and \NII\textrm{ }lines, we follow two different approaches depending on the data analysis. In Section \ref{abel_dispersion}, integrated intensities are obtained by simply adding the measured intensities in the observed spectra (in units of K \KMS) over different LSR radial velocity ranges, which were then converted into units of ergs s$^{-1}$ cm$^{-2}$. In this way, the overall behavior of the I(\CII) versus I(\NII) relationship in the AF region can be explored. For the data analysis in Section \ref{hii_contribution}, we follow the discussion in Section \ref{arches:selected_positions_sec}, focusing on identifying the radial velocity components traced by the \CII\textrm{ }spectra, using the optically thin \COOLINEAI\textrm{ }line, and to a minor extent the \CILINEAI\textrm{ }line, as a template. The procedure is outlined here: 

\begin{enumerate}
\item We identify all velocity components traced by the optically thin \COOLINEAI\textrm{ }transition. Broad (\DELTAV\textrm{ }$\geq$ 10 \KMS) and high S/N ($>$ 5) detected components are assumed to originate within the GC, given the large column densities and turbulence of the gas in the region  \citep{mezger1996,guesten2004}. Narrow features with radial velocities close to those of gas associated with spiral arms along the LOS are assumed to be in the foreground and are excluded from the analysis. With this information, Gaussian profiles (assumed for simplicity) are fit to obtain the central radial velocity and velocity width of each component. 
\item The radial velocity information is used to perform Gaussian fits to identify the same components in the \CII\textrm{ }spectrum. Due to the complexity of the spectra, usually multi-Gaussian profiles are fit, as the figures in Appendix \ref{appendix:gaussian_fits} show. Given that the \COOLINEAI\textrm{ }and \CII\textrm{ }lines trace gas under very different physical conditions, it is not always the case that each detected component (narrow or broad) in either line has a corresponding counterpart in the other. Therefore, in order to minimize misidentifications, the \CILINEAI\textrm{ }and \CILINEAII\textrm{ }data in \citet{garcia2016} are used to check the consistency of the identified components in the \CII\textrm{ }spectra. From this approach, three I(\CII) integrated intensities are obtained for each \CII\textrm{ }component: (1) the integral over the fit Gaussian profile I(\CII)$_{g}$; (2) the sum of measured intensities (in K) multiplied by the channel width (in \KMS) over the relevant spectral channel range listed in Table \ref{tab:data_summary} where the component is found I(\CII)$_{s}$ and indicated by the vertical black dotted lines in the figures of Appendix \ref{appendix:gaussian_fits}; and (3) the integrated intensity in the LSR radial velocity $-$90 \KMS\textrm{ }to $+$25 \KMS\textrm{ }range where most of the emission in the AF region is found I(\CII)$_{i}$ \citep{simpson2007,lang2001,garcia2016}, calculated in the same way as I(\CII)$_{s}$. A graphic description of the approach followed to derive I(\CII)$_{g}$, I(\CII)$_{s}$, and I(\CII)$_{i}$ is shown in Figure \ref{appA:spectra6}. After these definitions, the greatest differences between I(\CII)$_{g}$ and I(\CII)$_{s}$ appear when radial velocity components are largely blended.
\item Given the different nature of the \NII\textrm{ }line originating exclusively inside the \HII\textrm{ }region (i.e., no emission contamination of spiral arms along the LOS is expected) and the poorer S/N of the \NII\textrm{ }data set, only I(\NII)$_{s}$ integrated intensities are calculated over the same radial velocity ranges in which I(\CII)$_{s}$ intensities were obtained (vertical dotted black lines for figures in Appendix \ref{appendix:gaussian_fits}) as the Gaussian fitting approach in this case is likely not reliable. Finally, the integrated intensity I(\NII)$_{i}$ is obtained in the same way as for I(\CII)$_{i}$. 

\end{enumerate}

All derived I$_{g}$, I$_{s}$, and I$_{i}$ integrated intensities were then converted into units of ergs s$^{-1}$ cm$^{-2}$. A total of twelve radial velocity components are identified and listed in Table \ref{tab:data_summary} within the nine selected positions in Table \ref{arches:tab_positions}. The name of each radial velocity component consists of the position's name followed by the radial velocity obtained from the Gaussian fit to the \CII\textrm{ }spectrum. All details pertaining to deriving integrated intensities used for the data analysis are given in Appendix \ref{appendix:gaussian_fits}. 

%%%%%%%%%%%%%%%%%%%%
% INTEGRATED INTENSITIES PDR %
%%%%%%%%%%%%%%%%%%%%
\afterpage{
\begin{table*}
\centering
\caption{I(\CII) and I(\NII) integrated intensities in units of ergs s$^{-1}$ cm$^{-2}$ for the velocity components identified at the positions in Table \ref{arches:tab_positions}. I$_{g}$ intensities are derived from Gaussian fits to the \CII\textrm{ }spectra. I$_{s}$ are obtained by directly adding measured intensities to the corresponding spectra within the radial velocity range where each component is found. The corresponding initial (V$_{lsr,i}$) and final (V$_{lsr,f}$) radial velocities used to derive the I$_{s}$ integrated intensities are listed in the last two columns of the table. I$_{i}$ is obtained by adding measured intensities within the radial velocity range $-$90 \KMS\textrm{ }to $+$25 \KMS, as described in Section \ref{abel_dispersion}. The name of each radial velocity component is composed by the position's name followed by the radial velocity obtained from the Gaussian fit to the \CII\textrm{ }spectrum. \label{tab:data_summary}}
\bgroup
\def\arraystretch{1}
\begin{tabular}{lcccccrr}
\toprule
\multicolumn{1}{c}{\textbf{Source}}          &
\multicolumn{1}{c}{\textbf{I(\CII)$_{g}$}}  &
\multicolumn{1}{c}{\textbf{I(\CII)$_{s}$}}  &
\multicolumn{1}{c}{\textbf{I(\CII)$_{i}$}}   &
\multicolumn{1}{c}{\textbf{I(\NII)$_{s}$}}  &
\multicolumn{1}{c}{\textbf{I(\NII)$_{i}$}}   &
\multicolumn{1}{c}{\textbf{V$_{lsr,i}$}}      &
\multicolumn{1}{c}{\textbf{V$_{lsr,f}$}}      \\

\multicolumn{1}{c}{}                 &
\multicolumn{1}{c}{10$^{-4}$} &
\multicolumn{1}{c}{10$^{-4}$} &
\multicolumn{1}{c}{10$^{-4}$} &
\multicolumn{1}{c}{10$^{-4}$} &
\multicolumn{1}{c}{10$^{-4}$} &
\multicolumn{1}{c}{\KMS} &
\multicolumn{1}{c}{\KMS} \\

\midrule
%source                         CIIg                            CIIs                              CIIi                             NIIs                            NIIi       
%--------------------------------------------------------------------------------------------------------------------------------------------------------------
  P1-46\KMS   &  7.65  $\pm$  0.17  &  9.00  $\pm$  0.16  & 16.22  $\pm$  0.32  &  0.90  $\pm$  0.06  &  2.15  $\pm$  0.13  &  $-$59   & $-$32  \\
  P1-13\KMS   &  3.19  $\pm$  0.20  &  3.81  $\pm$  0.18  & 16.22  $\pm$  0.32  &  0.60  $\pm$  0.07  &  2.15  $\pm$  0.13  &  $-$32   & $+$1   \\
  P2-15\KMS   &  4.52  $\pm$  0.13  &  5.55  $\pm$  0.14  & 16.77  $\pm$  0.33  &  0.64  $\pm$  0.04  &  2.28  $\pm$  0.11  &  $-$24   & $-$4    \\
  W1-24\KMS  &  8.95  $\pm$  0.34  &  9.01  $\pm$  0.24  & 10.07  $\pm$  0.32  &  2.83  $\pm$  0.10  &  3.67  $\pm$  0.13  &  $-$54   & $+$10 \\
  W2-29\KMS  & 10.26  $\pm$  0.38  &  9.65  $\pm$  0.23  & 14.00  $\pm$  0.34  &  2.24  $\pm$  0.08  &  3.52  $\pm$  0.12 &  $-$53   & $-$6    \\
G010-17\KMS &  7.70  $\pm$  0.23  &  7.68  $\pm$  0.15  & 18.53  $\pm$  0.29  &  1.44  $\pm$  0.06  &  3.51  $\pm$  0.12  &  $-$33   & $+$1   \\
G007-16\KMS &  8.69  $\pm$  0.26  &  9.76  $\pm$  0.15  & 19.14  $\pm$  0.30  &  1.14  $\pm$  0.06  &  2.91  $\pm$  0.12  &  $-$58   & $-$26  \\
G007-41\KMS &  4.52  $\pm$  0.15  &  4.91  $\pm$  0.15  & 19.14  $\pm$  0.30  &  0.58  $\pm$  0.06  &  2.91  $\pm$  0.12  &  $-$26   & $+$2   \\
 E2N-18\KMS  & 14.87 $\pm$  0.42  & 14.46 $\pm$  0.27  & 14.26  $\pm$  0.28  &  2.18  $\pm$  0.11  &  2.08  $\pm$  0.11  &  $-$72   & $+$33  \\
  E2S-17\KMS &  8.14  $\pm$  0.24  &  8.44  $\pm$  0.25  &  8.44  $\pm$  0.27   &  2.56  $\pm$  0.11  &  2.69  $\pm$  0.12  &  $-$64   & $+$30  \\
  E1-40\KMS   &  5.49  $\pm$  0.61  &  4.83  $\pm$  0.19  & 15.27  $\pm$  0.32  &  0.82  $\pm$  0.07  &  2.22  $\pm$  0.11  &  $-$66   & $-$27   \\
  E1-16\KMS   &  5.90  $\pm$  0.53  &  6.75  $\pm$  0.15  & 15.27  $\pm$  0.32  &  0.66  $\pm$  0.05  &  2.22  $\pm$  0.11  &  $-$27   & $-$2     \\

\bottomrule
\end{tabular}
\egroup
\end{table*}}

%%%%%%%%%%%%%%%%%%%%%%%%%%%%%%%%%%%%%%%%%%%%%%%%%%%%%%%%%%%%%%%%%%%%%%%%%%%%%%%%%%%%%%%%%%%%%%%%%%%%
%%%%%%%%%%%%%%%%%%%%%%%%%%%%%%%%%%%%%%%%%%%%%%%%%%%%%%%%%%%%%%%%%%%%%%%%%%%%%%%%%%%%%%%%%%%%%%%%%%%%
%%%%%%%%%%%%%%%%%%%%%%%%%%%%%%%%%%%%%%%%%%%%%%%%%%%%%%%%%%%%%%%%%%%%%%%%%%%%%%%%%%%%%%%%%%%%%%%%%%%%

\section{Observational and theoretical I([CII]) versus I([NII]) relationship}\label{abel_dispersion}
The spatial distribution of the I(\CII)$_{i}$/I(\NII)$_{i}$ integrated intensity ratio in the AF region is shown in Figure \ref{arches:abel_ratio}. 
For each ratio, only integrated intensities with detections above the 3$\sigma$ significance level were considered, where $\sigma$ is the propagated noise in the radial velocity integration interval ($\sigma =$ \TRMS\textrm{ }$\times$ $\Delta V\times\sqrt{N}$, with $\Delta V$ the radial velocity resolution and $N$ the number of spectral channels). Overplotted on Figure \ref{arches:abel_ratio} is the 20 cm continuum emission of the region and the selected positions from Table \ref{arches:tab_positions}. \\

%Integrated Intensity Map
There are several interesting features in Figure \ref{arches:abel_ratio}. First, the spatial distribution of the integrated intensity ratios is not homogeneous, with the bulk of the values in the range 0 $<$ I(\CII)$_{i}$/I(\NII)$_{i}$ $<$ 10. These ratios are enclosed between two prominent lanes of larger ratios with I(\CII)$_{i}$/I(\NII)$_{i}$ $>$ 10 that are close (at least in projection) to the southern edge of the radio arc and the northern edge of the Sgr A-east supernova remnant (SNR).  Second, at G0.07$+$0.04 there is a local enhancement of  the integrated intensity ratios with regard to its surrounding area. While there is no definitive evidence for the interaction of G0.07$+$0.04 with the NTF tracing the local magnetic field in the GC \citep{lang1999b}, such a localized enhancement suggests that the gas is undergoing a different physical process at this location than in the rest of the region. In the opposite sense, the northern part of the W1 filament demonstrates a notorious decrease in the I(\CII)$_{i}$/I(\NII)$_{i}$ ratio, where W1 intersects the large ratio lane south of the radio arc. At this location, there are no evident structures tracing the local magnetic field, either in the 20 cm or 6 cm continuum observations \citep{lang1999b}, contrary to what is observed at the G0.07$+$0.04 location. Therefore, it is likely that the physical process affecting the I(\CII)$_{i}$/I(\NII)$_{i}$ ratio at the intersection is not strongly related to a local enhancement of the magnetic field. Third, the same lane of high I(\CII)$_{i}$/I(\NII)$_{i}$ values intersects the E1, E2S, and E2N filaments without altering the spatial continuity of such large ratios. This could be indicative of a geometric effect in which the plasma responsible for the continuum emission \citep{pauls1976,lang1999b} is spatially decoupled from the gas emitting in these submillimeter spectral lines. The prominent lane of I(\CII)$_{i}$/I(\NII)$_{i}$ ratios north of the Sgr A-east SNR seems to follow the edge of its continuum emission over the entire map, with a decrease in the ratio toward a local 20 cm continuum maximum at (+50'', +300'') and the location of the H3 region (shown in Figure \ref{intro:filaments_figure}), similarly to what is seen toward W1. The upper panel in Figure \ref{arches:abel_ratio} shows the histogram of the I(\CII)$_{i}$/I(\NII)$_{i}$ ratios in the map. A Gaussian profile was fit to characterize the distribution, from which the mean ratio R$_{c}$ $=$ 5.1 $\pm$ 0.1 and its standard deviation R$_{\sigma}$ $=$ 2.0 $\pm$ 0.1 were derived. The vast majority of the I(\CII)$_{i}$/I(\NII)$_{i}$ ratios seem to be relatively consistent, with a Gaussian-like distribution suggesting that there are no dramatic changes in the physical conditions of the bulk of the gas. Nonetheless, a more detailed  interpretation of the I(\CII)$_{i}$/I(\NII)$_{i}$ ratio spatial distribution in Figure \ref{arches:abel_ratio} is challenging as the PDR-\HII\textrm{ }contribution to the observed \CII\textrm{ }emission (or from foreground gas along the LOS) is not disentangled and the \NII\textrm{ }emission traces only the \HII\textrm{ }region. \\

%%%%%%%%%%%%%%%%%%%%%%%%%%%%%%%%%%%%%%%%%%%%%%%%%%%%%%%%%%%%%%%%%%%%%%%%%%%%%%%%%%%%%%%%%%%%%%%%%%%%
\afterpage{
\begin{figure}
\centering
\includegraphics[angle=0, width=\hsize]{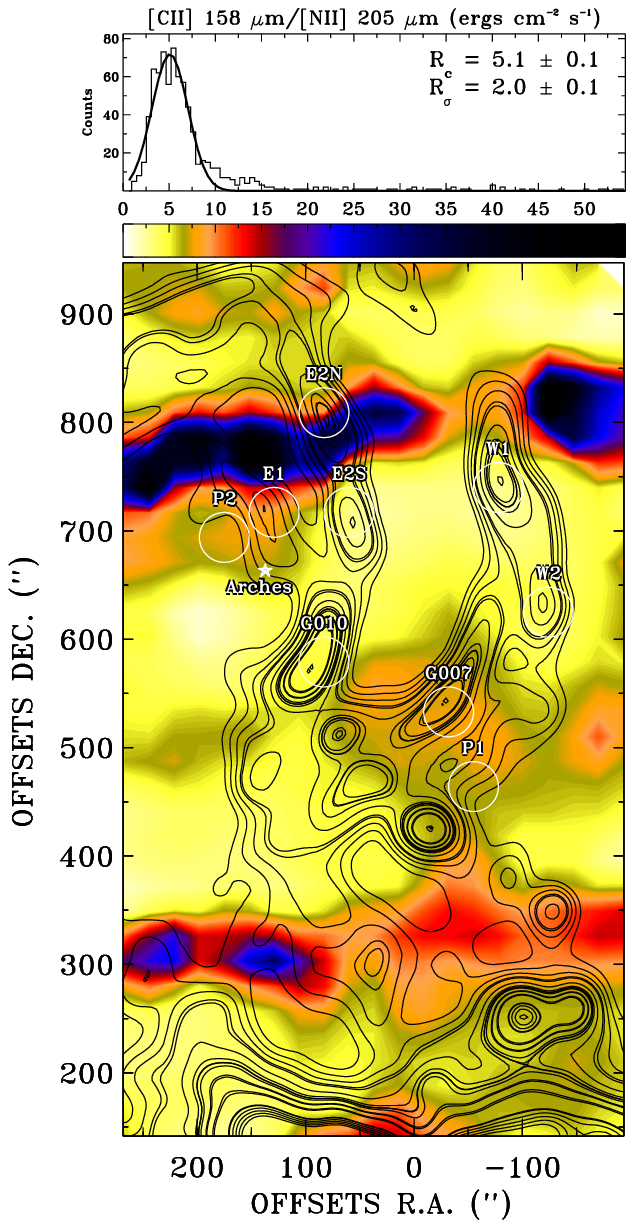}
\caption{
  \textbf{Bottom:} Spatial distribution of the I(\CII)$_{i}$/I(\NII)$_{i}$ integrated intensity ratio calculated for the radial velocity range $-$90 \KMS\textrm{ }to $+$25 \KMS. The AF traced by 20 cm continuum observations and the selected positions in Table \ref{arches:tab_positions} are shown as black contours and white circles, respectively. The white $\bigstar$ symbol represents the location of the Arches Cluster. \textbf{Top:} Histogram of the spatial distribution of I(\CII)$_{i}$/I(\NII)$_{i}$ integrated intensity ratios shown in the map. A Gaussian fit was performed to the histogram to characterize the central ratio (R$_{c}$) and standard deviation (R$_{\sigma}$) of the distribution.\label{arches:abel_ratio}} 
\end{figure}
}
%%%%%%%%%%%%%%%%%%%%%%%%%%%%%%%%%%%%%%%%%%%%%%%%%%%%%%%%%%%%%%%%%%%%%%%%%%%%%%%%%%%%%%%%%%%%%%%%%%%%

%Scatter Figure: Evolution
Models for the relative contribution of \HII\textrm{ }regions to the observed \CII\textrm{ }emission toward PDRs have been presented previously in the literature \citep{heiles1994,abel2006a}. In the models of \citet{abel2006a}, the \CIIl\textrm{ }emission coming from \HII\textrm{ }regions is traced by \NIIl\textrm{ }line observations \citep{heiles1994}, as the power-law fit in their Equation 1 shows. This reflects the amount of \CII\textrm{ }emission expected to arise from an \HII\textrm{ }region, while the rest of the emission above the prediction is attributed to the adjacent PDR, under the assumption of pressure equilibrium between both regions. To investigate the relationship between the \NII\textrm{ }and \CII\textrm{ }emission in the AF, Figure \ref{arches:abel_evolution} shows the evolution of the I(\CII) versus I(\NII) integrated intensities as a function of radial velocity integration interval, increasing from the upper-left to the bottom-right panel, where I(\CII)$_{i}$ versus I(\NII)$_{i}$ values are plotted. Given the complexity of the line profiles, the selection of any particular radial velocity interval to explore the behavior of I(\CII) versus I(\NII) integrated intensities could potentially affect the true scatter relationship in the GC, as different contributions to the \CII\textrm{ }emission (e.g., from foreground sources such as spiral arms) could be included in (or excluded from) the derived values. In Figure \ref{arches:abel_evolution}, the \citet{abel2006a} model predictions of the amount of \CII\textrm{ }emission expected to originate within \HII\textrm{ }regions for metallicities Z $=$ 0.1 - 1.0 Z\SOLAR\textrm{ }are overplotted as dotted and dashed lines, respectively. The vast majority of the I(\CII) versus I(\NII) data points move along the \citet{abel2006a} predictions for Z $=$ 0.1 Z\SOLAR. The few positions where the \NII\textrm{ }emission decreases with increasing radial velocity range are dominated by the rms noise in the spectra, artificially altering I(\NII). The overall behavior of the I(\CII) versus I(\NII) scatter relationships in Figure \ref{arches:abel_evolution} shows that the emission from both species is correlated (independently of the radial velocity range) as predicted by the \citet{abel2006a} models, but it does not seem to be ``calibrated'' for the GC in the sense that most of the data points fall below or along solar and subsolar metallicities, respectively, while in the GC super-solar metallicities from Z $=$ 1-3 Z\SOLAR\textrm{ }are measured \citep{mezger1979,cox1989,shields1994,mezger1996,yusef2007}. In the following, we argue against some of the reasons that could be responsable for the observed discrepancy between model and observations such as:\\

\textbf{Choice of antenna temperature scale:} \citet{garcia2016} showed that the most reasonable antenna temperature scale to apply to their data is \TA, in which the calibrated data have been divided by the telescope's forward efficiency (F$_{eff}$). Nonetheless, this represents a lower limit to the true convolved antenna temperatures. Since the main beam efficiencies (B$_{eff}$) for the \CII\textrm{ }and \NII\textrm{ }line transitions observed with the Herschel-HIFI satellite are virtually identical around $\sim$ 0.58 \citep{mueller2014}, scaling the integrated intensities of both lines by B$_{eff}$/F$_{eff}$ to change the antenna temperature scale to main beam (\TMB), which is an upper limit to the true convolved antenna temperatures, has no impact on the overall trend in Figure \ref{arches:abel_evolution}.

\textbf{Underestimation of I(\CII) integrated intensities:} the \CII\textrm{ }emission at the G0.07$+$0.04 position was measured by \citet{genzel1990} as $\sim$ 1.2$\times$10$^{-3}$ ergs s$^{-1}$ cm$^{-2}$ within a 55'' beam (relative calibration error 30\%). This is somewhat lower than what we estimate by integrating the spectrum in the velocity range $-$90 \KMS\textrm{ }$<$ \VLSR\textrm{ }$<$ $+$25 \KMS\textrm{ }in a 46'' beam, yielding $\sim$ 1.9$\times$10$^{-3}$ ergs s$^{-1}$ cm$^{-2}$ (relative calibration error of 25\%, \citet{garcia2015}). Nonetheless, and despite  the slightly different beam sizes, both estimates are consistent within the error uncertainties, which rules out a systematic underestimation of I(\CII) integrated intensities. \\

\textbf{Overestimation of I(\NII) integrated intensities:} in the AF, the \NII\textrm{ }emission is much weaker than the \CII\textrm{ }emission, yielding a lower S/N in the \NII\textrm{ }data than that in the \CII\textrm{ }line observations. For example, at G0.07$+$0.04 the rms noise of the \NII\textrm{ }spectrum ($\sim$ 0.34 K) yields S/N $\sim$ 5, while for the \CII\textrm{ }spectrum, the rms noise ($\sim$ 0.42 K) yields S/N $\sim$ 18. Since integrated intensities are calculated mostly for positions where S/N $>$ 3 for each line, overestimating I(\NII) is rather unlikely. \\
 
\textbf{C$^{++}$ instead of C$^{+}$:} \citet{simpson2007} showed that strong shocks from the Quintuplet Cluster Wolf-Rayet stars can ionize oxygen to $[$OIV$]$ (ionization potential $\sim$ 54.9 eV, \citet{draine2011}) and destroy dust grains. If this were also the case for carbon in the AF and most of the ionized carbon were in the form of C$^{++}$ (ionization potential $\sim$ 24.4 eV, \citet{draine2011}) rather than in C$^{+}$ due to a strong ultra-violet (UV) field from the Arches Cluster, the low \CII\textrm{ }line intensities with respect to the predicted ones from the \NII\textrm{ }line intensities could be explained. Nonetheless, this would not explain why this is still the case at positions far away from the Arches' location such as W1 and W2 where, given their large projected distances to the cluster ($>$ 10 pc), the UV radiation field should be significantly reduced. Therefore, this possibility seems rather unlikely. \\

%%%%%%%%%%%%%%%%%%%%%%%%%%%%%%%%%%%%%%%%%%%%%%%%%%%%%%%%%%%%%%%%%%%%%%%%%%%%%%%%%%%%%%%%%%%%%%%%%%%%
\afterpage{
\begin{figure*}
\centering
\includegraphics[angle=0, width=\hsize]{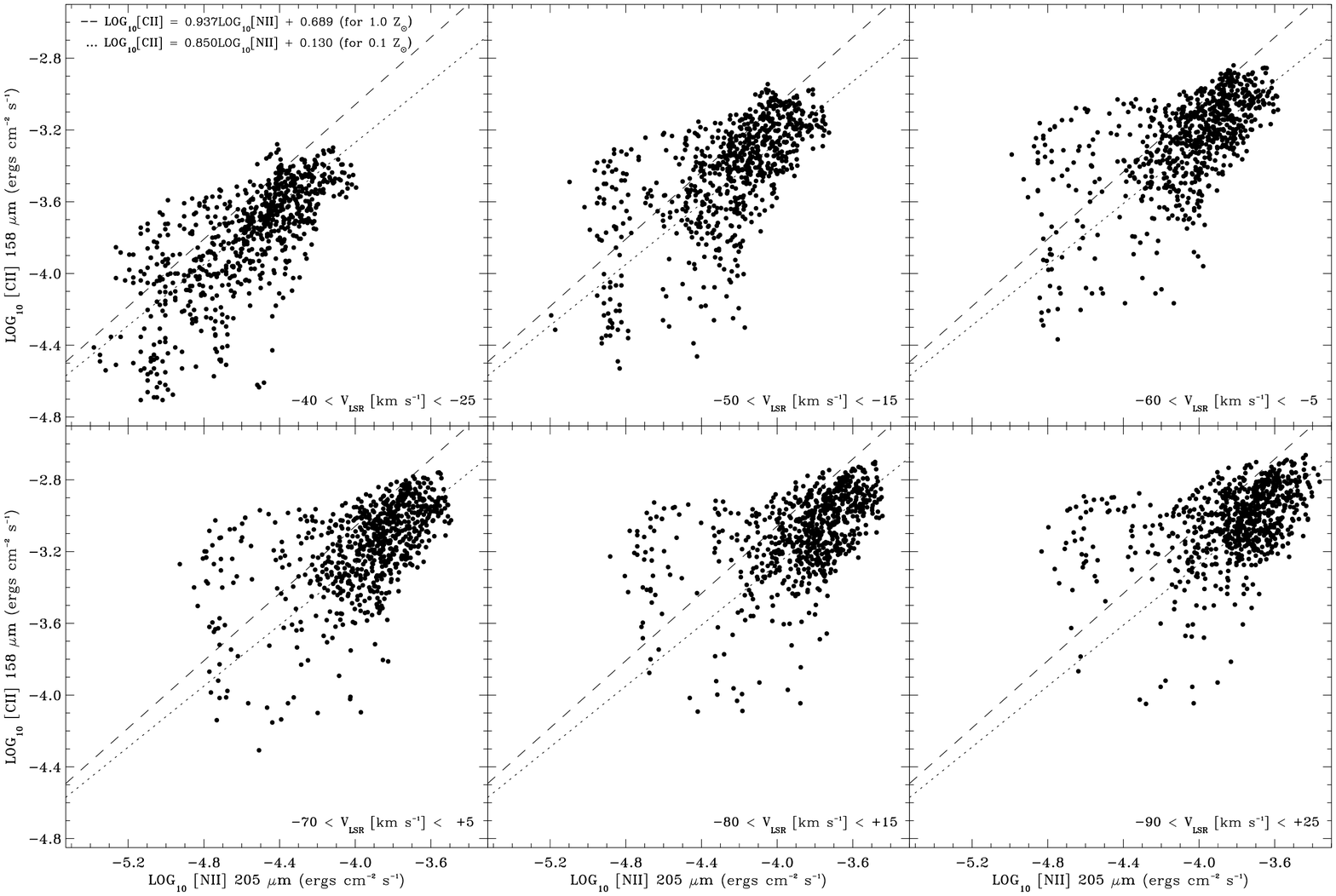}
\caption{Scatter plot of  I(\CII) versus I(\NII) integrated intensities calculated for increasingly larger radial velocity ranges from $-$40 \KMS\textrm{ }to $-$25 \KMS,$-$50 \KMS\textrm{ }to $-$15 \KMS, $-$60 \KMS\textrm{ }to $-$5 \KMS, $-$70 \KMS\textrm{ }to $+$5 \KMS,$-$80 \KMS\textrm{ }to $+$15 \KMS, and $-$90 \KMS\textrm{ }to $+$25 \KMS\textrm{ }(corresponding to I(\CII)$_{i}$ versus I(\NII)$_{i}$). The I(\CII) versus I(\NII) relationship in \citet{abel2006a} for metallicities 1.0 Z\SOLAR\textrm{ }and 0.1 Z\SOLAR\textrm{ }are shown as dashed and dotted straight lines in each panel. As the radial  velocity range increases, the bulk of the data points move along the \citet{abel2006a} prediction scale for the \CII\textrm{ }emission contribution of \HII\textrm{ }regions. \label{arches:abel_evolution} }
\end{figure*}}
%%%%%%%%%%%%%%%%%%%%%%%%%%%%%%%%%%%%%%%%%%%%%%%%%%%%%%%%%%%%%%%%%%%%%%%%%%%%%%%%%%%%%%%%%%%%%%%%%%%%

%Scatter Figure: Models
With none of the aforementioned reasons able to fully account for the behavior of the I(\CII) versus I(\NII) scatter relationship in Figure \ref{arches:abel_evolution}, we explore the validity of the main underlying assumptions in the \citet{abel2006a} theoretical work to account for it. There, the simplified case of an \HII\textrm{ }region and PDR in pressure equilibrium, for a single massive star, elemental abundances for the Orion Nebula, and a constant [C/N] ratio, are assumed. By the time \citet{abel2006a} was published, their model was quite advanced in the sense that no other theoretical code could handle both \HII\textrm{ }and PDR regions. In terms of their assumed elemental abundances, variations in the [C/N] ratio were not investigated. Following \citet{heiles1994}, it is possible to write I(\CII)$_{i}$/I(\NII)$_{i}$ as a function of the [C/N] abundance ratio, assumed to be $\sim$ [N(C$^{+}$)/N(N$^{+}$)], and the stellar temperature T$_{\star}$ as follows:

\begin{equation}\label{eq:cii_nii_relation}
\frac{I(\CII)_{i}}{I(\NII)_{i}} \approx 4.8\times\Bigg[\frac{N(C^{+})}{N(N^{+})}\Bigg]\times\Bigg(\frac{T_{\star}}{10^{4}}\Bigg)^{0.15}.
\end{equation}
From Equation \ref{eq:cii_nii_relation}, it is clear that the [C/N] abundance ratio is the dominant factor controlling the expected I(\CII)$_{i}$/I(\NII)$_{i}$ ratio, while it depends only marginally on the assumed stellar temperature. Moreover, \citet{abel2006a} showed that the intensity ratio is relatively insensitive to ionization parameter U (defined as the dimensionless ratio of hydrogen-ionizing flux to hydrogen density), and as a consequence to the assumed stellar continuum, which is a function of U and T$_{\star}$. \\

The abundance ratio dependence in Equation \ref{eq:cii_nii_relation} is specially interesting since it suggests that decoupled carbon and nitrogen elemental abundances in the GC could be driving the observed scatter relationship in Figure \ref{arches:abel_evolution}. For instance, the carbon abundance is thought to be between 3 \citep{arimoto1996} and 10 times the solar value \citep{sodroski1995,oka2005} in the GC, while the nitrogen elemental abundance is poorly constrained  \citep{rodriguez2005,simpson1995}. At the same time, secondary nitrogen production has been found in other Galactic nuclei \citep{liang2006,perezmontero2011}. \citet{liang2006} examined the oxygen abundance and [N/O] abundance ratio in nearly 40.000 metal-rich (8.4 $<$ 12 + log (O/H) $<$ 9.3 or equivalently 0.6 $<$ $Z$ $<$ 4.5) star-forming galaxies and their nuclei. They reported an increasing [N/O] abundance ratio with increasing 12 $+$ $\log$(O/H) values, interpreting such a trend as being consistent with a combination of primary nitrogen production (from massive stars forming in primordial gas) and secondary (from intermediate to low-mass stars), the latter being the dominant process in high-metallicity environments. \citet{perezmontero2011} found that nitrogen is overabundant in blue compact dwarf galaxies, which is something that cannot be explained solely by the presence of WR stars, pointing to a secondary production of nitrogen. \citet{florido2015} found observational evidence that bars in galaxies enhance the [N/O] abundance ratio, with $\log$(N/O) $\sim$ 0.09 dex larger than in unbarred galaxies. The presence of a bar also tends to enhance dust concentrations, star formation, and electron densities in the central regions of galaxies. In their work, they suggest that secondary nitrogen production could explain, at least qualitatively, the higher [N/O] abundance ratio observed in bulges less massive than $\sim$ 10$^{10}$ M\SOLAR, which puts the central bar of the Milky Way into this category \citep{mezger1996}. \\

%New Abel Models
Since the \citet{abel2006a} calculations did not investigate the variation in \NABUNDANCE\textrm{ }abundance ratio and did not include SEDs consistent with stellar clusters, we broadened their calculations to investigate the model sensitivities to these parameters. The revised calculations use a Starburst99 SED \citep{leitherer2010}, which is more appropriate for gas irradiated by a stellar cluster, and they use a range of ionization parameters varying from $\log$ U $=$ $-$4 to $-$1, in increments of 1 dex. Our choice of SED is a 4 Myr instantaneous starburst.  Our calculations include two metallicities (1 Z\SOLAR\textrm{ }and 2 Z\SOLAR) and two different hydrogen densities (2.5 and 3.5 in $\log$ n(H) units).  Elemental abundances in the model range from [C/H]  $=$ (0.13 $-$ 1.00)$\times$10$^{-3}$ and [N/H]  $=$ (0.40 $-$ 3.16)$\times$10$^{-4}$ which include the corresponding Orion Nebula (ON) values in \citet{simondiaz2011} [C/H]$_{\textrm{ON}}$ $=$ 2.334$\times$10$^{-4}$ and [N/H]$_{\textrm{ON}}$ $=$ 0.832$\times$10$^{-4}$, which are taken (in this work) as the standard elemental abundances for the Galactic disk. Thus, the model [C/N] ratios vary between 0.84 and 12.65, whereas  [C/N]$_{\textrm{ON}}$ $=$ 2.81 is obtained for the Galactic disk. Carbon and nitrogen abundances for each model are scaled between 0.5 and 2, in increments of 0.25.  Scaling them in this way implies that our models vary the [C/N] abundance ratio by a factor of 64. Our calculations include only the \HII\textrm{ }region, which we consider to end when the hydrogen ionization fraction falls below 0.01. Our results are shown in Figure \ref{arches:abel_models}, where the model [C/N] abundance ratios for the AF ([C/N]$_{\textrm{AF}}$) have been normalized to the Galactic disk value. For comparison, the (extrapolated) \citet{heiles1994} linear correlation between I(\CII) versus I(\NII) derived for the extended low-density warm ionized medium (ELDWIM) (Equation 12b in their work) is shown. A direct comparison between the measurements obtained toward the AF and the results of \citet{heiles1994} is difficult as the latter were derived from measurements in the $-$4.70 $<$ $\log$ I(\CII) $<$ $-$3.74 range, much lower than the overall integrated intensities in the AF. The scatter in Figure \ref{arches:abel_models} can be largely explained by abundance variations in the \NABUNDANCE\textrm{ }ratio from our model, which mimic the scatter in the I(\CII)$_{i}$ versus I(\NII)$_{i}$ relationship reasonably well. Only a small amount of scatter can be attributed to variations in the hardness of the radiation field due to changes in the ionization parameter U, as the I(\CII)$_{i}$/I(\NII)$_{i}$ ratio is largely insensitive to it over the parameter space covered by our calculations \citep{abel2006a}. The models most consistent with the general behavior of the measurements have metallicities between 1 Z\SOLAR\textrm{ }and 2 Z\SOLAR, a volume density of $\log$ n(H) $=$ 3.5, and an ionization parameter of $\log$ U $=$ $-$1 to $-$2. These values are fully consistent with average conditions of the gas in the GC \citep{mezger1996,guesten2004,rodriguez2005}. The vast majority of data points are well reproduced by models that satisfy [C/N]$_{\textrm{AF}}$ $<$ [C/N]$_{\textrm{ON}}$. Given the large carbon elemental abundances found toward the GC ranging from 3 to 10 times the solar value \citep{arimoto1996,sodroski1995,oka2005}, the model results generally suggest the presence of a nitrogen overabundance in the GC with regard to the value in the Galactic disk \citep{simondiaz2011} that keeps [C/N]$_{\textrm{AF}}$ $<$ [C/N]$_{\textrm{ON}}$ in the AF region. Moreover, these results suggest that [N/H] does not scale in the same way as [C/H] when moving from the Galactic disk toward the GC, but that it must rise faster in order to keep [C/N]$_{\textrm{AF}}$ $<$ [C/N]$_{\textrm{ON}}$. If this is indeed the case, any combination of physical processes responsible for this would have to satisfy at least these two conditions: (1) to enrich the GC ISM with nitrogen more efficiently than with carbon, and (2) to transport this nitrogen-enriched gas toward the inner parts of the GC, where the AF are thought to be located. \\

%%%%%%%%%%%%%%%%%%%%%%%%%%%%%%%%%%%%%%%%%%%%%%%%%%%%%%%%%%%%%%%%%%%%%%%%%%%%%%%%%%%%%%%%%%%%%%%%%%%%
\afterpage{
\begin{figure*}
\centering
\includegraphics[angle=0, width=\hsize]{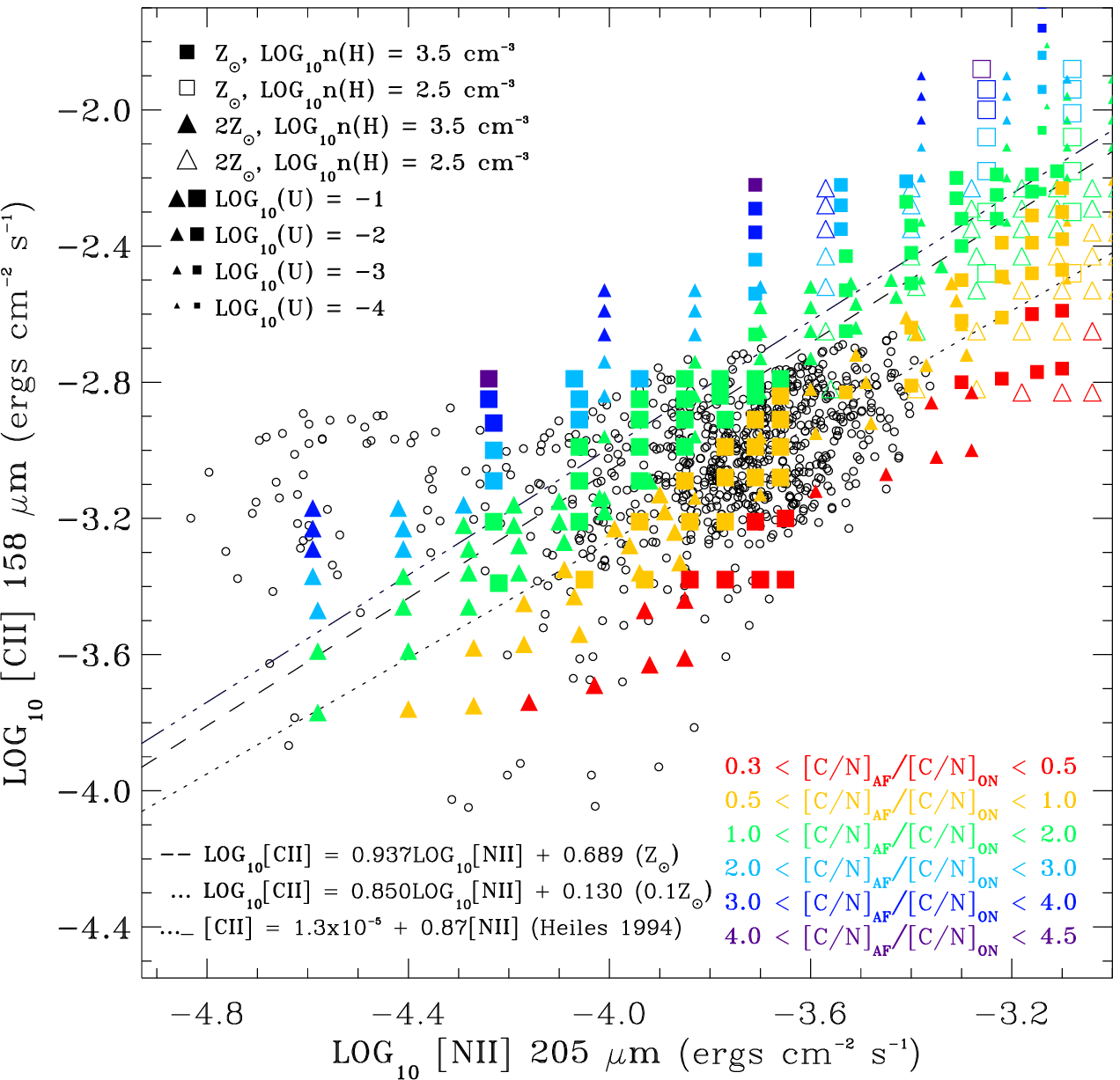}
\caption{Scatter plot of the I(\CII)$_{i}$ versus I(\NII)$_{i}$ integrated intensities. Data points are shown as open black circles. The \citet{abel2006a} predictions for the \CII\textrm{ }emission contribution from  \HII\textrm{ }regions, for metallicities 1.0 Z\SOLAR\textrm{ }and 0.1 Z\SOLAR\textrm{ }(Equations 1 and 2 in their work), and the \citet{heiles1994} linear correlation between I(\CII) and I(\NII) for the extended low-density warm ionized medium (ELDWIM) (Equation 12b in their work) are shown as dashed, dotted, and dotted-dashed lines, respectively. The model predictions in this work, using different [C/N] abundance ratios normalized to the Galactic disk value in \citet{simondiaz2011} are displayed in colors. Metallicities 1.0 Z\SOLAR\textrm{ }and 2.0 Z\SOLAR\textrm{ }are shown as filled and open $\blacksquare$ and $\blacktriangle$ symbols for $\log$ n(H) $=$ 3.5 and 2.5, respectively. The symbol size is proportional to the ionization parameter $\log_{10}$U $=$ $-$1, $-$2,$-$3, and $-$4.\label{arches:abel_models}}
\end{figure*}}
%%%%%%%%%%%%%%%%%%%%%%%%%%%%%%%%%%%%%%%%%%%%%%%%%%%%%%%%%%%%%%%%%%%%%%%%%%%%%%%%%%%%%%%%%%%%%%%%%%%%

For the first condition, it would be necessary for carbon and nitrogen production to be decoupled. \citet{garnett1995} found that, in irregular galaxies, the trend of the \NABUNDANCE\textrm{ }abundance ratio increases with [O/H] abundance, while for solar abundance stars and \HII\textrm{ }regions, it abruptly decreases, suggesting that the bulk production of nitrogen might indeed be decoupled from the carbon production in those galaxies. In terms of nitrogen formation in the CNO cycle, the precursor agent for $^{14}$N is the $^{13}$C carbon isotope via the $^{13}$C $+$ H $\rightarrow$ $^{14}$N $+$ $\gamma$ reaction. The relative amount of $^{13}$C in the GC is about three times that available in the solar neighborhood with [$^{12}$C/$^{13}$C] $\sim$ 25 \citep{guesten2004}, so stars forming from non-primordial gas in the GC will already contain a larger amount of $^{13}$C available for the formation of nitrogen than in other parts of the Galaxy. In this case, the secondary production of nitrogen from low- to intermediate-mass stars in the GC bulge could be a plausible origin of such an elemental abundance increase. Moreover, secondary nitrogen production has been reported for other Galactic nuclei similar to our GC \citep{liang2006,perezmontero2011,florido2015}, and in our own Galaxy the synthesis of nitrogen via  secondary production has been found to be a very important contributor to the present-day [N/H], especially for \HII\textrm{ }regions in the inner 5 kpc of the Galactic disk \citep{martin_hernandez2002}. Concerning the transport of nitrogen-enriched gas, the mass supply mechanism to the GC has to be investigated. Galactic bar models suggest that ejected gas from stars in the bulge can replenish the gas in the outer X$_{1}$ orbits within 1 to 2 orbital periods of the stellar bar rotation \citep{jenkins1994,mezger1996,mertel2018}. This gas would then migrate to the inner X$_{2}$ orbits via loss of angular momentum, mixing with the gas already in these orbits, keeping [N/H] high with regard to the Galactic disk value. As a result, a large fraction of the metals in the GC (carbon, nitrogen, and oxygen mainly) would not have originated within the GC. They would have been produced along the entire length of the Galactic bulge ($\sim$ few kpc) and then transported to the GC by gas flows \citep{mertel2013}. If this scenario is correct,  we expect elemental abundances in the GC to be consistent with values found at somewhat larger galactocentric radii (R$_{\rm{GAL}}$). Given that secondary nitrogen production in galactic nuclei of barred galaxies with physical conditions similar to the Milky Way's GC has been claimed to be responsible for the enhancement of the [N/H] abundance, we suggest that the scatter in Figure \ref{arches:abel_models} is consistent with the same [N/H] enhancement mechanism in the Milky Way's GC. If the gas supply mechanism for the GC is mainly due to the mass loss from low- to intermediate-mass stars in the Galactic bulge, this would not only impact the [N/H] abundance via secondary nitrogen production, but it would alter the elemental abundances of other species such as  the $^{17}$O and $^{18}$O oxygen isotopes as well, lowering the [$^{18}$O/$^{17}$O] ratio with regard to the Galactic value $\gtrapprox$ 4 \citep{wouterloot2008}. Consistently, lower values around  [$^{18}$O/$^{17}$O] $\sim$ 3 have been reported in a few places within the GC volume  \citep{penzias1981,zhang2015}. Moreover, our ongoing work on the [$^{18}$O/$^{17}$O] abundance ratio in the AF region (Steinke et al. 2021+, in prep.) has preliminary results consistent with the values previously reported for the GC. \\

%%%%%%%%%%%%%%%%%%%%%%%%%%%%%%%%%%%%%%%%%%%%%%%%%%%%%%%%%%%%%%%%%%%%%%%%%%%%%%%%%%%%%%%%%%%%%%%%%%%%
%%%%%%%%%%%%%%%%%%%%%%%%%%%%%%%%%%%%%%%%%%%%%%%%%%%%%%%%%%%%%%%%%%%%%%%%%%%%%%%%%%%%%%%%%%%%%%%%%%%%
%%%%%%%%%%%%%%%%%%%%%%%%%%%%%%%%%%%%%%%%%%%%%%%%%%%%%%%%%%%%%%%%%%%%%%%%%%%%%%%%%%%%%%%%%%%%%%%%%%%%

\section{[CII] emission from \HII\textrm{ }regions}\label{hii_contribution}

The \CII\textrm{ }and \NII\textrm{ }emission distributions in the AF are spatially and spectrally very closely related \citep{garcia2016}. Therefore, given their different physical origin, it is expected that part of the observed \CII\textrm{ }emission does not originate from PDRs, but instead from the \HII\textrm{ }region component. In order to disentangle the latter from the total observed \CII\textrm{ }emission, we  \emph{\emph{re-calibrated}} the I(\CII) versus I(\NII) relationship in \HII\textrm{ }regions \citep{abel2006a} for the physical conditions of the gas in the AF and applied this to the radial velocity components listed in Table \ref{tab:data_summary}. This is shown in Figure \ref{arches:abel_3}. In the figure, the model data points satisfy  0.3 < [C/N]$_{\textrm{AF}}$/[C/N]$_{\textrm{ON}}$ $<$ 0.5,  while measurement points  represent  I(\CII)$_{g}$ versus I(\NII)$_{s}$ integrated intensities. A least-square fit  to the model data points yields 
\begin{equation}\label{eq:c_n_fit}
\log I(\CII_{\rm{HII}}) = 1.068\times \log I(\NII) + 0.645.
%\log I(\CII) = (1.068 \pm 0.056)\times \log I(\NII) + (0.645 \pm 0.199) 
\end{equation}
% Explanation how results were calculated
Equation \ref{eq:c_n_fit} allows us to predict the \CII\textrm{ }emission contribution from the \HII\textrm{ }region I(\CII$_{\rm{HII}}$) in the AF based on our \NII\textrm{ }line observations for each of the identified radial velocity components. Then, by subtracting \CII$_{\rm{HII}}$ from the total \CII\textrm{ }emission, the amount of \CII\textrm{ }emission originating only in the PDR component I(\CII$_{\rm{PDR}}$), was derived. Following our integrated intensity definitions in Section \ref{arches:selected_positions_sec}, I(\NII)$_{s}$ and I(\NII)$_{i}$ were used in Equation \ref{eq:c_n_fit}, yielding the I(\CII$_{\rm{HII}}$) values I$_{s}$ and I$_{i}$ listed in Table \ref{tab:hii_contribution}, respectively. Then, I(\CII$_{\rm{PDR}}$) was obtained by subtracting the I(\CII$_{\rm{HII}}$) estimations from the I(\CII)$_{g}$, I(\CII)$_{s}$,  and I(\CII)$_{i}$  integrated intensities. These results are represented by the values D$_{g,s}$ $=$ I(\CII)$_{g}$ $-$ I$_{s}$, D$_{s,s}$ $=$ I(\CII)$_{s}$ $-$ I$_{s}$, and D$_{i,i}$   $=$ I(\CII)$_{i}$ $-$ I$_{i}$ in same table, with subindexes $g$, $s$, and $i$ referring to the integrated intensities defined in Section \ref{intensities}. Finally, the relative \CII\textrm{ }emission contribution from PDRs to the total observed emission contained in I(\CII)$_{g}$, I(\CII)$_{s}$, and I(\CII)$_{i}$, is estimated as R$_{gs,g}$ $=$ D$_{g,s}$/I(\CII)$_{g}$, R$_{ss,s}$ $=$ D$_{s,s}$/I(\CII)$_{s}$, and R$_{ii,i}$ $=$ D$_{i,i}$/I(\CII)$_{i}$, respectively. \\

%%%%%%%%%%%%%%%%%%%%%%%%%%%%%%%%%%%%%%%%%%%%%%%%%%%%%%%%%%%%%%%%%%%%%%%%%%%%%%%%%%%%%%%%%%%%%%%%%%%%
\afterpage{
\begin{figure}
\centering
\includegraphics[angle=0, width=\hsize]{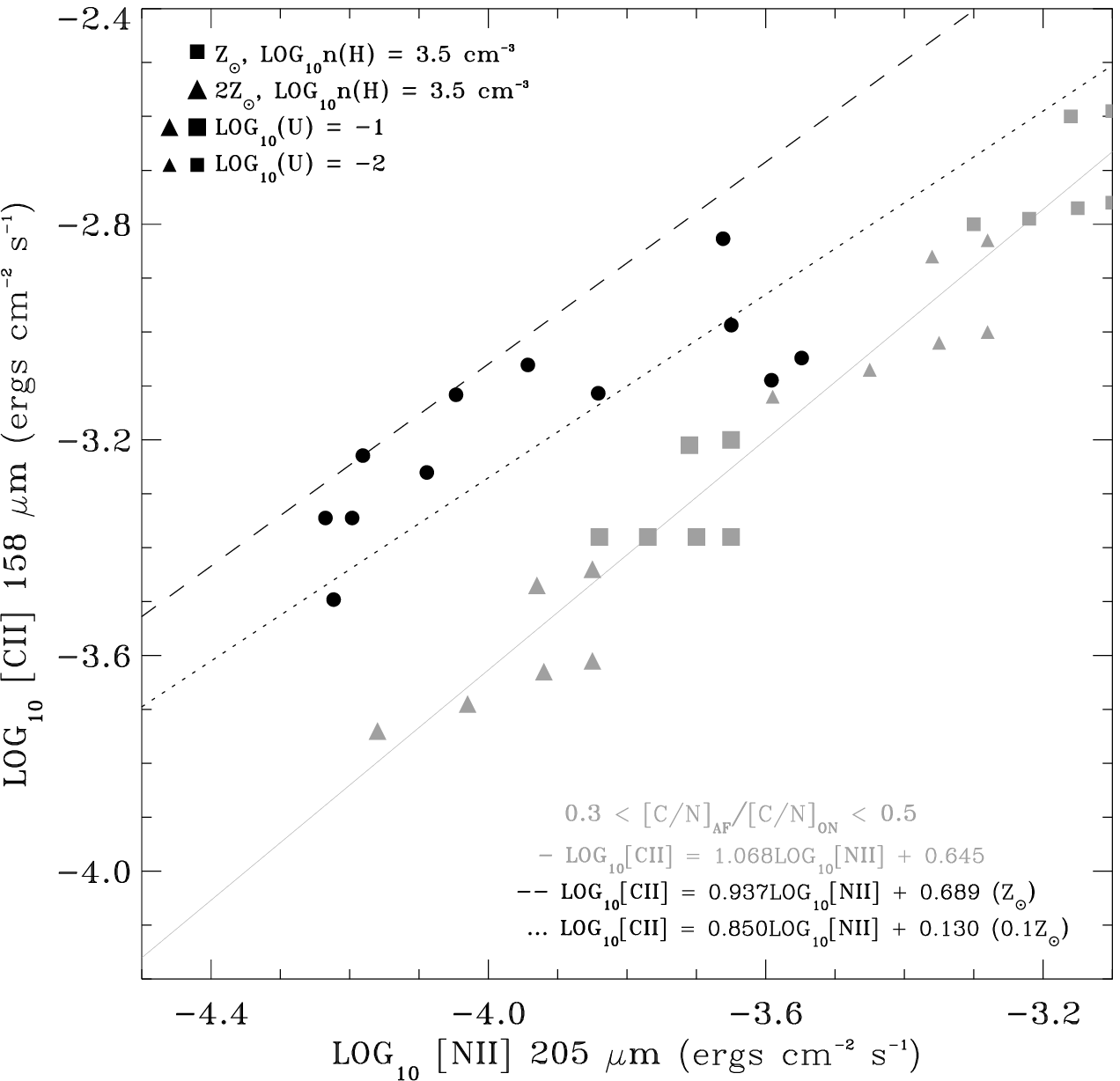}
\caption{Scatter plot of I(\CII)$_{g}$ versus I(\NII)$_{s}$ integrated intensities for radial velocity components listed in Table \ref{tab:data_summary}. Data points are shown as filled black circles, while model points satisfying 0.3 $<$ [C/N]$_{\textrm{AF}}$/[C/N]$_{\textrm{ON}}$ $<$ 0.5 are shown in gray, with symbol shape and size depending on the gas physical parameters, as listed in the upper left corner of the figure. A least-squares fit to the model predictions yields $\log$ I(\CII$_{\rm{HII}}$) $=$ 1.068$\times \log$ I(\NII) $+$ 0.645, shown by the gray straight line. For comparison, the expected \CII\textrm{ }emission contribution from \HII\textrm{ }regions in the \citet{abel2006a} models for 1.0 Z\SOLAR\textrm{ }and 0.1 Z\SOLAR\textrm{ }are shown as dashed and dotted black straight lines, respectively.\label{arches:abel_3}}
\end{figure}
}
%%%%%%%%%%%%%%%%%%%%%%%%%%%%%%%%%%%%%%%%%%%%%%%%%%%%%%%%%%%%%%%%%%%%%%%%%%%%%%%%%%%%%%%%%%%%%%%%%%%%

% Explanation of results in Table 3 and the explicit/implicit assumptions in deriving these numbers\\
To obtain the results in Table  \ref{tab:hii_contribution}, two main assumptions were made. First, we assumed that all the features present in the observed \CII\textrm{ }spectra were the result of the superposition of emission and absorption features arising from foreground sources in the Galactic disk (e.g., spiral arms whose emission contribution was properly removed by fitting Gaussian profiles to the observed spectra), gas associated with an \HII\textrm{ }region component, and gas associated with a PDR component. We did not include the possibility of \CII\textrm{ }emission arising from diffuse CO$-$dark H$_{2}$ clouds, which has been claimed to be the case for Galactic disk sources \citep{langer2014,langer2017}. This is justified by the large average volume densities ($\sim$ 10$^{4}$ cm$^{-3}$) in the GC \citep{guesten2004}, which make such a contribution small, if any at all. Second, the results shown in Table \ref{tab:hii_contribution} rely entirely on Equation \ref{eq:c_n_fit} being a reasonable model to account for the \CII\textrm{ }emission contribution from \HII\textrm{ }regions to the observed \CII\textrm{ }emission in the AF. Models satisfying 0.3 $<$ [C/N]$_{\textrm{AF}}$/[C/N]$_{\textrm{ON}}$ $<$ 0.5 can explain the I(\CII)$_{i}$ versus I(\NII)$_{i}$ scatter relationship in Figure \ref{arches:abel_models}, in the same way shown in Figure 6 of \citet{abel2006a} for the Orion Nebula. For [C/N]$_{\textrm{ON}}$ $=$ 2.81 \citep{simondiaz2011}, our results imply that the [C/N]$_{\textrm{AF}}$ abundance ratio has to satisfy
\begin{equation}\label{eq:c_n_for_PDR}
0.84 < \textrm{[C/N]}_{\textrm{AF}} < 1.41.
\end{equation}
We defer the discussion on the feasibility for the gas in the AF to have [C/N]  abundance ratios in the range defined by the condition in Equation \ref{eq:c_n_for_PDR} to Section \ref{c_n_ratio_af}. \\

%%%%%%%%%%%%%%%%%%%%
% INTEGRATED INTENSITIES PDR %
%%%%%%%%%%%%%%%%%%%%
\afterpage{
\begin{sidewaystable*}
\centering
\caption{\CII\textrm{ }emission contribution from \HII\textrm{ }regions (\CII$_{\rm{HII}}$) and PDRs (\CII$_{\rm{PDR}}$) in the Arched Filaments (in units of ergs s$^{-1}$ cm$^{-2}$) for the radial velocity components listed in Table \ref{tab:data_summary}. The name of each radial velocity component is listed in column 1. The predicted \CII$_{\rm{HII}}$ emission contribution obtained using Equation \ref{eq:c_n_fit} and either the I(\NII)$_{s}$ or the I(\NII)$_{i}$ integrated intensities in Table \ref{tab:data_summary} is listed in columns 2 and 3, respectively. Columns 4 to 6 contain the \CII$_{\rm{PDR}}$ emission contribution obtained by subtracting the \CII$_{\rm{HII}}$ predictions (listed in columns 2 and 3) from the integrated intensities I(\CII)$_{g}$, I(\CII)$_{s}$, and I(\CII)$_{i}$ listed in Table \ref{tab:data_summary}, where D$_{g,s}$ $=$ I(\CII)$_{g}$ $-$ I$_{s}$, D$_{s,s}$ $=$ I(\CII)$_{s}$ $-$ I$_{s}$, and D$_{i,i}$   $=$ I(\CII)$_{i}$ $-$ I$_{i}$, respectively. Columns 7 to 9 contain the relative emission contribution from PDRs wrt. the total \CII\textrm{ }integrated intensities in Table \ref{tab:data_summary}, with R$_{gs,g}$ $=$ D$_{g,s}$/I(\CII)$_{g}$, R$_{ss,s}$ $=$ D$_{s,s}$/I(\CII)$_{s}$, and R$_{ii,i}$ $=$ D$_{i,i}$/I(\CII)$_{i}$, respectively.\label{tab:hii_contribution}}
\bgroup
\def\arraystretch{1} 
\begin{tabular}{l|cc|ccc|ccc}
\toprule
\multicolumn{1}{c}{\textbf{Source}}                                   &
\multicolumn{2}{c}{\textbf{\CII$_{\rm{HII}}$}}                   &
\multicolumn{3}{c}{\textbf{\CII$_{\rm{PDR}}$ (D)}}          &
\multicolumn{3}{c}{\textbf{\CII$_{\rm{PDR}}$/\CII (R)}}  \\

\multicolumn{1}{c}{}          &
\multicolumn{1}{c}{I$_{s}$}   &
\multicolumn{1}{c}{I$_{i}$}   &
\multicolumn{1}{c}{D$_{g,s}$} &
\multicolumn{1}{c}{D$_{s,s}$} &
\multicolumn{1}{c}{D$_{i,i}$} &
\multicolumn{1}{c}{R$_{gs,g}$} &
\multicolumn{1}{c}{R$_{ss,s}$} &
\multicolumn{1}{c}{R$_{ii,i}$} \\

\multicolumn{1}{c}{}          &
\multicolumn{1}{c}{10$^{-4}$}  &
\multicolumn{1}{c}{10$^{-4}$}  &
\multicolumn{1}{c}{10$^{-4}$}  &
\multicolumn{1}{c}{10$^{-4}$}  &
\multicolumn{1}{c}{10$^{-4}$}  &
\multicolumn{1}{c}{}          &
\multicolumn{1}{c}{}          &
\multicolumn{1}{c}{}          \\
\midrule

  P1-46\KMS   &    2.11  $\pm$    0.15  &      5.34  $\pm$    0.34  &      5.55  $\pm$    0.23  &      6.89  $\pm$    0.22  &     10.87  $\pm$    0.47  &      0.72  $\pm$    0.03  &      0.77  $\pm$    0.03  &      0.67  $\pm$    0.03 \\  
  P1-13\KMS   &    1.36  $\pm$    0.16  &      5.34  $\pm$    0.34  &      1.83  $\pm$    0.26  &      2.44  $\pm$    0.24  &     10.87  $\pm$    0.47  &      0.57  $\pm$    0.09  &      0.64  $\pm$    0.07  &      0.67  $\pm$    0.03 \\  
  P2-15\KMS   &    1.46  $\pm$    0.11  &      5.68  $\pm$    0.28  &      3.06  $\pm$    0.17  &      4.09  $\pm$    0.18  &     11.08  $\pm$    0.43  &      0.68  $\pm$    0.04  &      0.74  $\pm$    0.04  &      0.66  $\pm$    0.03 \\  
  W1-24\KMS  &    7.18  $\pm$    0.26  &      9.47  $\pm$    0.36  &      1.77  $\pm$    0.43  &      1.83  $\pm$    0.35  &      0.60  $\pm$    0.48  &      0.20  $\pm$    0.05  &      0.20  $\pm$    0.04  &      0.06  $\pm$    0.05 \\  
  W2-29\KMS  &    5.58  $\pm$    0.21  &      9.05  $\pm$    0.33  &      4.68  $\pm$    0.44  &      4.07  $\pm$    0.31  &      4.95  $\pm$    0.48  &      0.46  $\pm$    0.05  &      0.42  $\pm$    0.03  &      0.35  $\pm$    0.04 \\  
G010-17\KMS &    3.47  $\pm$    0.17  &      9.02  $\pm$    0.32  &      4.23  $\pm$    0.28  &      4.20  $\pm$    0.23  &      9.51  $\pm$    0.43  &      0.55  $\pm$    0.04  &      0.55  $\pm$    0.03  &      0.51  $\pm$    0.02 \\  
G007-16\KMS &    2.73  $\pm$    0.16  &      7.38  $\pm$    0.33  &      5.96  $\pm$    0.31  &      7.04  $\pm$    0.22  &     11.76  $\pm$    0.44  &      0.69  $\pm$    0.04  &      0.72  $\pm$    0.03  &      0.61  $\pm$    0.02 \\  
G007-41\KMS &    1.32  $\pm$    0.15  &      7.38  $\pm$    0.33  &      3.20  $\pm$    0.21  &      3.59  $\pm$    0.21  &     11.76  $\pm$    0.44  &      0.71  $\pm$    0.05  &      0.73  $\pm$    0.05  &      0.61  $\pm$    0.02 \\  
 E2N-18\KMS  &    5.42  $\pm$    0.29  &      5.17  $\pm$    0.30  &      9.45  $\pm$    0.51  &      9.05  $\pm$    0.40  &      9.09  $\pm$    0.41  &      0.64  $\pm$    0.04  &      0.63  $\pm$    0.03  &      0.64  $\pm$    0.03 \\  
 E2S-17\KMS &    6.43  $\pm$     0.28  &      6.80  $\pm$    0.31  &      1.70  $\pm$    0.37  &      2.01  $\pm$    0.38  &      1.64  $\pm$    0.41  &       0.21  $\pm$    0.05  &      0.24  $\pm$    0.05  &      0.19  $\pm$    0.05 \\  
  E1-40\KMS   &    1.90  $\pm$    0.17  &      5.52  $\pm$    0.30  &      3.59  $\pm$    0.63  &      2.94  $\pm$    0.25  &      9.75  $\pm$    0.44  &      0.65  $\pm$    0.14  &      0.61  $\pm$    0.06  &      0.64  $\pm$    0.03 \\  
  E1-16\KMS   &    1.51  $\pm$    0.13  &      5.52  $\pm$    0.30  &      4.39  $\pm$    0.54  &      5.24  $\pm$    0.20  &      9.75  $\pm$    0.44  &      0.74  $\pm$    0.11  &      0.78  $\pm$    0.04  &      0.64  $\pm$    0.03 \\  

\bottomrule
\end{tabular}
\egroup
\end{sidewaystable*}}

Calculating integrated intensities for both species in three different ways (I$_{g}$, I$_{s}$, and I$_{i}$) allows us to check the consistency between different \CII$_{\rm{HII}}$ and \CII$_{\rm{PDR}}$ estimates, since the complexity of the line profiles shown in Appendix \ref{appendix:gaussian_fits} could easily yield misleading results. The relative \CII\textrm{ }emission contributions from PDRs to the observed total emission R$_{gs,g}$, R$_{ss,s}$, and R$_{ii,i}$ are remarkably similar. This is particularly true for R$_{gs,g}$ and R$_{ss,s}$, which are fully consistent within the error uncertainties. In the case of W1-24\KMS, the derived R$_{ii,i}$ value is much lower than both R$_{gs,g}$ and R$_{ss,s}$, outside error estimates. From the W1-24\KMS\textrm{ }spectra in Appendix \ref{appendix:gaussian_fits}, it seems that \NII\textrm{ }emission unrelated to the source might be included in the  $-$90 \KMS\textrm{ }to $+$25 \KMS\textrm{ }radial velocity range, artificially increasing the I(\CII$_{\rm{HII}}$) prediction via Equation \ref{eq:c_n_fit}, decreasing the PDR contribution for this component, and consequently yielding an artificially low R$_{ii,i}$ value. The R$_{gs,g}$ values in Table \ref{tab:hii_contribution} are the closest to the true fraction of \CII\textrm{ }emission originating from PDRs in the AF as they are derived from the I(\CII)$_{g}$ and I(\NII)$_{s}$ integrated intensities, which are the most accurate approximation of the true emission of the radial velocity components in both spectral lines. The fraction of \CII\textrm{ }emission in PDRs varies significantly among the different radial velocity components, from $\sim$ 0.20 up to $\sim$ 0.75, with no apparent spatial trend with regard to the position of the Arches Cluster, shown in Figure \ref{intro:filaments_figure}. Our results are in agreement with the predictions in \citet{abel2006a}  showing that between 10\% and 60\% of the observed \CII\textrm{ }emission can be expected to arise from \HII\textrm{ }regions in the Orion Nebula. \citet{langer2017} used \CII\textrm{ }line observations from the GOT C$^{+}$ Survey \citep{langer2010,pineda2013,langer2014} and \NII\textrm{ }line observations from a Herschel-HIFI survey of ten GOT C$^{+}$ lines \citep{goldsmith2015,langer2016} to disentangle the \CII\textrm{ }emission contribution from the fully ionized and neutral (defined as the combination of PDR and CO$-$dark H$_{2}$) gas, toward the LOS ($l$, $b$) $=$ (0,0), and within the GC volume. They reported that the fraction of \CII\textrm{ }emission arising from the fully ionized and neutral gas toward this LOS is $\sim$ 0.30 and $\sim$ 0.70 of the total observed \CII\textrm{ }emission, respectively. Our results in Table \ref{tab:hii_contribution} show that at least 5 out of the 12 radial velocity components are fully consistent with their results, within error uncertainties, despite the fact that they are found toward different LOS. Since \citet{langer2017} showed that the \CII\textrm{ }emission in the central molecular zone (CMZ) arises primarily from PDRs and highly ionized gas, we have excluded the contribution of any CO$-$dark H$_{2}$ gas in the AF from the comparison. For the other components in Table \ref{tab:hii_contribution}, the fraction of \CII\textrm{ }emission found toward \HII\textrm{ }regions and PDRs changes substantially. This reinforces the idea that extrapolating these results to different LOS, or even to other locations within the GC volume, has to be taken with caution. Moreover, such large variations could have a strong impact when trying to model the \CII\textrm{ }emission from the AF, as \CII\textrm{ }emission is a standard probe of the FUV fields in PDRs, dominating the cooling of atomic gas at low FUV fields (G$_{\circ}$ $<$ 10$^{3}$) and low densities (n$_{crit}$(\CII) $\sim$ 3$\times$10$^{3}$ cm$^{-3}$) \citep{hollenbach1999}. Hence, any PDR modeling attempt that does not properly disentangle the \CII\textrm{ }emission contributions from \HII\textrm{ }regions and PDRs from the total \CII\textrm{ }emission could give misleading results. \\

\section{[C/N] elemental abundance ratio in the AF}\label{c_n_ratio_af}

The results in Section \ref{hii_contribution} suggest that a lower [C/N] abundance ratio than that found toward the Galactic disk \citep{simondiaz2011} is needed to explain the I(\CII)$_{i}$ versus I(\NII)$_{i}$ observations in Figure \ref{arches:abel_models} seen toward the AF in the GC. This can be tested in a relatively independent way by combining our \NII\textrm{ }line observations and the I(\CII) integrated intensity model predictions in Equation \ref{eq:c_n_fit} with estimates for the electron density (n$_{e}$) and electron temperature (T$_{e}$) in each radial velocity component. Together, these observational/empirical constraints allow us to derive the [C/N] abundance ratio  directly, under a few reasonable assumptions, so the consistency of the [C/N] abundance ratio range in Equation \ref{eq:c_n_for_PDR} estimated for the AF region can be tested. \\

% N(N+) column densities
Following \citet{roellig2016}, under the assumption of local thermodynamic equilibrium (LTE) and optically thin \NII\textrm{ }emission \citep{goldsmith2015}, the ionized nitrogen column density N(N$^{+}$) for any given radial velocity component in Table \ref{tab:abundance_results} is derived as
\begin{equation}\label{eq:n_abundance}
N(N^{+}) = \frac{I(\NII)}{c(n_{e},T_{e})} [cm^{-2}],
\end{equation}
where I(\NII) is the integrated intensity I(\NII)$_{s}$ or I(\NII)$_{i}$ (in units of K \KMS) listed in Table \ref{tab:data_summary}. The c(n$_{e}$,T$_{e}$) coefficients are scaling factors between I(\NII) and N(N$^{+}$) as a function of the adopted n$_{e}$ and T$_{e}$, that take into account the relative population of the three fine-structure levels of the nitrogen ion ($^{3}$P$_{0}$, $^{3}$P$_{1}$, and $^{3}$P$_{2}$) due to collisions with electrons and spontaneous decay. They are calculated for the  \NII\textrm{ }205 $\mu m$ and  122 $\mu m$ transitions in \citet{roellig2016}. These coefficients are almost constant for 5000 K $<$ T$_{e}$ $<$ 8000 K  and vary around 15\% for 100 cm$^{-3}$ $<$ n$_{e}$ $<$ 500 cm$^{-3}$ \citep{roellig2016}, which are the relevant ranges for the gas physical conditions in the AF region and for the observed \NII\textrm{ }line transition at 205 $\mu m$ used in this work. The optically thin  assumption for the \NII\textrm{ }line emission is well justified as \citet{goldsmith2015} showed that, for regions with n$_{e}$ $\geq$ 100 cm$^{-3}$, column densities up to N(N$^{+}$) $\sim$ 10$^{18}$ cm$^{-2}$ still yield optically thin  emission. This is in agreement with our nitrogen column density results in Table \ref{tab:abundance_results}, which are larger than the average in the Galactic plane N(N$^{+}$) $\sim$ 4.7 $\times$ 10$^{16}$ cm$^{-2}$  \citep{goldsmith2015}, which is expected given the  \NII\textrm{ }emission concentration toward the GC.\\

% N(C+) column densities
The ionized carbon column density N(C$^{+}$) is calculated following Equation 1 in \citet{langer2017}, under the assumption of the EOT limit for the \CII\textrm{ }line transition \citep{goldsmith2012}:
\begin{equation}\label{eq:c_abundance}
N(C^{+}) = 2.92\times 10^{15}\Bigg[1 + \frac{e^{\Delta E/kT_{kin}}}{2}   \Bigg(1+\frac{n_{cr,e}}{n_{e}}\Bigg) \Bigg] I(\CII)[cm^{-2}],
\end{equation}
where I(\CII) is the \CII$_{\rm{HII}}$ integrated intensity (in units of K \KMS) listed in Table \ref{tab:hii_contribution} for each radial velocity component. In Equation \ref{eq:c_abundance}, a critical electron density n$_{cr,e}$ $\sim$ 50 cm$^{-3}$ at kinetic temperature \TKIN\textrm{ }$\sim$ 8000 K is adopted. These values are typical for \HII\textrm{ }regions and are close to the average T$_{e}$ $\sim$ 7000 K found toward the AF \citep{pauls1976,goldsmith2012,langer2015}. The fact that collision partners other than electrons are not considered in Equation \ref{eq:c_abundance} is based on the assumption that electrons dominate collisions over H and H$_{2}$ in regions where the \NII\textrm{ }line is detected \citep{langer2015}. For the \CII\textrm{ }line transition at 158 $\mu m$, the term $\Delta$E/k $=$ 91.21 K. The factor 2.92$\times$10$^{15}$  is obtained  from a combination of fundamental constants (e.g., $h$, $c$, $k$), the frequency of the \CII\textrm{ }line transition, and the adaptation of Equation \ref{eq:c_abundance} to K \KMS\textrm{ }units for the integrated intensities \citep{langer2015}. Finally, we approximate \TKIN\textrm{ }$\sim$ T$_{e}$\textrm{ }in Equation \ref{eq:c_abundance} for each radial velocity component  in Table \ref{tab:abundance_results}. \\

% Input n_e, T_e , and c(n_e,T_e) values for positions in our work.
Under the aforementioned assumptions, Equations \ref{eq:n_abundance} and \ref{eq:c_abundance} are basically a function of I(\CII) and I(\NII) integrated intensities, and the gas physical parameters n$_{e}$ and T$_{e}$.  The latter two were derived by \citet{lang2001} for the entire extension of the AF using high-resolution 8.3 GHz continuum and H92$\alpha$ recombination line observations. We assigned n$_{e}$ and T$_{e}$ estimates listed in their Tables 4 and 7 to our positions W1, W2, G010, G007, E2N, E2S, and E1 based on spatial coincidence and assuming a single value for each parameter along the LOS. For position E1, we adopted the electron density derived for their source D, which is  the spatially closest source to E1 in their data. At the location of P1 and P2, just outside the AF as shown in Figure \ref{intro:filaments_figure}, no prominent 8.3 GHz continuum emission was detected in \citet{lang2001}. Given that their measured I(\NII) integrated intensities in Table \ref{tab:data_summary} are not particularly lower than the ones detected toward the other locations within the AF, the lack of  prominent continuum emission might indicate an LOS distance effect.  Therefore, we adopted the upper limit n$_{e}$ $\sim$ 100 cm$^{-3}$ derived for molecular clouds located far from thermal continuum sources in the GC derived by \citet{rodriguez2005} and the lowest electron temperature found by \citet{lang2001} toward the AF with T$_{e}$ $\sim$ 5000 K. The c(n$_{e}$,T$_{e}$) coefficients for each position were derived by interpolating in n$_{e}$ parameter space the values in \citet{roellig2016}, as c(n$_{e}$,T$_{e}$) is constant over the relevant T$_{e}$ range in the present work. \\

% IFC and independent assumption of these results wrt. Section 6 assumptions + Discussion of results in Table 4 and comparison with model prediction
From Equations \ref{eq:n_abundance} and \ref{eq:c_abundance}, the [C/N] abundance ratio is obtained by first assuming that the gas is fully single ionized \citep{langer2017}, and secondly, that the ratio of column densities reflects the ratios of the carbon and nitrogen elemental abundances, similarly to our model assumptions in Section \ref{abel_dispersion}. Then, it follows:
\begin{equation}\label{eq:c_n_abundance_assumption}
\Bigg[\frac{ N(C^{+})}{N(N^{+})} \Bigg] \sim \Bigg[\frac{C}{N} \Bigg]. \\
\end{equation}
The first assumption in Equation \ref{eq:c_n_abundance_assumption} is analogous to assuming that the carbon and nitrogen ionization correction factors (ICFs), defined as semiempirical correction factors that scale any given ionic elemental abundance to derive the corresponding elemental abundance of the neutral species, and usually used in observations of FIR to optical lines from \HII\textrm{ }regions \citep{simpson1995,martin_hernandez2002,rodriguez2005}, are around unity for the \CII\textrm{ }line transition at 158 $\mu m$ and \NII\textrm{ }line transition at 205 $\mu m$. The second assumption  in Equation \ref{eq:c_n_abundance_assumption} implies that the depletion of carbon and nitrogen atoms onto dust grains is low. There is some observational evidence suggesting that, in Nebulae, a significant fraction of carbon atoms might be trapped in carbonaceous grains and in polycyclic aromatic hydrocarbons (PAHs), although it is rather unknown how severe these effects can be \citep{draine2003,jenkins2009,simondiaz2011}. Given that [C/H] elemental abundances in the GC are rather large in relation to Galactic disk values (\citet{arimoto1996,sodroski1995,oka2005,simondiaz2011}), and that there is mounting evidence of the presence of smaller dust grain sizes in the GC compared to the Galactic disk  \citep{nishiyama2006,nishiyama2008,nishiyama2009}, probably pointing to efficient large-scale dust destruction mechanisms in the GC, it seems reasonable to expect that most of the carbon within the harsh environmental conditions in the GC is found in the gas phase rather than in the solid phase. In the nitrogen case, \citet{jenkins2009} conducted an extensive study toward several LOS in the Milky Way characterizing the depletion of C, O, and N, among other elements, onto dust grains, finding no evidence for nitrogen depletion. A possible cause for this could be related to the formation of N$_{2}$, which is very stable, and that has been suggested to inhibit the depletion of nitrogen into any solid compound in the ISM \citep{gail1986}. The only exception for nitrogen to be found in the solid phase could be related to the presence of ices \citep{hilyblant2010,simondiaz2011}, which are not expected to live long within the GC environment containing the ionized gas. To summarize, we consider the assumptions yielding Equation \ref{eq:c_n_abundance_assumption} a reasonable first-order approximation to describe the relationship between ionized and neutral carbon and nitrogen gas in the AF. The values used in Equations \ref{eq:n_abundance} and \ref{eq:c_abundance} to derive the column densities N(N$^{+}$) and N(C$^{+}$), and the [C/N] abundance ratios in the AF via Equation \ref{eq:c_n_abundance_assumption}, are summarized in Table \ref{tab:abundance_results}. For the calculations, integrated intensities I(\NII)$_{s}$ and I(\NII)$_{i}$ in Table \ref{tab:data_summary} and I(\CII)$_{\rm{HII}}$ values in Table \ref{tab:hii_contribution} were used to derive [C/N]$_{s,s}$ and [C/N]$_{i,i}$, where subindexes identify the corresponding N$^{+}$ and C$^{+}$ column densities involved in the [C/N] abundance ratio calculation. Error uncertainties were obtained by propagating the experimental errors in I$_{s}$ and I$_{i}$, through Equations \ref{eq:n_abundance}, \ref{eq:c_abundance}, and  \ref{eq:c_n_abundance_assumption}. The derived [C/N]$_{s,s}$ and [C/N]$_{i,i}$  abundance ratios are fully consistent within error uncertainties. Therefore, in the following, we focus our discussion only on the [C/N]$_{s,s}$ results. \\

%%%%%%%%%%%%%%%%%
%%% C/N ABUNDANCES %%%
%%%%%%%%%%%%%%%%%

\afterpage{
\begin{table*}
\centering
\small
\caption{[C/N] abundance ratios derived from \CII\textrm{ }and \NII\textrm{ }line observations of the AF. Radial velocity components are listed in the first column. Electron temperature T$_{e}$, electron density n$_{e}$, and c(n$_{e}$,T$_{e}$) coefficients relating ionized nitrogen integrated intensities to ionized nitrogen column densities are listed in columns 2 to 4, respectively. N$^{+}$ and C$^{+}$ column densities N(N$^{+}$)$_{s}$, N(N$^{+}$)$_{i}$, N(C$^{+}$)$_{s}$, and N(C$^{+}$)$_{i}$ estimated from I(\NII)$_{s}$ and I(\NII)$_{i}$ integrated intensities in Table \ref{tab:data_summary}, in combination with the model prediction in Equation \ref{eq:c_n_fit}, are listed in columns 5 to 8, respectively. The [C/N] abundance ratios derived from the N$^{+}$ and C$^{+}$ column densities, [C/N]$_{s,s}$ and [C/N]$_{i,i}$ are listed in columns 9 and 10, where subindexes identify the corresponding N$^{+}$ and C$^{+}$ column densities involved in the [C/N] abundance ratio calculation. At the bottom of the table, average abundance ratios for the AF [C/N]$_{\textrm{AF}}$ are listed.\label{tab:abundance_results}}
\bgroup
\begin{tabular}{lllccccccc}
\toprule
  
\multicolumn{1}{c}{\textbf{Source}}                     &
\multicolumn{1}{c}{\textbf{T$_{e}$}}                    &
\multicolumn{1}{c}{\textbf{n$_{e}$}}                    &
\multicolumn{1}{c}{\textbf{c(n$_{e}$,T$_{e}$)}} &
\multicolumn{1}{c}{\textbf{N(N$^{+}$)$_{s}$}}   &
\multicolumn{1}{c}{\textbf{N(N$^{+}$)$_{i}$}}   &
\multicolumn{1}{c}{\textbf{N(C$^{+}$)$_{s}$}}   &
\multicolumn{1}{c}{\textbf{N(C$^{+}$)$_{i}$}}   &
\multicolumn{1}{c}{\textbf{[C/N]$_{s,s}$}}      &
\multicolumn{1}{c}{\textbf{[C/N]$_{i,i}$}}      \\

\multicolumn{1}{c}{}                    &
\multicolumn{1}{c}{}                    &
\multicolumn{1}{c}{}                    &
\multicolumn{1}{c}{10$^{-16}$}  &
\multicolumn{1}{c}{10$^{17}$}   &
\multicolumn{1}{c}{10$^{17}$}   &
\multicolumn{1}{c}{10$^{17}$}   &
\multicolumn{1}{c}{10$^{17}$}   &
\multicolumn{1}{c}{}                    &
\multicolumn{1}{c}{}                    \\

\multicolumn{1}{c}{}            &
\multicolumn{1}{c}{(K)}         &
\multicolumn{1}{c}{}                    &
\multicolumn{1}{c}{(cm$^{-3}$)} &
\multicolumn{1}{c}{(cm$^{-2}$)} &
\multicolumn{1}{c}{(cm$^{-2}$)} &
\multicolumn{1}{c}{(cm$^{-2}$)} &
\multicolumn{1}{c}{(cm$^{-2}$)} &
\multicolumn{1}{c}{}            &
\multicolumn{1}{c}{}            \\

\midrule 
%source                Te                                               ne                                     N(N+)_s                      N(N+)_i                        N(C+)hii_s               N(C+)hii_i                    [C/N]Hii,ss                 [C/N]Hii,ii     
  P1-46\KMS   & 5000\tablefoottext{a}           & 100\tablefoottext{b}           &   2.29        &       1.23  $\pm$  0.08  &  2.93  $\pm$  0.17  &    1.54  $\pm$  0.11  &  3.91  $\pm$  0.25  &    1.26  $\pm$  0.12  &  1.33  $\pm$  0.12 \\
  P1-13\KMS   & 5000\tablefoottext{a}   & 100\tablefoottext{b}          &   2.29  &       0.82  $\pm$  0.09  &  2.93  $\pm$  0.17  &    1.00  $\pm$  0.12  &  3.91  $\pm$  0.25  &    1.22  $\pm$  0.20  &  1.33  $\pm$  0.12 \\
  P2-15\KMS   & 5000\tablefoottext{a}   & 100\tablefoottext{b}          &   2.29          &       0.87  $\pm$  0.06  &  3.11  $\pm$  0.14  &    1.07  $\pm$  0.08  &  4.16  $\pm$  0.21  &    1.23  $\pm$  0.13  &  1.34  $\pm$  0.09 \\
  W1-24\KMS  & 5500                             & 180                           &   2.22          &       3.99  $\pm$  0.13  &  5.17  $\pm$  0.18  &    4.91  $\pm$  0.18  &  6.49  $\pm$  0.25  &    1.23  $\pm$  0.06  &  1.25  $\pm$  0.07 \\
  W2-29\KMS  & 5500                             & 290                           &   2.15  &       3.25  $\pm$  0.12  &  5.11  $\pm$  0.18  &    3.70  $\pm$  0.14  &  6.00  $\pm$  0.22  &    1.14  $\pm$  0.06  &  1.17  $\pm$  0.06 \\
G010-17\KMS & 5000                              & 290                           &   2.15  &       2.08  $\pm$  0.09  &  5.10  $\pm$  0.17  &    2.30  $\pm$  0.11  &  5.98  $\pm$  0.21  &    1.10  $\pm$  0.07  &  1.17  $\pm$  0.06 \\
G007-16\KMS & 6700                              & 270                           &   2.19          &       1.64  $\pm$  0.09  &  4.16  $\pm$  0.17  &    1.81  $\pm$  0.11  &  4.91  $\pm$  0.22  &    1.11  $\pm$  0.09  &  1.18  $\pm$  0.07 \\
G007-41\KMS & 6700                              & 270                           &   2.19          &       0.83  $\pm$  0.09  &  4.16  $\pm$  0.17  &    0.88  $\pm$  0.10  &  4.91  $\pm$  0.22  &    1.06  $\pm$  0.16  &  1.18  $\pm$  0.07 \\
 E2N-18\KMS  & 5700                             & 310                           &   2.15          &       3.16  $\pm$  0.16  &  3.03  $\pm$  0.16  &    3.58  $\pm$  0.19  &  3.41  $\pm$  0.20  &    1.13  $\pm$  0.08  &  1.13  $\pm$  0.09 \\
 E2S-17\KMS  & 5600                             & 250                           &   2.19          &       3.66  $\pm$  0.15  &  3.85  $\pm$  0.16  &    4.30  $\pm$  0.19  &  4.54  $\pm$  0.21  &    1.18  $\pm$  0.07  &  1.18  $\pm$  0.07 \\ 
  E1-40\KMS   & 7500                            & 460\tablefoottext{c}           &   2.05        &       1.24  $\pm$  0.10  &  3.37  $\pm$  0.17  &    1.23  $\pm$  0.11  &  3.58  $\pm$  0.19  &    0.99  $\pm$  0.12  &  1.06  $\pm$  0.08 \\
  E1-16\KMS   & 7500                            & 460\tablefoottext{c}           &   2.05        &       1.00  $\pm$  0.08  &  3.37  $\pm$  0.17  &    0.98  $\pm$  0.09  &  3.58  $\pm$  0.19  &    0.98  $\pm$  0.12  &  1.06  $\pm$  0.08 \\
  
\midrule
\multicolumn{8}{l}{Average [C/N]$_{\textrm{AF}}$:} & 1.13  $\pm$  0.09 &  1.20  $\pm$  0.09 \\
\bottomrule 
\end{tabular}
\egroup
\tablefoot{\\
\tablefoottext{a}{Lowest value found in the AF \citep{lang2001}.}\\
\tablefoottext{b}{Upper limit for molecular clouds far from thermal sources in the CMZ \citep{rodriguez2005}.}\\
\tablefoottext{c}{Valued adopted from source D in \citet{lang2001}.}\\
}
\end{table*}}

% [C/N] abundance ratios
The [C/N] abundance ratios found toward the AF are surprisingly homogeneous, with all estimates lying within a narrow 0.95 $\leq$  [C/N] $\leq$ 1.25 range.  This result is in full agreement with the [C/N] range shown in the condition defined in Equation \ref{eq:c_n_for_PDR}. This shows that using the derived \CII\textrm{ }and observed \NII\textrm{ }emission from the \HII\textrm{ }region to derive the [C/N] abundance ratios in the AF yields consistent results with two almost independent approaches. The only caveat in the direct comparison between both [C/N] estimates is that the N(C$^{+}$) and N(N$^{+}$) column densities in Equations \ref{eq:n_abundance} and \ref{eq:c_abundance} are not fully independent as the I(\CII)$_{\rm{HII}}$ integrated intensities used in Equation \ref{eq:c_abundance} are derived from I(\NII) integrated intensities via Equation \ref{eq:c_n_fit}. Nonetheless, when considering all the variables involved in the [C/N] calculations, the results in the present work strongly suggest that the [C/N] abundance ratio in the AF differs substantially from that measured in the Galactic disk \citep{simondiaz2011}, pointing to a different ISM enrichment process in the GC. Using the average [C/N]$_{\textrm{AF}}$ $=$ 1.13 $\pm$ 0.09 in Table \ref{tab:abundance_results} and the abundance ratio derived for the Orion Nebula [C/N]$_{\textrm{ON}}$ $=$ 2.81 $\pm$ 0.01 \citep{simondiaz2011}, we obtain [C/N]$_{\textrm{AF}}$/[C/N]$_{\textrm{ON}}$ $=$ 0.40 $\pm$ 0.03, in full agreement with what is expected from our model results in Section \ref{hii_contribution}. \\

% [C/H] and [N/H] elemental ratios
The results in Table \ref{tab:abundance_results} allow us to predict the average nitrogen elemental abundance toward the AF [N/H]$_{\textrm{AF}}$, as long as the average carbon elemental abundance [C/H]$_{\textrm{AF}}$ in the GC is provided. A simple expression to estimate [N/H]$_{\textrm{AF}}$ as a function of our results and the abundance derived in the Galactic disk for the Orion Nebula \citep{simondiaz2011} is given by
\begin{equation}\label{eq:n_abundance_prediction}
\Bigg[\frac{N}{H} \Bigg]_{\textrm{AF}} = \alpha \times \Bigg( \frac{[C/N]_{\textrm{ON}}}{ [C/N]_{\textrm{AF}}}  \Bigg)  \times  \Bigg[\frac{N}{H} \Bigg]_{\textrm{ON}},  \\
\end{equation}
where $\alpha > 1$ is a scaling factor between the carbon elemental abundance in the GC and the one found in the Galactic disk so that [C/H]$_{\textrm{AF}} = \alpha\times$[C/H]$_{\textrm{ON}}$, with $\alpha$ ranging between 3 and 10  in the GC \citep{arimoto1996,sodroski1995,oka2005}. In a conservative estimate with $\alpha = 3$, adopting [C/N]$_{\textrm{ON}}$ $=$ 2.81 \citep{simondiaz2011},  [C/N]$_{\textrm{AF}}$ $=$ 1.13 from Table \ref{tab:abundance_results}, and [N/H]$_{\textrm{ON}}$ $=$ 8.32$\times$10$^{-5}$ for the Orion Nebula \citep{simondiaz2011}, Equation \ref{eq:n_abundance_prediction} yields an average nitrogen elemental abundance in the AF of [N/H]$_{\textrm{AF}}$ $=$ 6.21$\times$10$^{-4}$. This prediction can be compared with the values derived by \citet{simpson1995} for G0.10$+$0.02 ([N/H]$_{\textrm{S95}}$ $=$ 2.66 $\pm$ 0.64 $\times$10$^{-4}$) and G1.13$-$0.11 ([N/H]$_{\textrm{S95}}$ $=$ 4.39 $\pm$ 1.00 $\times$10$^{-4}$) also located in the GC. The average [N/H]$_{\textrm{AF}}$ value is a factor of 2.3  and 1.4 greater than those derived by \citet{simpson1995}, respectively. Considering that their results rely heavily on a rather uncertain ICF(N$^{+2}$) value for their [NIII] observations at 57 $\mu m$, such a discrepancy is not surprising. We also notice that the same factor of 2.3 is obtained for G0.10$+$0.02 (G010-17\KMS\textrm{ }in Table \ref{tab:abundance_results}) if its individual [C/N]$_{s,s}$ and [C/N]$_{i,i}$ values are used to calculate the [C/N] average in the GC instead of using [C/N]$_{\textrm{AF}}$ in Equation \ref{eq:n_abundance_prediction}. \\

The production of nitrogen and its evolution in Galaxies is a highly nontrivial process compared to other species such as oxygen and carbon. Nitrogen can be produced by high-, intermediate-, and low-mass stars, enriching the ISM on different timescales, while oxygen is produced mostly by massive, short-lived stars. Hence, the [N/O] abundance ratio gives important information on the primary or secondary nature of nitrogen. \citet{esteban2018} derived the radial distribution of the 12$+\log$(N/H) elemental abundance and the [N/O] abundance ratio as a function of R$_{\rm{GAL}}$ down to 5.7 kpc. In their Figure 4, a steadily increasing [N/H] with decreasing R$_{\rm{GAL}}$ can be seen, yielding a least$-$squares fit 12$+ \log$(N/H) $=$ 8.21 $-$ 0.059 $\times$ R$_{\rm{GAL}}$. Very recently, \citet{arellano2020} reassessed these results taking advantage of the large improvement in the accuracy of heliocentric distances for \HII\textrm{ }regions in the Galactic disk obtained from Gaia parallaxes of the second data release (DR2)  \citep{gaia2018}, confirming the general trend of the nitrogen elemental abundance radial distribution found by \citet{esteban2018}. Nonetheless, whether  the [N/H] radial distribution holds for smaller R$_{\rm{GAL}}$ is unclear. An extrapolation of the \citet{esteban2018} results toward the GC yields [N/H] $\sim$ 1.6$\times$10$^{-4}$ which is a factor of $\sim$ 4 lower than our [N/H]$_{\textrm{AF}}$ prediction. \citet{esteban2018} and \citet{arellano2020} also showed that the [N/O] ratio is almost flat for R$_{\rm{GAL}}$ $>$ 8 kpc, suggesting that the bulk of nitrogen in the Galaxy is not formed by standard secondary processes above this R$_{\rm{GAL}}$, but for the \HII\textrm{ }region S2-61 located at R$_{\rm{GAL}}$ $\sim$ 5.7 kpc, the largest [N/O] ratio in their sample is found, showing a large enhancement in the inner parts of the Galactic disk, and suggesting that the secondary production of nitrogen plays an important role for small R$_{\rm{GAL}}$. \citet{stanghellini2018} showed than nitrogen can be highly enhanced at the expense of carbon in high-mass stars, suggesting that a nitrogen-rich shell is ejected by hot-bottom burning (HBB) stars, transforming most of their carbon into nitrogen. \citet{colzi2018} studied the fractionation of nitrogen in the Galaxy by measuring the isotopic [$^{14}$N/$^{15}$N] abundance ratio  as a function of R$_{\rm{GAL}}$. Both isotopes have different origins, with massive stars in their He-burning phase as the primary contributors of $^{14}$N to the ISM, while $^{15}$N is produced mostly in the hot CNO cycle of novae outbursts by low-mass stars. They reported a decrease in the [$^{14}$N/$^{15}$N] ratio toward the GC, sampled down to R$_{\rm{GAL}}$ $\sim$ 2 kpc, implying that the secondary production of nitrogen dominates over its primary production. \\

Our results suggest that the nitrogen elemental abundance in the GC might be underestimated, and that the nitrogen enrichment of the ISM increases much faster toward smaller R$_{\rm{GAL}}$ than the carbon elemental abundance. A plausible  explanation for this behavior is the secondary production of nitrogen from low- to intermediate-mass stars in the Galactic bulge, whose mass loss would act as an important mass supply source to the inner parts of the GC where the AF are located.

%%%%%%%%%%%%%%%%%%%%%%%%%%%%%%%%%%%%%%%%%%%%%%%%%%%%%%%%%%%%%%%%%%%%%%%%%%%%%%%%%%%%%%%%%%%%%%%%%%%%
%%%%%%%%%%%%%%%%%%%%%%%%%%%%%%%%%%%%%%%%%%%%%%%%%%%%%%%%%%%%%%%%%%%%%%%%%%%%%%%%%%%%%%%%%%%%%%%%%%%%
%%%%%%%%%%%%%%%%%%%%%%%%%%%%%%%%%%%%%%%%%%%%%%%%%%%%%%%%%%%%%%%%%%%%%%%%%%%%%%%%%%%%%%%%%%%%%%%%%%%%

\section{Summary and conclusions}\label{conclusions}

% Summary of the work and observations used
In this work, we studied the relative contributions of \HII\textrm{ }regions and PDRs to the total \CII\textrm{ }emission observed toward the AF and nearby positions, and their implications for the ISM composition in the GC. Our analysis is based on \CII\textrm{ }and \NII\textrm{ }line observations at 158 $\mu m$ and 205 $\mu m$, respectively, published in \citet{garcia2016}  (publicly available), complemented with observations of the optically thin \COOLINEAI\textrm{ }transition to be published in detail elsewhere. The \CII\textrm{ }and \NII\textrm{ }data sets in \citet{garcia2016} represent the most spatially extended velocity-resolved observations available for the entire Sgr A complex (containing the AF) at the time of publication. Our main results can be summarized as follows: 

\begin{itemize}
% Spatial distribution of I([CII])/I([NII]) ratios
\item The spatial distribution of the I(\CII)/I(\NII) integrated intensity ratios (in units of ergs s$^{-1}$ cm$^{-2}$) toward the AF is relatively homogeneous. The bulk of the ratios is confined between two prominent lanes of high I(\CII)/I(\NII) values, with one located south of the radio arc and the other found to the north of the Sgr A-east SNR, outlined by 20 cm radio continuum observations \citep{yusef1987}. Some striking features in the I(\CII)/I(\NII) spatial distribution are:  (1) a strong decrease in the high ratio lanes at the intersections (at least in projection) with W1 (south of the radio arc) and  the location of the H3 \HII\textrm{ }region (north of Sgr A-east), while no decrease is seen at the intersection with the E1, E2S, and E2N (south of the radio arc); and (2) a local enhancement of the I(\CII)/I(\NII) ratio at the location of G0.07$+$0.04, where it has been suggested that the gas is interacting with the NTF, tracing the local magnetic field in the GC \citep{lang1999b}. The features in the I(\CII)/I(\NII) spatial distribution strongly suggest that geometry effects in the form of the distance component along the LOS must play a role in the coupling between the thermal plasma responsible for the radio continuum emission and the ionized gas responsible for the observed I(\CII)/I(\NII) ratios in the AF, while localized physical phenomena might be driving the high ratios close to major nonthermal sources such as the radio arc and the Sgr A-east SNR. A histogram of the I(\CII)/I(\NII) ratios in the AF can be well described by a Gaussian profile, with a mean of R$_{c}$ $=$ 5.1 $\pm$ 0.1 and standard deviation R$_{\sigma}$ $=$ 2.0 $\pm$ 0.1.

%  I([CII]\textrm{ }) vs. I([NII]) scatter relationship and model predictions for [C/N]
\item The I(\CII) versus I(\NII) relationship in the AF region was studied by expanding the models in \citet{abel2006a} for the relative \CII\textrm{ }emission contribution from \HII\textrm{ }regions to the total observed \CII\textrm{ }emission toward PDRs. Different values of the [C/N] abundance ratio, SEDs consistent with stellar clusters, and a parameter grid for the hydrogen density n(H), metallicity Z, and ionization parameter U, suitable for the physical conditions of the ISM at the GC, were included. Our models show that the large scatter in the I(\CII) versus I(\NII) relationship can be largely explained by variations in the [C/N] abundance ratio in the AF. The models satisfying \mbox{0.3 $<$ [C/N]$_{\textrm{AF}}$/[C/N]$_{\textrm{ON}}$ $<$ 0.5}, with the abundance ratio [C/N]$_{\textrm{AF}}$ predicted for the AF and [C/N]$_{\textrm{ON}}$ derived for the Orion Nebula in the Galactic disk \citep{simondiaz2011} best disentangle the \CII\textrm{ }emission contribution from  \HII\textrm{ }regions and PDRs toward the AF in the GC, similarly to the predictions in \citet{abel2006a} for the Orion Nebula. We interpret the fact that carbon-to-nitrogen abundance ratios in the AF and in the Galactic disk satisfy \mbox{[C/N]$_{\textrm{AF}}$/[C/N]$_{\textrm{ON}}$ $<$ 1} as a strong indication of a large nitrogen enrichment of the ISM when moving from the Galactic disk to the GC. Carbon elemental abundances have already been\textbf{} estimated at between 3 and 10 times larger in the GC than the values measured in the Galactic disk \citep{arimoto1996,sodroski1995,oka2005}.

%  [CII]\textrm{ } emission prediction from HII Regions 
\item We derive the \CII\textrm{ }emission contribution from \HII\textrm{ }regions traced by \NII\textrm{ }line observations toward nine selected positions in and around the AF: seven positions trace the filaments' segments E1, E2N, E2S, G0.10$+$0.02, W1, W2, and G0.07$+$0.04, and two positions, P1 and P2, trace nearby high-density gas outside them \citep{serabyn1987}. By combining the optically thin \CILINEAI\textrm{ }and \COOLINEAI\textrm{ }transitions, we identify a total of twelve radial velocity components within the nine selected positions in \CII\textrm{ }for which we derive the \HII\textrm{ }region emission contribution. For this purpose, we select the models satisfying \mbox{0.3 $<$ [C/N]$_{\textrm{AF}}$/[C/N]$_{\textrm{ON}}$ $<$ 0.5}, with model parameter ranges consistent with the ISM characteristics in the GC, that is, metallicities from 1 Z\SOLAR\textrm{ }to 2 Z\SOLAR, hydrogen volume density $\log$ n(H) $=$ 3.5, and ionization parameters $\log$ U $=$ $-$1 to $-$2. A least-squares fit to the model data points yields \mbox{$\log$ I(\CII) $=$ 1.068$\times \log$ I(\NII) $+$ 0.645}. Then, assuming that the total observed \CII\textrm{ }emission toward the AF originates either in the \HII\textrm{ }region or PDR only, we show that the fraction of \CII\textrm{ }emission in PDRs can vary dramatically between locations in the AF, with a fraction ranging from $\sim$ 0.20 to $\sim$ 0.75. Therefore, any \CII\textrm{ }modeling attempt in the region has to properly disentangle the contribution from \HII\textrm{ }regions and PDRs in order to avoid the production of misleading results. 

 %  [N/H] elemental abundance prediction from [CII]\textrm{ } and [NII] observations tracing the H II region.
 \item The results for the carbon-to-nitrogen abundance ratio in the AF are fully consistent with an (almost) independent derivation of the same ratio by calculating N(C$^{+}$) and N(N$^{+}$) column densities, and assuming that they directly mirror the values of [C/N]. For the twelve radial velocity components identified in this work, we obtain abundance ratios in the range 0.95 $\leq$  [C/N]$_{\textrm{AF}}$ $\leq$ 1.25 with an average value of [C/N]$_{\textrm{AF}}$ $=$ 1.13 $\pm$ 0.09 for the AF, in full agreement with the range previously derived by our models. By combining this result with the carbon-to-nitrogen abundance ratio measured in the Galactic disk [C/N]$_{\textrm{ON}}$ $\sim$ 2.81 \citep{simondiaz2011} and adopting a three-times-higher carbon elemental abundance in the GC than in the solar vicinity, we predict an average nitrogen elemental abundance for the AF of  [N/H]$_{\textrm{AF}}$ $=$ 6.21$\times$10$^{-4}$, which is a factor of 2.3 higher than the one derived by \citet{simpson1995} for G0.10$+$0.02, relying on a rather poorly constrained ionization correction factor ICF(N$^{+2}$) for their [NIII] observations at 57 $\mu m$. Nonetheless, given the vastly different approaches between the present work and the one in  \citet{simpson1995}, this relatively small discrepancy is remarkable. 
 
  \item We propose that our [N/H]$_{\textrm{AF}}$ prediction is close to the real average value in the GC. If this is indeed the case, it would imply that the nitrogen elemental abundance in the GC is  a factor of $\sim$ 7.5 larger than that of the Galactic disk. We suggest that a plausible explanation for such a scenario could be the nitrogen enrichment of the ISM in the GC  via  the secondary production of nitrogen from low- to intermediate-mass stars in the Galactic bulge. If their mass loss through stellar winds acts as the main gas supply source for the gas in the inner orbits of the GC, a large nitrogen elemental abundance in the AF located within the GC is expected. Overall, our results show that tight constraints on the [C/N] abundance ratio for the GC are needed, significantly tighter than previous abundance measurements have discerned. 

\end{itemize}

Our results have shown that a large fraction of the observed \CII\textrm{ }158 $\mu m$ emission in the AF region does not originate within PDRs but in \HII\textrm{ }regions. This can have a considerable impact on PDR model results used to derived the gas physical conditions such as impinging FUV fields, hydrogen densities, hydrogen ionization rates due to cosmic-rays, and so on, because the gas cooling rate traced by this ionized carbon fine-structure line can be largely overestimated and give misleading results. Moreover, these results show that the [C/N] abundance ratio in the AF is considerably smaller than the Galactic plane average value pointing to a nitrogen-enriched ISM in the GC and suggesting a secondary origin of the nitrogen production. The impact of these results on the PDR modeling of the AF region will be investigated in a follow-up publication.\\

%%%%%%%%%%%%%%%%%%%%%%%%%%%%%%%%%%%%%%%%%%%%%%%%%%%%%%%%%%%%%%%%%%%%%%%%%%%%%%%%%%%%%%%%%%%%%%%%%%%%
%%%%%%%%%%%%%%%%%%%%%%%%%%%%%%%%%%%%%%%%%%%%%%%%%%%%%%%%%%%%%%%%%%%%%%%%%%%%%%%%%%%%%%%%%%%%%%%%%%%%
%%%%%%%%%%%%%%%%%%%%%%%%%%%%%%%%%%%%%%%%%%%%%%%%%%%%%%%%%%%%%%%%%%%%%%%%%%%%%%%%%%%%%%%%%%%%%%%%%%%%

\begin{acknowledgements}
  %We thank the anonymous referee for his/her critical comments that have helped us to improve the paper
This work was sponsored (in part) by the Chinese Academy of Sciences (CAS), through a grant to the CAS South America Center for Astronomy (CASSACA) in Santiago, Chile, by the ``Comisi\'on Nacional de Ciencia y Tecnolog\'ia (CONICYT)'' via Project FONDECYT de Iniciaci\'on 11170551, and by the Collaborative Research Centre 956, sub-project A5, funded by the Deutsche Forschungsgemeinschaft (DFG). Robert Simon acknowledges support by the french Agence National de Recherche (ANR) and the german Deutsche Forschungsgemeinschaft (DFG) through the project "GENESIS" (project numbers ANR-16-CE92-0035-01/DFG1591/2-1). The data used in this work are based partially on observations carried out under project ID 020-16 with the IRAM 30m telescope. IRAM is supported by INSU/CNRS (France), MPG (Germany) and IGN (Spain).\\
\end{acknowledgements}

%%%%%%%%%%%%%%%%%%%%%%%%%%%%%%%%%%%%%%%%%%%%%%%%%%%%%%%%%%%%%%%%%%%%%%%%%%%%%%%%%%%%%%%%%%%%%%%%%%%%
%%%%%%%%%%%%%%%%%%%%%%%%%%%%%%%%%%%%%%%%%%%%%%%%%%%%%%%%%%%%%%%%%%%%%%%%%%%%%%%%%%%%%%%%%%%%%%%%%%%%
%%%%%%%%%%%%%%%%%%%%%%%%%%%%%%%%%%%%%%%%%%%%%%%%%%%%%%%%%%%%%%%%%%%%%%%%%%%%%%%%%%%%%%%%%%%%%%%%%%%%

%\Online
\begin{appendix} %First online appendix

\clearpage

%%%%%%%%%%%%%%%%%%%%%%%%%%%%%%%%%%%%%%%%%%%%%%%%%%%%%%%%%%%%%%%%%%%%%%%%%%%%%%%%%%%%%%%%%%%%%%%%%%%%
%%%%%%%%%%%%%%%%%%%%%%%%%%%%%%%%%%%%%%%%%%%%%%%%%%%%%%%%%%%%%%%%%%%%%%%%%%%%%%%%%%%%%%%%%%%%%%%%%%%%
%%%%%%%%%%%%%%%%%%%%%%%%%%%%%%%%%%%%%%%%%%%%%%%%%%%%%%%%%%%%%%%%%%%%%%%%%%%%%%%%%%%%%%%%%%%%%%%%%%%%

\section{Estimation of I([CII]) and I([NII]) integrated intensities}\label{appendix:gaussian_fits}

In this appendix, the procedure to obtain I(\CII) and I(\NII) integrated intensities for each position in Table \ref{arches:tab_positions} is outlined. The lack of detection of the optically thin \CCII\textrm{ }transition, the low brightness of both \CII\textrm{ }and \NII\textrm{ }lines, and the poor S/N in the \NII\textrm{ }data force us to use complementary observations acting as templates to identify the radial velocity components in both transitions, later used for the data analysis. Many of the existing \CCII\textrm{ }observations in massive star-forming regions have shown that its line profile is closely followed by the optically thin \COOLINEAI\textrm{ }transition \citep{guevara2019}. \citet{goldsmith2012} showed that the \CII\textrm{ }line falls into the so-called EOT limit for peak antenna temperatures $\leq$ 8 K, where opacity effects are not the dominant factor shaping the line profile. This can also be seen in the comparison between the \CII, \CILINEAI, and \COOLINEAI\textrm{ }observations in Figure \ref{intro:cii_ci_c18o} and in the \CILINEAII\textrm{ }spectra shown in the figures of this appendix. Despite the very different gas physical conditions the aforementioned transitions trace, they show relatively similar broad spectral features (assumed to originate in the GC). Therefore, though opacity effects may be present, they are unlikely to have a significant effect on shaping the observed line profiles. Instead, they are more likely dominated by the presence of different radial velocity components in the spectra. In this sense, by carrying out a multiwavelength comparison, we can estimate the corresponding \CII\textrm{ }emission contribution to all radial velocity components by fitting multi-Gaussian profiles (assumed for simplicity) in the proper radial velocity ranges, where the components are found. Then, for the \NII\textrm{ }line observations, the corresponding integrated intensities are calculated by simply adding the emission, multiplied by the channel spectral width, in the radial velocity range where the \CII\textrm{ }emission for each component is found. \\

Figures \ref{appA:spectra1} to \ref{appA:spectra5} show the \CII\textrm{ }(open blue), \NII\textrm{ }(filled black), \CILINEAII\textrm{ }(filled dark gray), \CILINEAI\textrm{ }(filled gray), and \COOLINEAI\textrm{ }(open red) color-coded spectra. Antenna temperatures have been scaled by the factor next to the line's name for display purposes. The red velocity components in the multi-Gaussian fits to the \COOLINEAI\textrm{ }lines are used to recover their counterpart radial velocity components in the \CII\textrm{ }spectra multi-Gaussian fits, shown in blue. Dashed-line curves are the radial velocity components identified in both \CII\textrm{ }and  \COOLINEAI\textrm{ }lines that are used for the analysis in this work, while dotted-line Gaussian components are needed to reproduce other relevant features in the spectra besides the GC-related emission. The vertical dotted lines mark the radial velocity range for which the I(\NII) integrated intensity, associated with each GC \CII\textrm{ }radial velocity component, was calculated. With this approach, five different integrated intensities (listed in Table \ref{tab:data_summary}) are calculated. Firstly, we have the I$_{g}$ values derived from the multi-Gaussian fits to the \CII\textrm{ }spectra. Then we have the I$_{s}$ values for both \CII\textrm{ }and \NII\textrm{ }line transitions obtained by directly adding the emission (multiplied by the spectral channel width) in the radial velocity channels marked by the vertical dotted black lines. Finally, we have the I$_{i}$ values obtained in the same way as I$_{s}$, but for the radial velocity range of $-$90 \KMS\textrm{ }to $+$25 \KMS, which is used in the data analysis described in Section \ref{abel_dispersion}. The approach to derive I$_{g}$, I$_{s}$, I$_{i}$ is graphically shown in Figure \ref{appA:spectra6}. Given the complexity of the \CII\textrm{ }spectra, there are several emission features that are not used in our analysis as they do not satisfy our requirements to be classified as a radial velocity component of the AF in the GC. Narrow radial velocity components obtained from the Gaussian fits are excluded from the analysis, as they most likely originate outside the GC, given their small line width and central radial velocities coincident with those of spiral arms along the LOS. Broad radial velocity components with positive central radial velocities (V$_{LSR}$ $>$ 0 \KMS) might be part of the GC but are most likely not associated with the AF because most of their gas is found toward negative radial velocities, which is shown in their I(\CII) and I(\NII) integrated intensity channel maps \citep{garcia2016}. \CII\textrm{ }and \NII\textrm{ }emission features at very negative radial velocities are seen at positions P1, W1, W2, G0.10$+$0.02, and G0.07$+$0.04. Since they have neither \CILINEAI\textrm{ }nor \COOLINEAI\textrm{ }evident counterparts detected above the rms noise level of the spectra, they are excluded from our analysis as no assessment can be made about opacity effects, origin (inside or outside the GC), or the number of radial velocity components that describe them. In the following, individual notes on the multi-Gaussian fitting procedure are given for each position in Table \ref{arches:tab_positions}: 

\begin{figure}
\centering
\begin{minipage}{\hsize}
\centering
\includegraphics[angle=-0, width=\hsize]{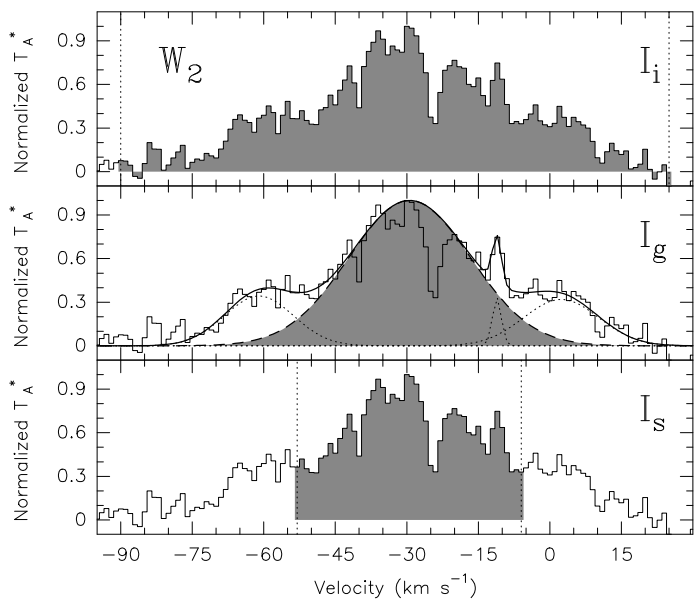}
\end{minipage}
\caption{Graphical description of the approach adopted to derived I$_{g}$, I$_{s}$, and I$_{i}$ integrated intensities for the \CII\textrm{ }and \NII\textrm{ }line observations, as described in Section \ref{intensities}. 
In the example, the spectra in the upper, middle, and lower panels correspond to the \CII\textrm{ }line at the W2 position. The spectra were normalized to the peak antenna temperature for display purposes. \label{appA:spectra6}
}
\end{figure}

\begin{itemize}
\item \textbf{P1:} The \COOLINEAI\textrm{ }multi-Gaussian fit identifies two broad components tracing the AF. They are blended with two narrow features assumed to be foreground components. The \CILINEAI\textrm{ }spectrum shows all features, while the \CILINEAII\textrm{ }transition only shows the broad components, reflecting different excitation conditions between the foreground and the GC gas. The multi-Gaussian \CII\textrm{ }fit is forced to reproduce the two broad components, which are blueshifted with regard to the radial velocity fit in the \COOLINEAI\textrm{ }spectrum. The narrow foreground feature at $-$35 \KMS\textrm{ }is detected at the same radial velocity as the one seen in the optically thin \COOLINEAI\textrm{ }transition. The \NII\textrm{ }spectrum has most of the emission toward the brightest \CII\textrm{ }component.

\item \textbf{P2:} The \COOLINEAI\textrm{ }multi-Gaussian fit identifies one broad component tracing the AF, blended with four narrow features assumed to be foreground components. The \CILINEAI\textrm{ }spectrum shows all the complex features in the \COOLINEAI\textrm{ }spectrum, while the \CILINEAII\textrm{ }transition shows mostly the broad component, reflecting different excitation conditions between the foreground and the GC gas. The multi-Gaussian \CII\textrm{ }fit is forced to reproduce the broad component at the center of the line, which is slightly blueshifted with regard to the radial velocity fit in the \COOLINEAI\textrm{ }spectrum. The emission in the \NII\textrm{ }spectrum is distributed between $-$60 \KMS\textrm{ }and $+$30 \KMS in nearly three major components, roughly correlated with the emission in the \CII\textrm{ }spectrum.

\item \textbf{W1:} The \COOLINEAI\textrm{ }multi-Gaussian fit identifies one broad component tracing the AF, blended with four narrow features assumed to be foreground components. The \CILINEAI\textrm{ }shows all features in the \COOLINEAI\textrm{ }spectrum, while the \CILINEAII\textrm{ }transition shows only the broad component and the two narrow features at $-$6 \KMS\textrm{ }and $+$2 \KMS, reflecting the changing excitation conditions of the gas along the same LOS. The multi-Gaussian \CII\textrm{ }fit is forced to reproduce the broad component, which is slightly redshifted in relation to the radial velocity fit in the \COOLINEAI\textrm{ }spectrum, as well as several less bright features. The \NII\textrm{ }spectrum has most of the emission in the same spectral range as the main \CII\textrm{ }component.

\item \textbf{W2:} The \COOLINEAI\textrm{ }multi-Gaussian fit identifies one broad component tracing the AF, blended with two narrow features assumed to be foreground components, and two other narrow features around radial velocity 0 \KMS, probably associated with cold gas along the LOS to the GC. The \CILINEAI\textrm{ }and \CILINEAII\textrm{ }transitions mainly show the broad component and the prominent feature near 0 \KMS, which are seen in the \COOLINEAI\textrm{ }spectrum. The multi-Gaussian \CII\textrm{ }fit is forced to reproduce the broad component at the center of the line, which is slightly redshifted with regard to the radial velocity fit in the \COOLINEAI\textrm{ }spectrum. The \NII\textrm{ }spectrum has most of the emission toward the main \CII\textrm{ }component.

\item \textbf{G0.10$+$0.02:} The \COOLINEAI\textrm{ }multi-Gaussian fit identifies one broad component tracing the AF, with four other narrow features in the spectrum assumed to be foreground components. The \CILINEAI\textrm{ }spectrum shows almost all the features of the \COOLINEAI\textrm{ }spectrum, while the \CILINEAII\textrm{ }transition mostly shows the broad component and the feature near 0 \KMS, reflecting different excitation conditions of the gas along the LOS. The multi-Gaussian \CII\textrm{ }fit is forced to reproduce the broad component at the center of the line, which is slightly redshifted in relation to the radial velocity fit in the \COOLINEAI\textrm{ }spectrum. The emission in the \NII\textrm{ }spectrum is mostly distributed between $-$75 \KMS\textrm{ }and $+$20 \KMS, and it is roughly correlated with the emission in the \CII\textrm{ }spectrum.

\item \textbf{G0.07$+$0.04:} The \COOLINEAI\textrm{ }multi-Gaussian fit identifies two broad components tracing the AF. They are blended with one small and two narrow components. The \CILINEAI\textrm{ }and \CILINEAII\textrm{ }transitions show all features seen in the \COOLINEAI\textrm{ }spectrum. The multi-Gaussian \CII\textrm{ }fit is forced to reproduce the two broad components, which are blueshifted with regard to the radial velocity fit in the \COOLINEAI\textrm{ }spectrum. The narrow foreground feature at $-$35 \KMS\textrm{ }is detected at the same radial velocity as the one seen in the optically thin \COOLINEAI\textrm{ }transition. The \NII\textrm{ }spectrum has most of the emission toward the brightest \CII\textrm{ }component, although a considerable fraction of the emission is found in a radial velocity range with no \COOLINEAI\textrm{ }counterpart emission, probably tracing gas deeper into the \HII\textrm{ }region.

\item \textbf{E2N:} The \COOLINEAI\textrm{ }multi-Gaussian fit identifies one broad component tracing the AF, blended with three narrow features assumed to be foreground components. The \CILINEAI\textrm{ }shows all features in the \COOLINEAI\textrm{ }spectrum, while the \CILINEAII\textrm{ }transition shows only the broad component, reflecting the changing excitation conditions of the gas along the same LOS. The multi-Gaussian \CII\textrm{ }fit is forced to reproduce the broad component, which is blueshifted in relation to the radial velocity fit in the \COOLINEAI\textrm{ }spectrum. The narrow feature in the center of the very broad velocity component is interpreted as a different source, and its emission contribution, though small, is fit away from the main component. The \NII\textrm{ }spectrum has most of the emission in the same spectral range as the main \CII\textrm{ }component.

\item \textbf{E2S:} The \COOLINEAI\textrm{ }multi-Gaussian fit identifies one very broad component tracing the AF, with four other narrow features assumed to be foreground components. The \CILINEAI\textrm{ }and the \COOLINEAI\textrm{ }transitions mostly trace the broad component identified in the \COOLINEAI\textrm{ }spectrum, while the narrow feature at $+$3 \KMS\textrm{ }is detected only in the \CILINEAI\textrm{ }transition, reflecting different excitation conditions of the gas along the LOS. The \CII\textrm{ }spectrum is extremely complex, with several narrow features at different radial velocities blended with a much broader emission component. Hence, the multi-Gaussian \CII\textrm{ }fit is forced to reproduce what we interpret as an underlying broad component at the center of the line, which is consistent with the one observed in the \COOLINEAI\textrm{ }line. This component is redshifted with regard to the radial velocity fit in the \COOLINEAI\textrm{ }spectrum. The emission in the \NII\textrm{ }spectrum is mostly distributed between $-$75 \KMS\textrm{ }and $+$30 \KMS, and it is roughly correlated with the emission of the underlying \CII\textrm{ }broad component.

\item \textbf{E1:} The \COOLINEAI\textrm{ }multi-Gaussian fit identifies two broad components tracing the AF. They are blended with three narrow features assumed to be foreground components. The \CILINEAI\textrm{ }spectrum shows emission at all features except for the narrow component at $-$35 \KMS, while the \CILINEAII\textrm{ }transition only shows the broad components, reflecting different excitation conditions between the foreground and the GC gas. The multi-Gaussian \CII\textrm{ }fit is forced to reproduce the two broad components, which are slightly blue-shifted with regard to the radial velocity fit in the \COOLINEAI\textrm{ }spectrum. The narrow feature at $-$42 \KMS\textrm{ }is fit to improve the overall fit of the broad components. The \NII\textrm{ }spectrum has most of the emission toward the identified \CII\textrm{ }broad components, with the rest found toward positive radial velocities, where a counterpart in the \CII\textrm{ }line is also detected.
  
\end{itemize}
% \CILINEAII\textrm{ }(filled dark-gray), \CILINEAI\textrm{ }(filled gray), and \COOLINEAI\textrm{ }(open red) 
\begin{figure*}
\centering
\begin{minipage}{\hsize}
\centering
\includegraphics[angle=-0, width=\hsize]{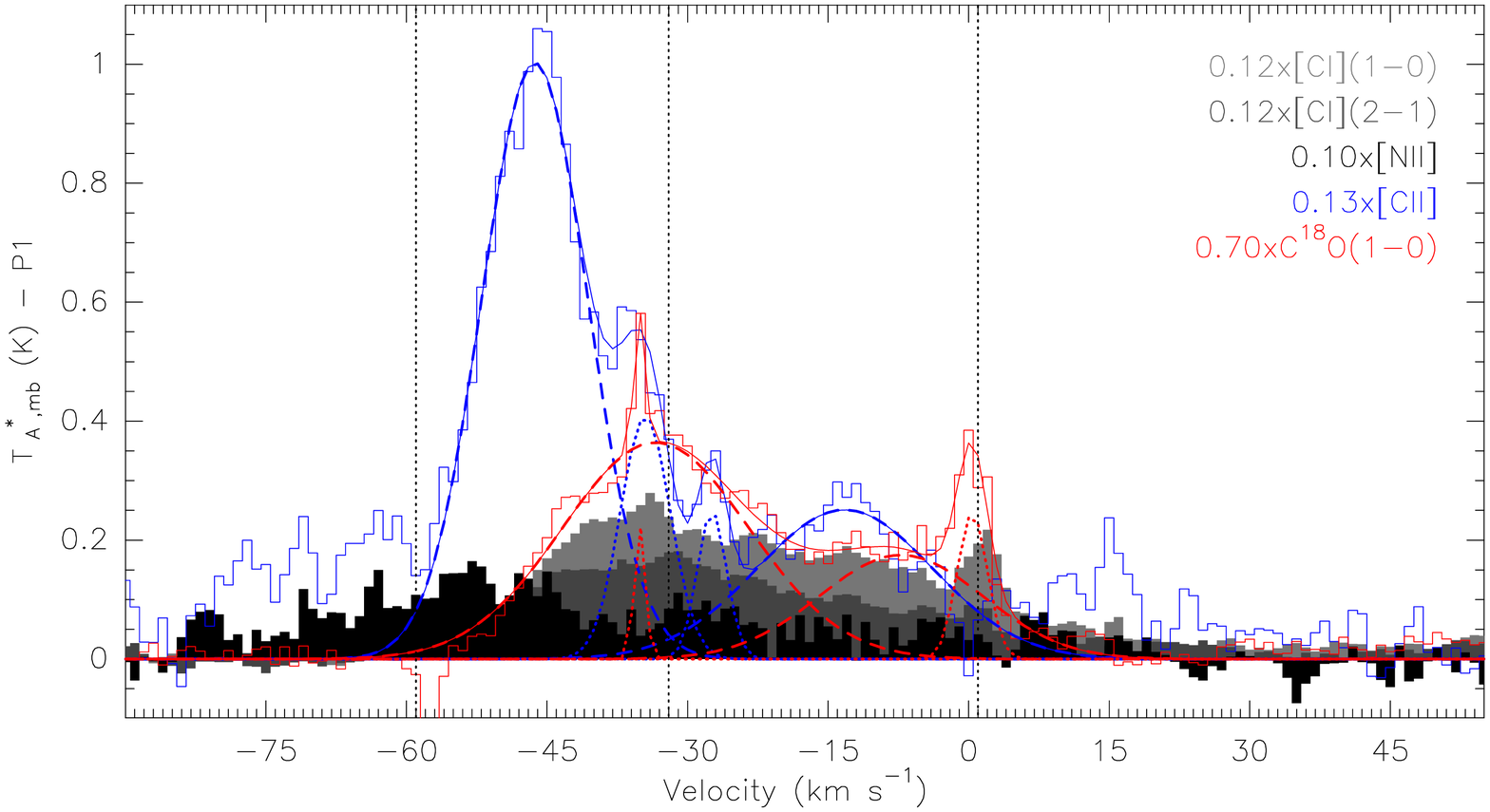}
\includegraphics[angle=-0, width=\hsize]{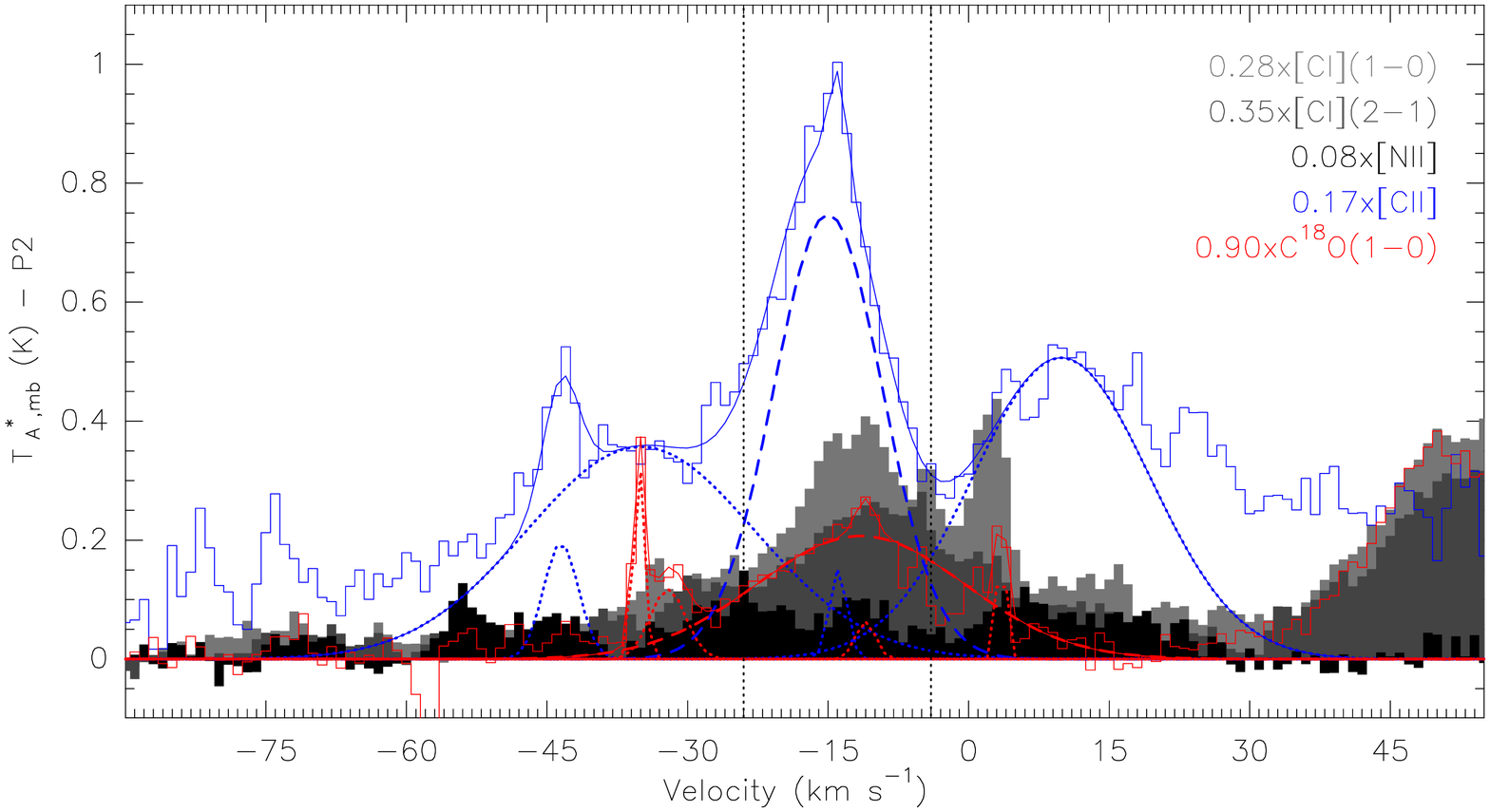}
\end{minipage}
\caption{Comparison of \CII\textrm{ }(open blue), \NII\textrm{ }(filled black), \CILINEAII\textrm{ }(filled dark gray), \CILINEAI\textrm{ }(filled gray), and \COOLINEAI\textrm{ }(open red) spectra for positions P1 and P2 in Table \ref{arches:tab_positions}. Antenna temperatures have been scaled by the factor shown next to the line's name for display purposes. Gaussian component fits to the \CII\textrm{ }and \COOLINEAI\textrm{ }spectra are shown in blue and red, respectively. Only dashed-line Gaussian components in the \CII\textrm{ }spectra are used for the analysis in this work as only for them can a counterpart in the \COOLINEAI\textrm{ }optically thin emission be identified. Vertical dotted black lines show the radial velocity limits for which integrated intensities were calculated for the \NII\textrm{ }line by directly adding the observed emission, multiplied by the spectral channel width, to that LSR velocity range. \label{appA:spectra1}
}
\end{figure*}

\begin{figure*}
\centering
\begin{minipage}{\hsize}
\centering
\includegraphics[angle=-0, width=\hsize]{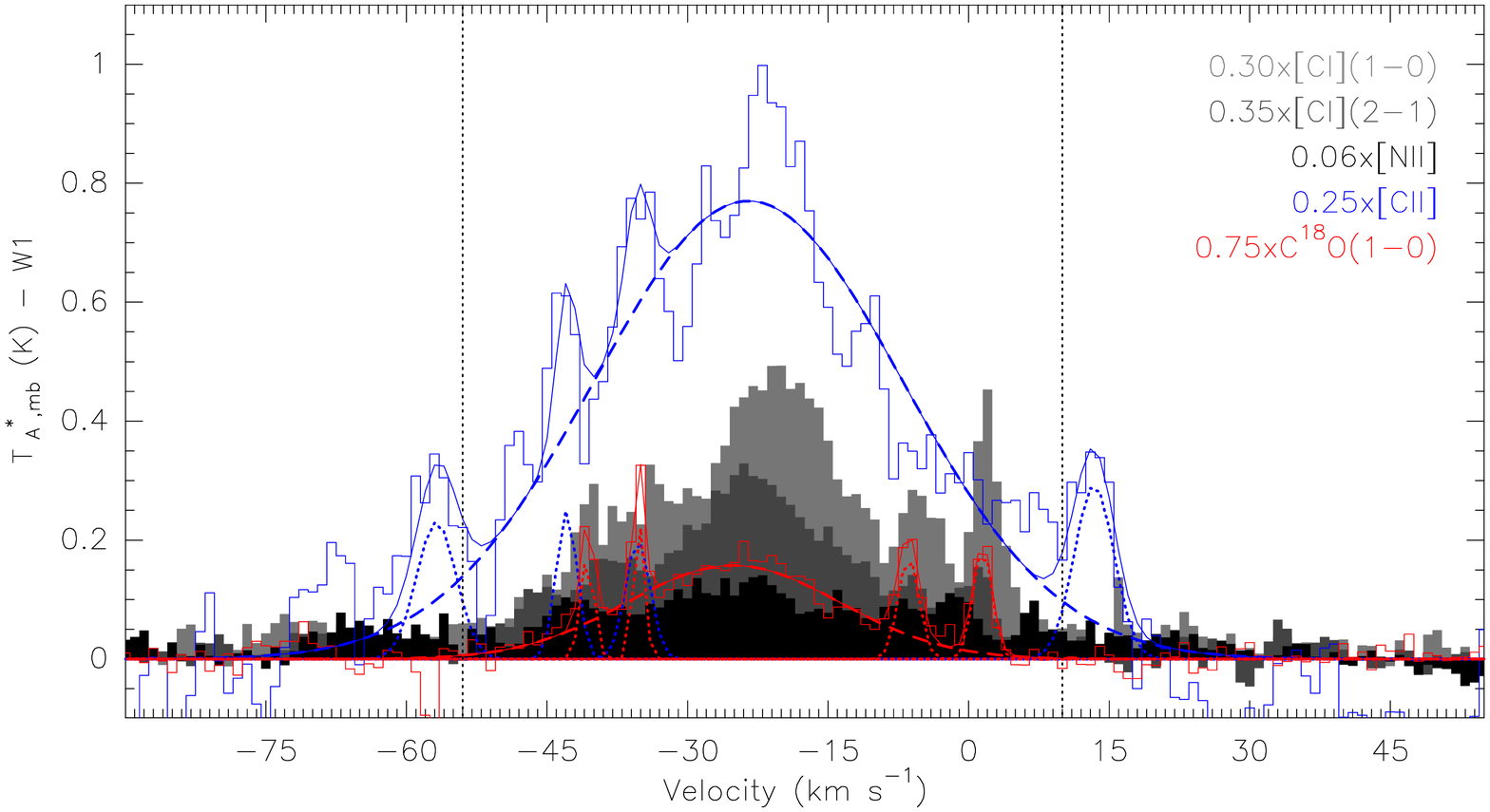}
\includegraphics[angle=-0, width=\hsize]{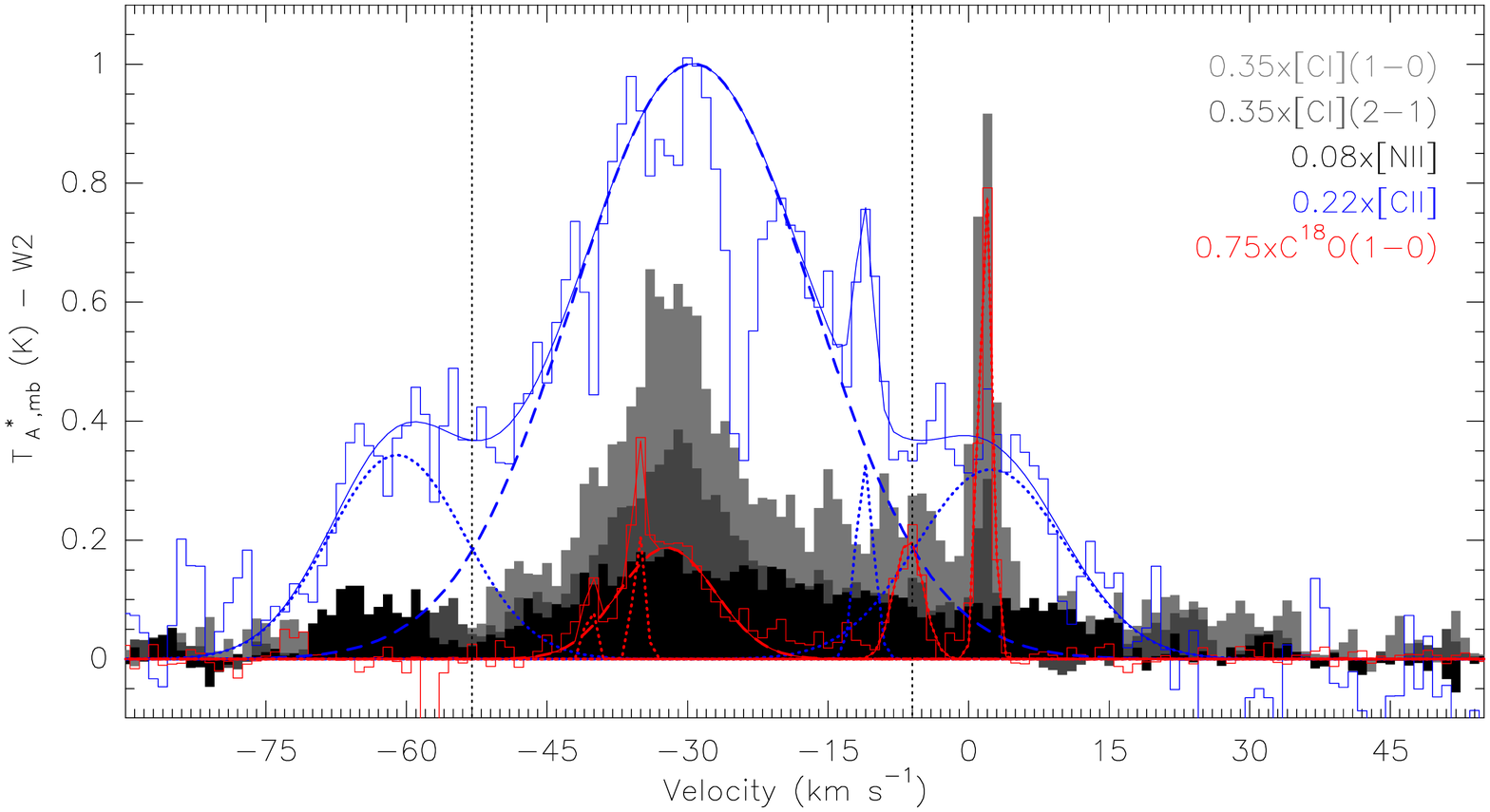}
\end{minipage}
\caption{Comparison of \CII\textrm{ }(open blue), \NII\textrm{ }(filled black), \CILINEAII\textrm{ }(filled dark gray), \CILINEAI\textrm{ }(filled gray), and \COOLINEAI\textrm{ }(open red) spectra for positions P1 and P2 in Table \ref{arches:tab_positions}. Antenna temperatures have been scaled by the factor shown next to the line's name for display purposes. Gaussian component fits to the \CII\textrm{ }and \COOLINEAI\textrm{ }spectra are shown in blue and red, respectively. Only dashed-line Gaussian components in the \CII\textrm{ }spectra are used for the analysis in this work as only for them can a counterpart in the \COOLINEAI\textrm{ }optically thin emission be identified. Vertical dotted black lines show the radial velocity limits for which integrated intensities were calculated for the \NII\textrm{ }line by directly adding the observed emission, multiplied by the spectral channel width, to that LSR velocity range. \label{appA:spectra2}
}
\end{figure*}

\begin{figure*}
\centering
\begin{minipage}{\hsize}
\centering
\includegraphics[angle=-0, width=\hsize]{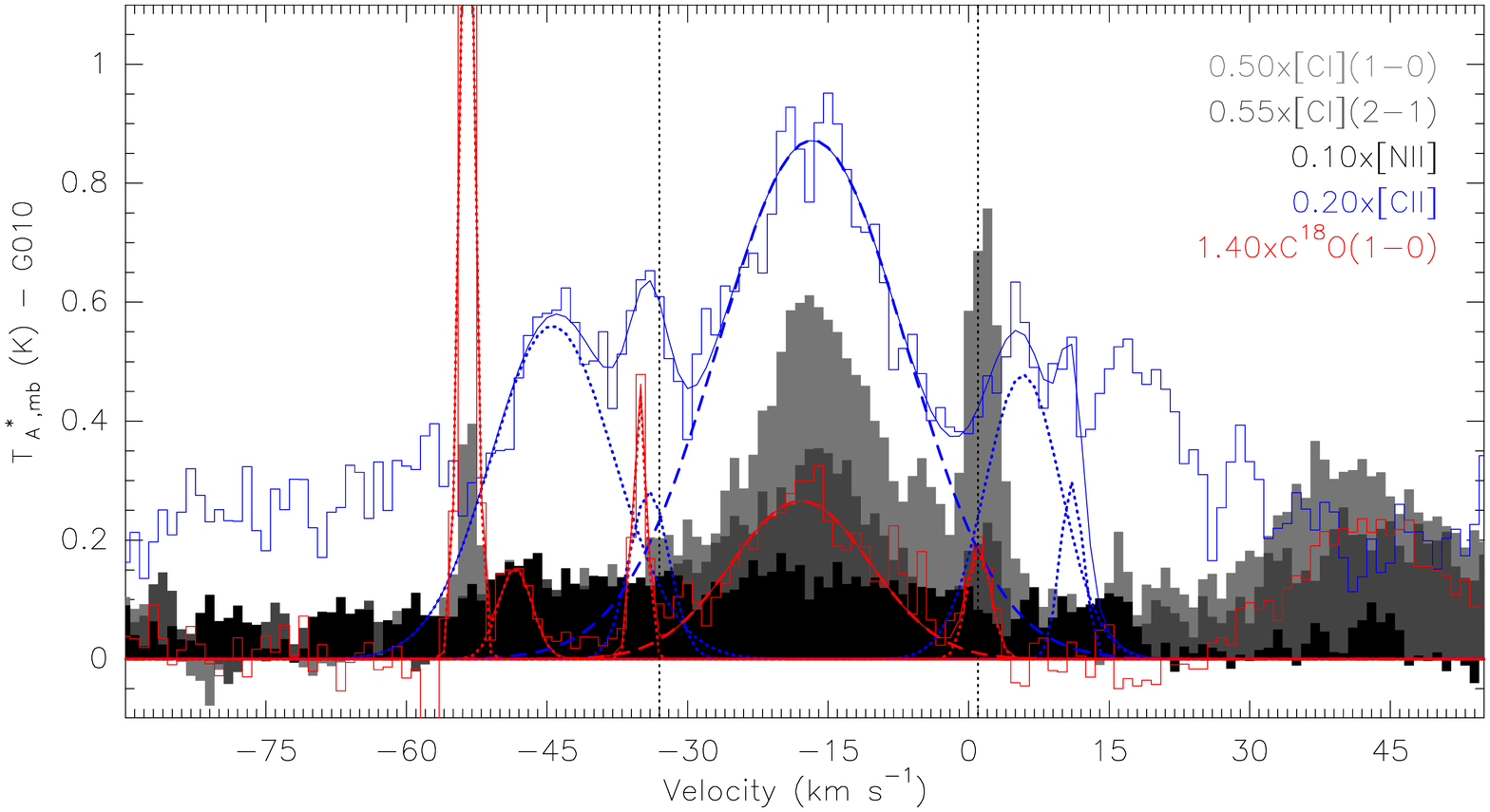}
\includegraphics[angle=-0, width=\hsize]{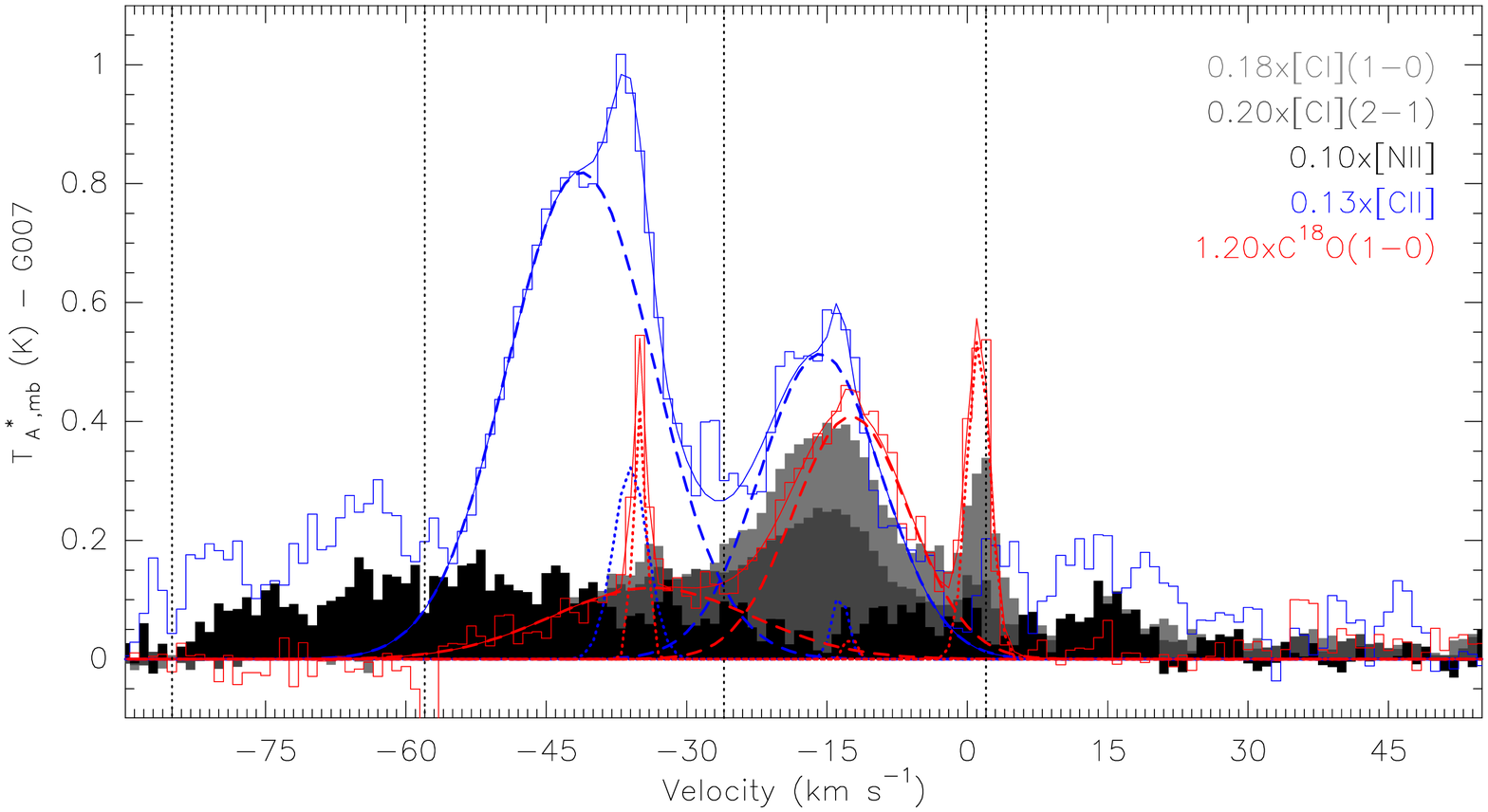}
\end{minipage}
\caption{Comparison of \CII\textrm{ }(open blue), \NII\textrm{ }(filled black), \CILINEAII\textrm{ }(filled dark gray), \CILINEAI\textrm{ }(filled gray), and \COOLINEAI\textrm{ }(open red) spectra for positions G0.10$+$0.02 and G0.07$+$0.04 in Table \ref{arches:tab_positions}. Antenna temperatures have been scaled by the factor shown next to the line's name for display purposes. Gaussian component fits to the \CII\textrm{ }and \COOLINEAI\textrm{ }spectra are shown in blue and red, respectively. Only dashed-line Gaussian components in the \CII\textrm{ }spectra are used for the analysis in this work as only for them can a counterpart in the \COOLINEAI\textrm{ }optically thin emission be identified. Vertical dotted black lines show the radial velocity limits for which integrated intensities were calculated for the \NII\textrm{ }line by directly adding the observed emission, multiplied by the spectral channel width, to that LSR velocity range. \label{appA:spectra3}
}
\end{figure*}

\begin{figure*}
\centering
\begin{minipage}{\hsize}
\centering
\includegraphics[angle=-0, width=\hsize]{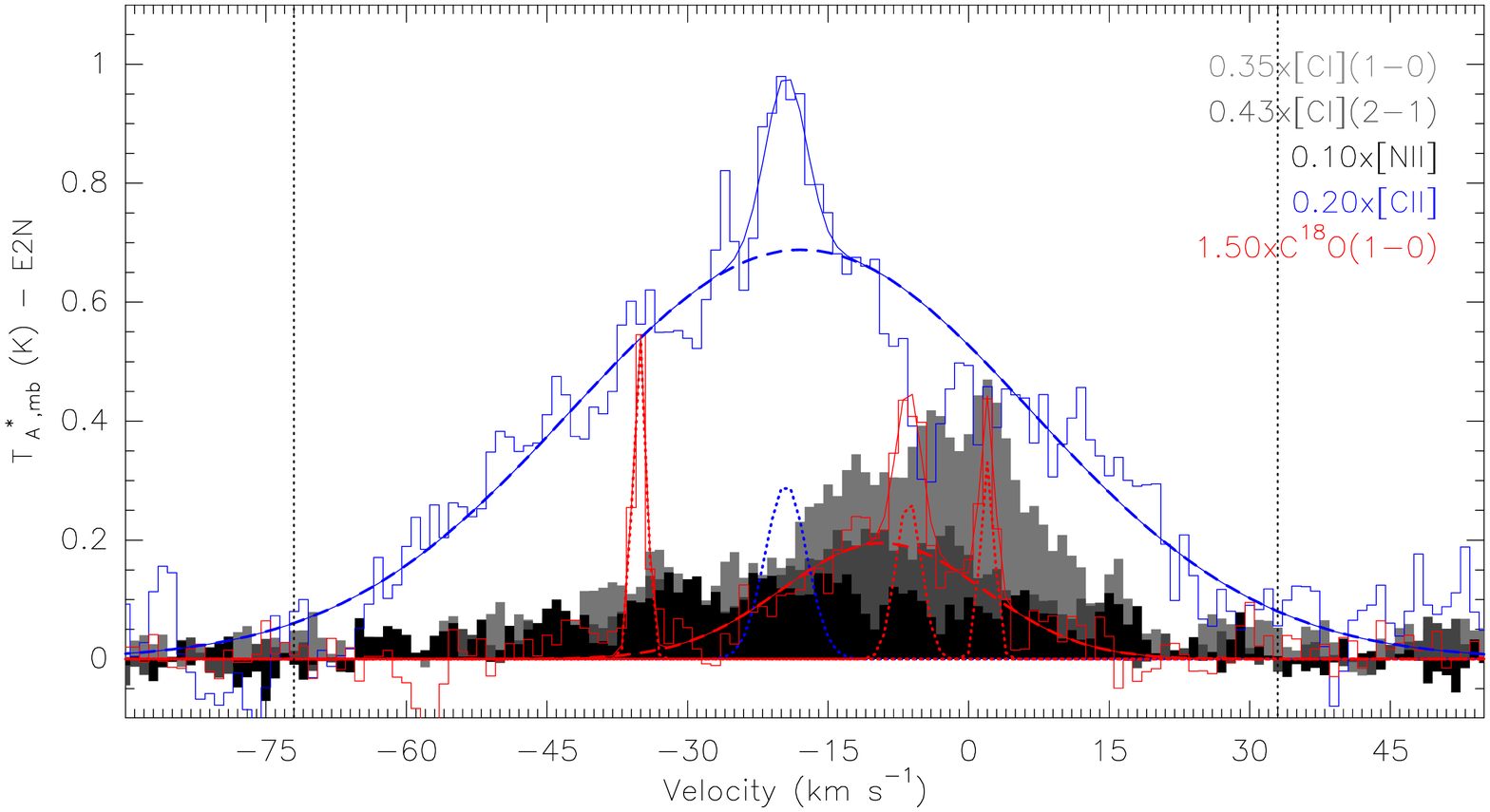}
\includegraphics[angle=-0, width=\hsize]{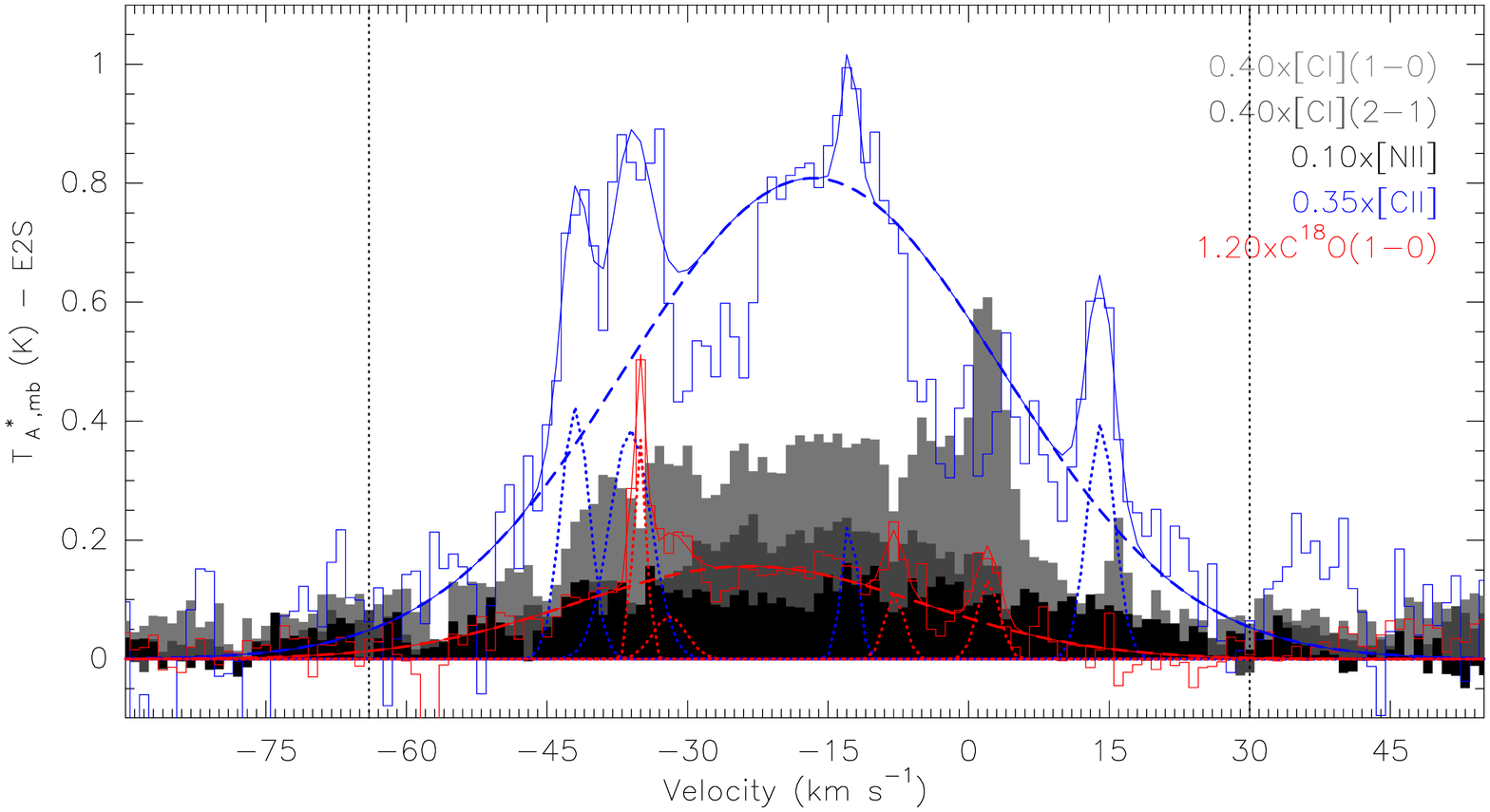}
\end{minipage}
\caption{Comparison of \CII\textrm{ }(open blue), \NII\textrm{ }(filled black), \CILINEAII\textrm{ }(filled dark gray), \CILINEAI\textrm{ }(filled gray), and \COOLINEAI\textrm{ }(open red) spectra for positions E2N and E2S in Table \ref{arches:tab_positions}. Antenna temperatures have been scaled by the factor shown next to the line's name for display purposes. Gaussian component fits to the \CII\textrm{ }and \COOLINEAI\textrm{ }spectra are shown in blue and red, respectively. Only dashed-line Gaussian components in the \CII\textrm{ }spectra are used for the analysis in this work as only for them can a counterpart in the \COOLINEAI\textrm{ }optically thin emission be identified. Vertical dotted black lines show the radial velocity limits for which integrated intensities were calculated for the \NII\textrm{ }line by directly adding the observed emission, multiplied by the spectral channel width, to that LSR velocity range. \label{appA:spectra4}
}
\end{figure*}

\begin{figure*}
\centering
\begin{minipage}{\hsize}
\centering
\includegraphics[angle=-0, width=\hsize]{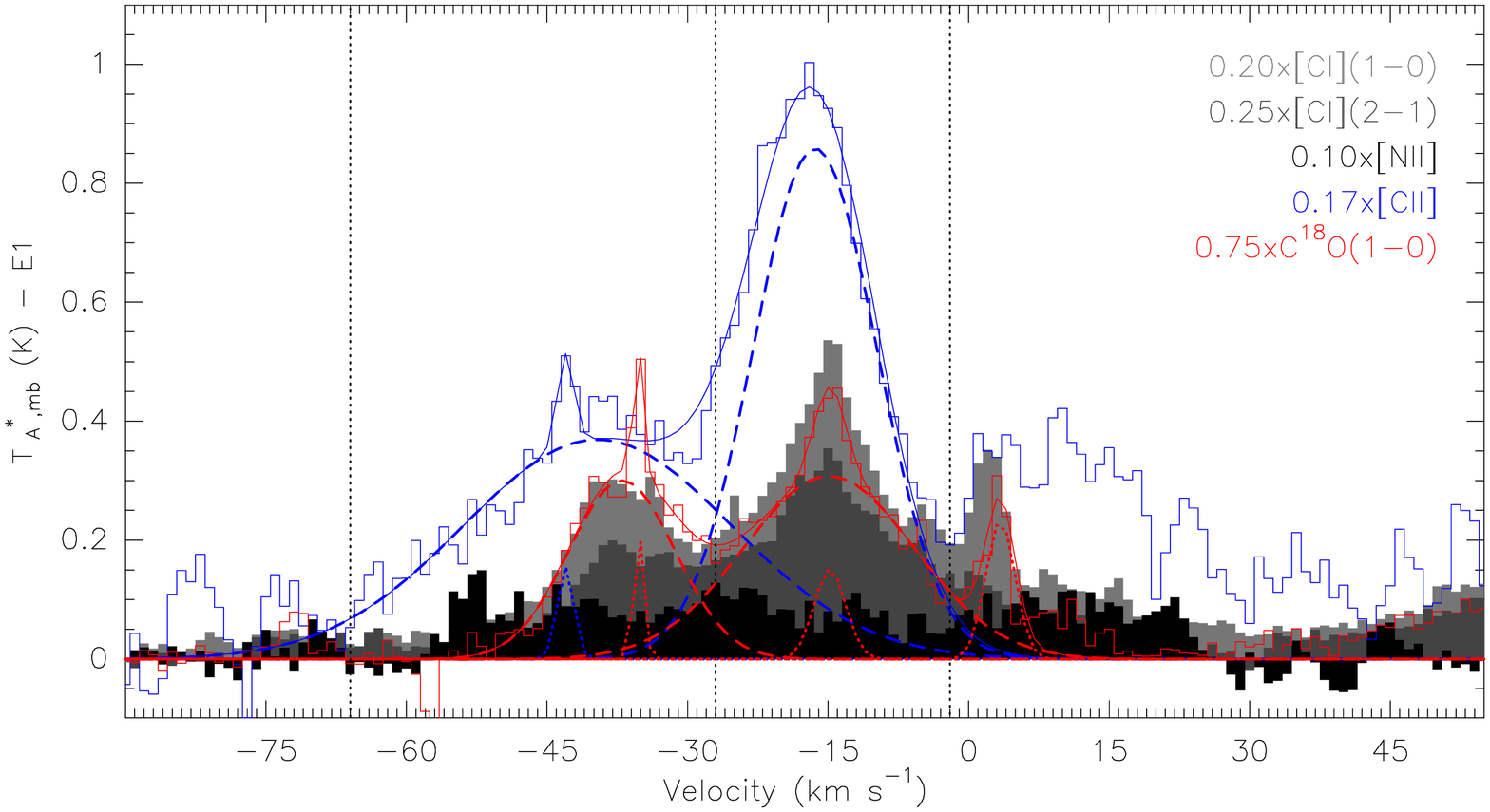}
\end{minipage}
\caption{Comparison of \CII\textrm{ }(open blue), \NII\textrm{ }(filled black), \CILINEAII\textrm{ }(filled dark gray), \CILINEAI\textrm{ }(filled gray), and \COOLINEAI\textrm{ }(open red) spectra for position E1 in Table \ref{arches:tab_positions}. Antenna temperatures have been scaled by the factor shown next to the line's name for display purposes. Gaussian component fits to the \CII\textrm{ }and \COOLINEAI\textrm{ }spectra are shown in blue and red, respectively. Only dashed-line Gaussian components in the \CII\textrm{ }spectra are used for the analysis in this work as only for them can a counterpart in the \COOLINEAI\textrm{ }optically thin emission be identified. Vertical dotted black lines show the radial velocity limits for which integrated intensities were calculated for the \NII\textrm{ }line by directly adding the observed emission, multiplied by the spectral channel width, to that LSR velocity range. \label{appA:spectra5}
}
\end{figure*}

\end{appendix}

\end{document}